\documentclass[fleqn,usenatbib]{mnras}

\usepackage{newtxtext,newtxmath}
\usepackage[T1]{fontenc}
\usepackage{ae,aecompl}

\usepackage{graphicx}	% Including figure files
\usepackage{amsmath}	% Advanced maths commands
\usepackage{comment}
\usepackage[dvipsnames]{xcolor}
\usepackage{url}
\usepackage{subcaption}
\usepackage{natbib}
\usepackage{threeparttablex}

\graphicspath{{figs/}}

\newcommand{\ha}{H$\upalpha$}
\newcommand{\hb}{H$\upbeta$}
\newcommand{\hg}{H$\upgamma$}
\newcommand{\hd}{H$\updelta$}
\newcommand{\he}{H$\upvarepsilon$}
\newcommand{\hi}{H\,\textsc{i}}
\newcommand{\hii}{H\,\textsc{ii}}

\newcommand{\hh}{H\textsubscript{2}}

\newcommand{\ppxf}{\textsc{ppxf}}
\newcommand{\logoh}{\ensuremath{\log\left( \rm O / H \right) + 12}}
\newcommand{\logu}{\ensuremath{\log\left(U\right)}}
\newcommand{\forb}[2]{[#1\,\textsc{#2}]}

\title[The ISM conditions in Minkowski's Object]{Unravelling the enigmatic ISM conditions in Minkowski's Object}

\author[H. R. M. Zovaro et al.]{Henry R. M. Zovaro$^{1}$\thanks{E-mail: henry.zovaro@anu.edu.au},
Robert Sharp$^{1}$,
Nicole P. H. Nesvadba$^{2}$,
Lisa Kewley$^{1,3}$,
\newauthor
Ralph Sutherland$^{1}$,
Philip Taylor$^{1,3}$,
Brent Groves$^{4}$,
Alexander Y. Wagner$^{5}$,
\newauthor
Dipanjan Mukherjee$^{6}$,
Geoffrey V. Bicknell$^{1}$
\\
$^{1}$Research School of Astronomy and Astrophysics, The Australian National University, Canberra, ACT 2611, Australia\\
$^{2}$Laboratoire Lagrange, Boulevard de l'Observatoire, CS 34229, 06304 Nice Cedex 4, France\\
$^{3}$ARC Centre of Excellence for All Sky Astrophysics in 3 Dimensions (ASTRO 3D)\\
$^{4}$International Centre for Radio Astronomy Research, University of Western Australia, 35 Stirling Highway, Crawley WA 6009, Australia\\
$^{5}$University of Tsukuba, Center for Computational Sciences, Tennodai 1-1-1, 305-0006, Tsukuba, Ibaraki, Japan\\
$^{6}$The Inter-University Centre for Astronomy and Astrophysics, Post Bag 4, Ganeshkhind, Pune, Maharashtra 411007, India\\	
}

\date{Accepted 2020 October 2. Received 2020 October 2; in original form 2020 August 4.}

\pubyear{2020}

\begin{document}
\label{firstpage}
\pagerange{\pageref{firstpage}--\pageref{lastpage}}
\maketitle

% This is a simple template for authors to write new MNRAS papers.
% The abstract should briefly describe the aims, methods, and main results of the paper.
% It should be a single paragraph not more than 250 words (200 words for Letters).
% No references should appear in the abstract.
\begin{abstract}
Local examples of jet-induced star formation lend valuable insight into its significance in galaxy evolution and can provide important observational constraints for theoretical models of positive feedback.
Using optical integral field spectroscopy, we present an analysis of the ISM conditions in Minkowski's Object ($z = 0.0189$), a peculiar star-forming dwarf galaxy located in the path of a radio jet from the galaxy NGC\,541. 
Full spectral fitting with \ppxf{} indicates that Minkowski's Object primarily consists of a young stellar population $\sim 10\,\rm Myr$ old, confirming that the bulk of the object's stellar mass formed during a recent jet interaction. 
Minkowski's Object exhibits line ratios largely consistent with star formation, although there is evidence for a low level ($\lesssim 15 \,\rm per \, cent$) of contamination from a non-stellar ionising source.
Strong-line diagnostics reveal a significant variation in the gas-phase metallicity within the object, with \logoh{} varying by $\sim 0.5\,\rm dex$, which cannot be explained by in-situ star formation, an enriched outflow from the jet, or enrichment of gas in the stellar bridge between NGC\,541 and NGC\,545/547.
We hypothesise that Minkowski's Object either (a) was formed as a result of jet-induced star formation in pre-existing gas clumps in the stellar bridge, or (b) is a gas-rich dwarf galaxy that is experiencing an elevation in its star formation rate due to a jet interaction, and will eventually redden and fade, becoming an ultra-diffuse galaxy as it is processed by the cluster. 
\end{abstract}

%%%%%%%%%%%%%%%%%%%%%%%%%%%%%%%%%%%%%%%%%%%%%%%%%%

%%%%%%%%%%%%%%%%% BODY OF PAPER %%%%%%%%%%%%%%%%%%
\defcitealias{Croft2006}{C06}
\defcitealias{Lacy2017}{L17}

\section{Introduction}

It has long been established that jets play a vital role in curbing star formation, by preventing cooling flows in cluster environments \citep{McNamaraNulsen2012}, by driving powerful outflows \citep{Morganti2005,Holt2008,Schulz2018} and by rendering the interstellar medium (ISM) of their host galaxies turbulent \citep{Nesvadba2010,Nesvadba2011,Ogle2010}.
In comparison, the phenomenon of \textit{positive feedback}, wherein jets enhance or trigger star formation, remains poorly understood, although it may make an important contribution to the star formation histories of galaxies across cosmic time \citep{Silk2005,Gaibler2012}.

The observed alignment of jets with stellar components and enhanced emission from dense gas is strong evidence of positive feedback.
Many radio galaxies at high redshifts exhibit extended stellar continua that are aligned with the axis of the radio jet, referred to as the `alignment effect' \citep{Chambers1987,McCarthy1987,Best1996}. Similar alignments have been observed in cluster environments at low redshift \citep{McNamara&O'Connell1993}.
More recent studies have also revealed alignment of molecular gas and radio jets in galaxies at high redshift \citep{Klamer2004,Emonts2014}. 
Hydrodynamical simulations have shown that shocks generated by the passage of a jet through an ionised medium may induce runaway cooling, leading to the formation of both atomic and molecular gas, which may collapse to form stars \citep{Fragile2004,Fragile2017}. 
However, the detailed workings of positive feedback remain elusive, largely due to the dearth of local examples of jet-induced star formation, severely limiting our ability to study this complex phenomenon with sufficient angular resolution to compare observations to detailed hydrodynamical simulations.

% Brief intro to MO
Minkowski's Object \citep[][hereafter MO]{Minkowski1958} is one of only three known examples of jet-induced star formation in the local Universe, the others being the string of OB associations in the jet of Centaurus A \citep{Blanco1975,Santoro2015,Santoro2016,Salome2016a,Salome2016b,Salome2017} and a star-forming region in the jet of 3C\,285 \citep{Salome2015}. 
Located in the cluster Abell\,194, MO is a star-forming dwarf galaxy located along the path of a jet from the nearby elliptical galaxy NGC\,541.
Due to its low redshift \citep[$z = 0.0189$,][hereafter C06]{Croft2006}, MO presents a critical opportunity to study an object formed by positive feedback in high resolution.

Whilst past studies of MO have focused on broad-band photometry or slit spectroscopy, this work utilises integral field spectroscopy, enabling us to spatially resolve the ISM conditions in this object for the first time.
% Roadmap
In this paper, we present a study of the stellar and gas properties of MO using optical integral field spectroscopy.
In Section~\ref{sec: paper 3: Minkowski's Object}, we summarise the properties of MO, and in Section~\ref{sec: paper 3: Properties of NGC 541 and Abell 194} we give an overview of NGC\,541 and the cluster environment of MO. 
We present our observations and data reduction in Section~\ref{sec: paper 3: Observations and data reduction}, before detailing the stellar population analysis in Section~\ref{sec: paper 3: Stellar population analysis} and our analysis of the emission-line gas in Section~\ref{sec: paper 3: Emission line gas}. We then consider the possible causes of the observed metallicity variation in MO in Section~\ref{sec: paper 3: The cause of the metallicity variation}, before discussing the origin of MO in Section~\ref{sec: paper 3: On the formation of Minkowski's Object} and summarising our findings in Section~\ref{sec: paper 3: Conclusion}.
For the remainder of this paper, we assume a cosmology with $H_0 = 70\rm\,km\,s^{-1}\,Mpc^{-1}$, $\Omega_M = 0.3$, and $\Omega_\Lambda = 0.7$.

%%%%%%%%%%%%%%%%%%%%%%%%%%%%%%%%%%%%%%%%%%%%%%%%%%
\section{Minkowski's Object}\label{sec: paper 3: Minkowski's Object}

\begin{figure*}
	\centering
	\includegraphics[width=0.75\linewidth]{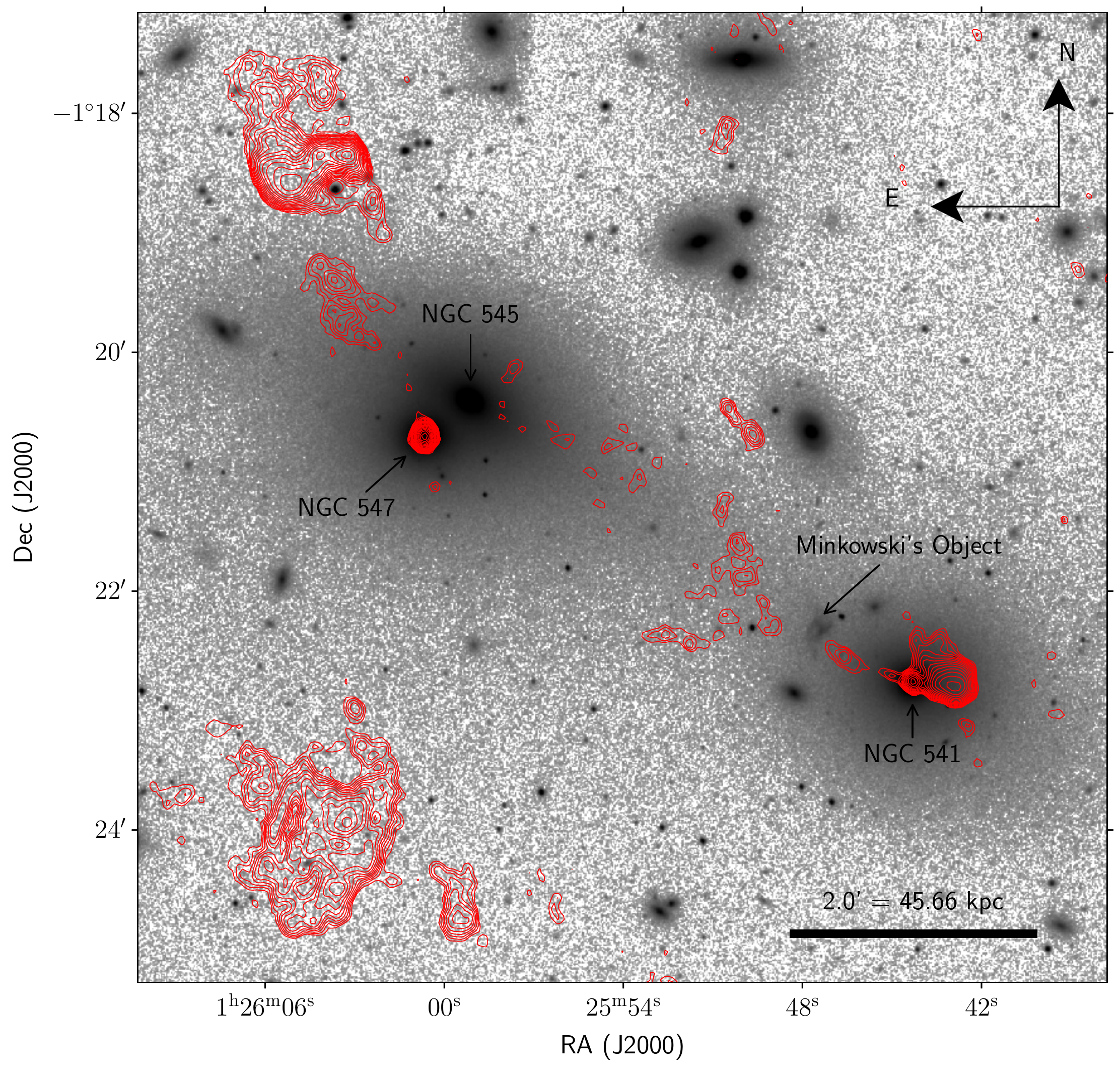}
	\caption{
			Sloane Digital Sky Survey (SDSS) $i$ band image of Abell 194, showing the relative positions of NGC\,541, NGC\,545, NGC\,547 and Minkowski's Object. The colour scale of the image has been adjusted to highlight the faint stellar bridge extending from NGC\,541 to NGC\,545/547. The red contours show the 1.4\,GHz continuum flux from VLA/FIRST~\citep{Becker1995} indicating the radio jets from NGC\,541 and NGC\,547. The contours represent 30 logarithmically spaced intervals from $3\sigma \,\rm Jy \, beam^{-1}$ to $0.1 \,\rm Jy \, beam^{-1}$, where $\sigma = 1.4 \times 10^{-4} \, \rm Jy \, beam^{-1}$ is the measured rms noise in the image.
	}
	\label{fig: paper 3: Abell 194}
\end{figure*}

MO is a peculiar dwarf galaxy located in the cluster Abell\,194, with properties summarised in Table~\ref{tab: paper 3: MO & NGC541 properties}. MO lies in the path of an FR\,I-type~\citep{Fanaroff&Riley1974} jet from the elliptical galaxy NGC\,541~\citep[$z=0.01809$,][]{Smith2000} which lies $18~\rm kpc$ to the South-West, as shown in Fig.~\ref{fig: paper 3: Abell 194}.

\begin{figure*}
	\centering
	\includegraphics[width=0.7\linewidth]{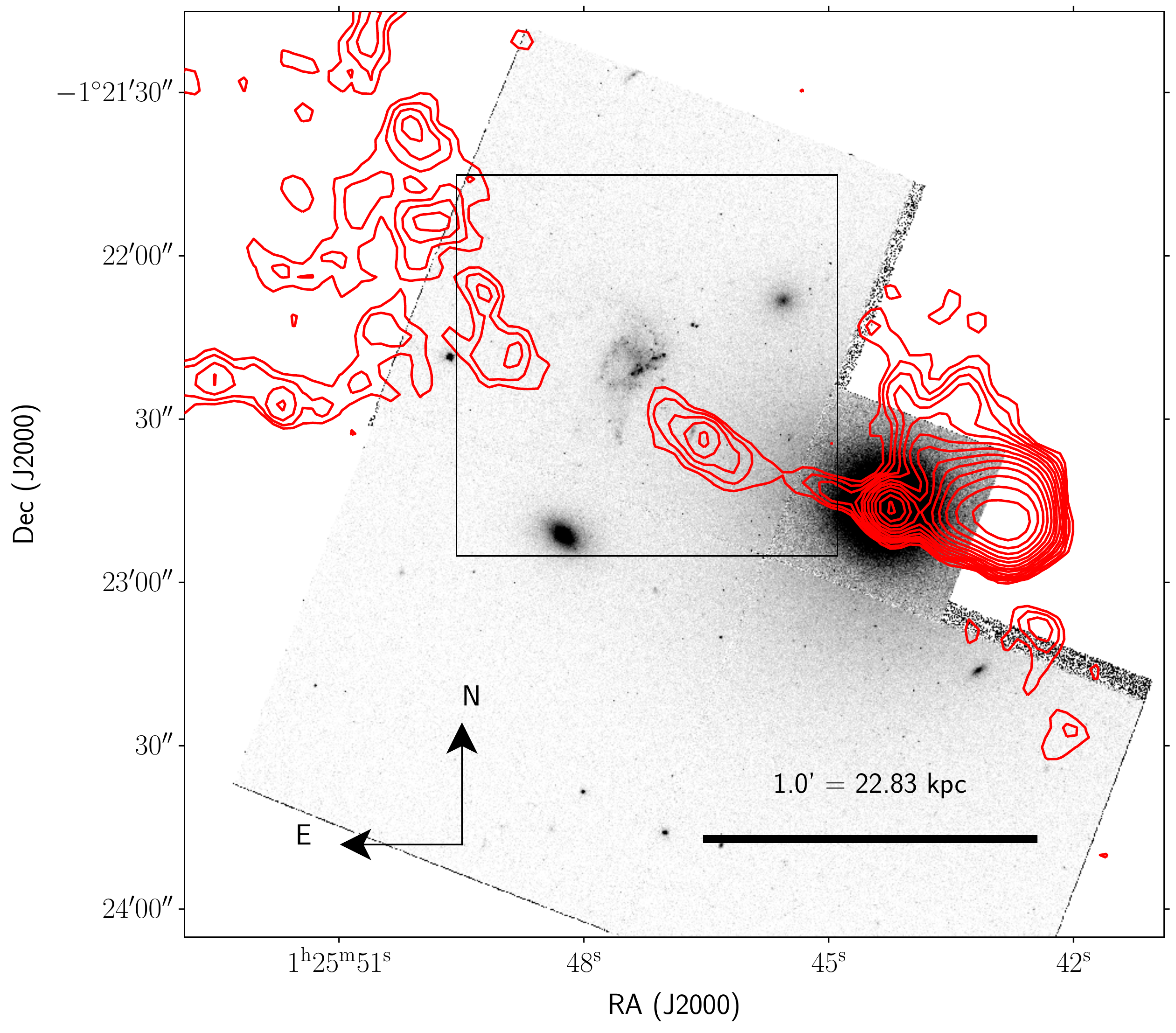}
	\includegraphics[width=0.7\linewidth]{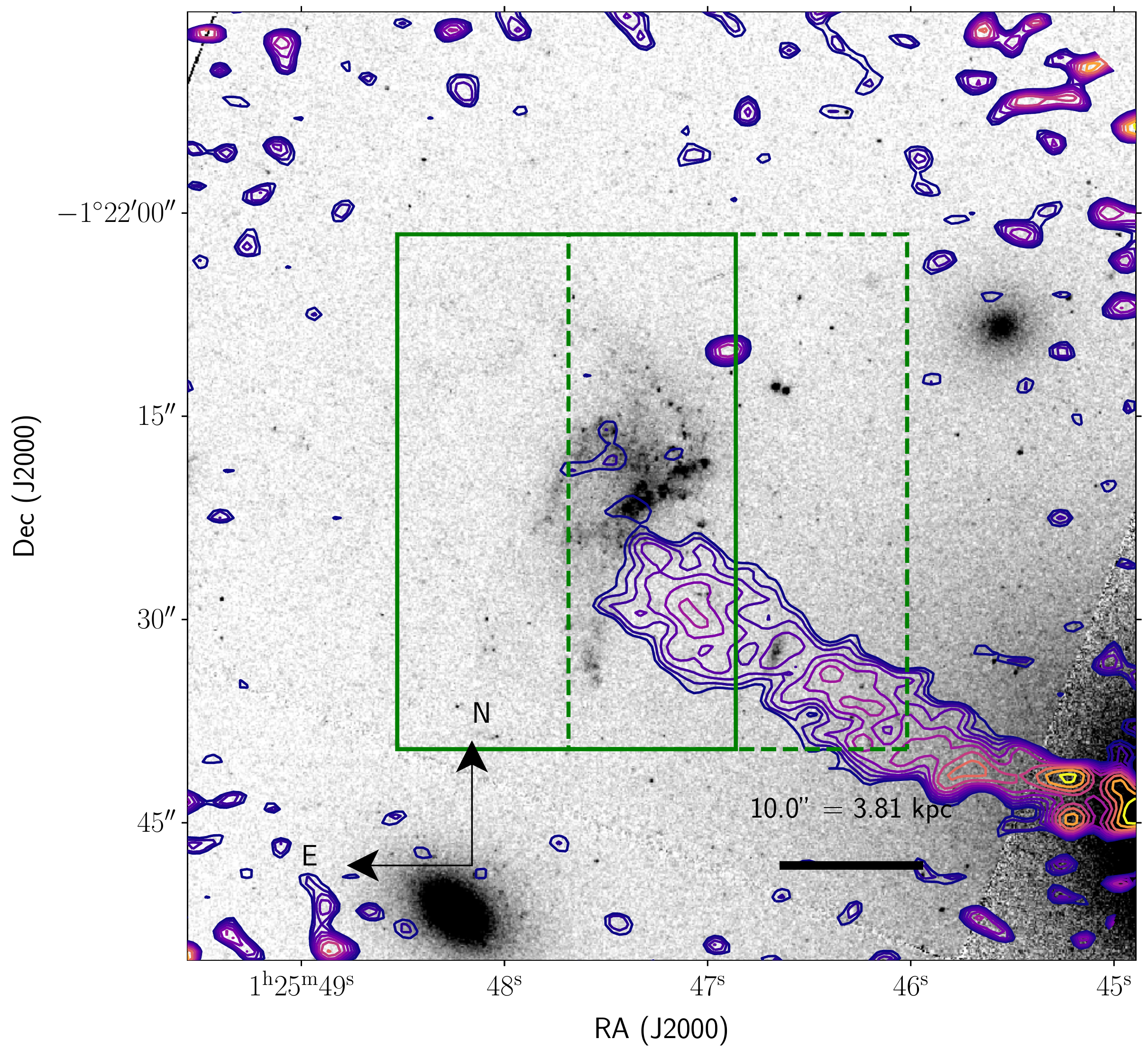}
	\caption{
		Top: archival \textit{HST} WFPC2 F555W image~\citep{VerdoesKleijn1999}, overlaid with contours showing the 1.4\,GHz continuum flux from VLA/FIRST indicating the path of the radio jet from NGC\,541. 
		The contours represent 20 logarithmically spaced intervals from $3\sigma \,\rm Jy \, beam^{-1}$ to $1\times 10^{-1} \,\rm Jy \, beam^{-1}$, where $\sigma = 1.4 \times 10^{-4} \, \rm Jy \, beam^{-1}$ is the measured rms noise in the image.
		Bottom: a magnified view of the region outlined by the black square in the top panel, overlaid with contours showing the ALMA 106\,GHz continuum from the jet \citepalias{Lacy2017}. The contours represent 10 logarithmically spaced intervals from $3\sigma \,\rm Jy \, beam^{-1}$ to $1.6 \times 10^{-4} \,\rm Jy \, beam^{-1}$, where $\sigma = 9.2 \times 10^{-6} \, \rm Jy \, beam^{-1}$ is the measured rms noise in the image. The green solid and dashed boxes indicate the two base pointings of our WiFeS integral field spectroscopy observations.
	}
	\label{fig: paper 3: HST overlay}
\end{figure*}

As can be seen in Fig.~\ref{fig: paper 3: HST overlay}, MO has an irregular and filamentary structure, with the brightest clumps forming a `bar' approximately perpendicular to the jet axis. 
\citetalias{Croft2006} estimated MO to have a stellar mass $M_* = 1.9 \times 10^{7} \,\rm M_\odot$, with a broad-band spectral energy distribution (SED) consistent with a stellar population resulting from a single starburst 7.5~Myr ago with a secondary population of stars older than 1~Gyr.

MO is embedded in a cloud of neutral gas with an estimated mass $M_{\rm H\,\textsc{i}} = 4.9 \times 10^{8} \,\rm M_\odot$ \citepalias{Croft2006} which is located along the path of the jet, extending downstream of MO (see their fig. 3).
Molecular gas has been detected in four distinct clumps with intervening diffuse emission \citep[][their fig. 2; hereafter L17]{Lacy2017} with a total mass $M_{\rm H_2} = 3.0 \times 10^7 \rm M_\odot$ assuming a standard Milky Way (MW) CO-to-\hh{} conversion factor $\alpha_{\rm CO}$, or $M_{\rm H_2} = 1.8 \times 10^8 \rm M_\odot$ assuming an $\alpha_{\rm CO}$ appropriate for a galaxy with half-solar gas-phase metallicity.
% Kinematics 
Both the \hi{} and molecular gas exhibit steep velocity gradients perpendicular to the jet axis, which \citetalias{Lacy2017} attribute to a jet interaction. The velocity dispersion of the CO ($\lesssim 10~\rm km~s^{-1}$) is much lower than that of the \hi{} in the same region ($25~\rm km~s^{-1}$); this may be caused by shocks generated by the jet passing through an inhomogeneous medium \citepalias{Lacy2017}.

\begin{table*}
	\begin{threeparttable}
		\centering
		\caption{Properties of MO and NGC\,541.}
		\begin{tabular}{cccc}
			\hline
			     \textbf{Property}       &     \textbf{Symbol}     &                           \textbf{MO}                            &                     \textbf{NGC\,541}                     \\ \hline
			        Stellar mass         &          $M_*$          &        $1.9 \times 10^7 \,\rm M_\odot$\textsuperscript{\textit{a}}         &   $4.7 \times 10^{11} \,\rm M_\odot$\textsuperscript{\textit{d}}  \\
			         \hi{} mass          & $M_{\rm H\,\textsc{i}}$ &         $4.9 \times 10^8 \rm\, M_\odot$\textsuperscript{\textit{a}}         &                            ---                            \\
			     Molecular gas mass      &      $M_{\rm H_2}$      &          $3.0 \times 10^7 \rm\,  M_\odot$\textsuperscript{\textit{b},\ensuremath{\dagger}}          &    $1.7 \times 10^8 \,\rm M_\odot$\textsuperscript{\textit{d},\ensuremath{\dagger}}    \\
			     & &  $1.8 \times 10^8 \rm \, M_\odot$\textsuperscript{\textit{b},\ensuremath{\ddagger}} & \\
			    Star formation rate      &        $\rm SFR$        & $(0.24 - 0.28) \pm 0.09 \,\rm M_\odot \, yr^{-1}$\textsuperscript{\textit{c}} &    $0.095 \,\rm M_\odot \, yr^{-1}$\textsuperscript{\textit{d}}    \\
			Specific star formation rate &       $\rm sSFR$        &             $13 - 15 \rm \, Gyr^{-1}$\textsuperscript{\textit{c}}          &        $0.0002 \,\rm Gyr^{-1}$\textsuperscript{\textit{d}}         \\
			       Depletion time (atomic gas)       &      $t_{\rm dep, \, H\,I}$      &             $1.7 - 2.0 \,\rm Gyr$\textsuperscript{\textit{c},\S}             & --- \\
			       Depletion time (molecular gas)     &      $t_{\rm dep, \,H_2}$      &            $0.11 - 0.12 \,\rm Gyr$\textsuperscript{\textit{c},\ensuremath{\dagger}}             & $1.79 \,\rm Gyr$\textsuperscript{\textit{d},\ensuremath{\dagger}}  \\
																	&      &                 $0.64 - 0.75 \,\rm Gyr$\textsuperscript{\textit{c},\ensuremath{\ddagger}}             &  \\ \hline
		\end{tabular}
		\begin{tablenotes}
			\footnotesize
			\item References: $^a$\citetalias{Croft2006}; $^b$\citetalias{Lacy2017}; $^c$This work; $^d$\citet{Salome2015}; \textsuperscript{\S}Based on \hi{} mass; \textsuperscript{\ensuremath{\dagger}}assuming a MW $\alpha_{\rm CO}$; \textsuperscript{\ensuremath{\ddagger}}assuming a half-solar metallicity $\alpha_{\rm CO}$.
		\end{tablenotes} 
		\label{tab: paper 3: MO & NGC541 properties}
	\end{threeparttable}
\end{table*}

%%%%%%%%%%%%%%%%%%%%%%%%%%%%%%%%%%%%%%%%%%%%%%%%%%
\section{Abell\,194 and NGC\,541}\label{sec: paper 3: Properties of NGC 541 and Abell 194}

Abell\,194 is a poor \citep{Abell1989} and cold \citep{Sakelliou2008} cluster with a linear morphology at $z = 0.018$~\citep{Struble1999}. It hosts two powerful radio galaxies close to the cluster centre, NGC\,541 and NGC\,547, which is part of the interacting pair NGC\,545/547~\citep[$z = 0.01781$ and $z = 0.01852$ respectively;][]{Smith2000}. 
The Southern jet from NGC\,547 is associated with an X-ray cavity \citep{Bogdan2011}.
As shown in Fig.~\ref{fig: paper 3: Abell 194}, NGC\,541 and NGC\,545/547 are separated by approximately 100\,kpc in projection, and are connected by a faint stellar bridge, a remnant of a previous or ongoing interaction. Both radio galaxies host bent jets, which \citet{Sakelliou2008} attribute to the motion of the galaxies through the cluster. Abell\,194 may be undergoing a significant merger event, in which both NGC\,541 and NGC\,545 are falling towards NGC\,547 \citep{Bogdan2011}. 

NGC\,541 is a massive elliptical galaxy, and is the brightest cluster galaxy (BCG) in Abell\,194. It is extremely gas-poor, with a gas fraction of $0.4$\,per\,cent and a star formation rate (SFR) of $0.095 \,\rm M_\odot \, yr^{-1}$ \citep{Salome2015}. NGC\,541 may harbour a disc or ring of molecular gas \citep{Salome2015}, possibly a remnant of a previous interaction.
The properties of NGC\,541 are summarised in Table~\ref{tab: paper 3: MO & NGC541 properties}.

\subsection{Properties of the radio jet}
NGC\,541 hosts an asymmetric FR\,I radio source with $P_{1.4~\textrm{GHz}} = 5.9 \times 10^{23}~\rm W~Hz^{-1}$ \citep{vanBreugel1985} which can be seen in Fig.~\ref{fig: paper 3: HST overlay}. The Eastern jet is aligned with the stellar bridge, and gradually broadens along its path from the host galaxy until it interacts with MO, at which point it decollimates. The Western jet is less extended and is sharply bent a short distance from the nucleus.

Using the ALMA 106\,GHz continuum image of \citetalias{Lacy2017}, we used the minimum energy method, as detailed in section 5.1 of \citet{Zovaro2019a}, to estimate the power of the Eastern jet. The jet volume was approximated as a truncated cone, with length 21\,kpc and end surface radii of 1.5\,kpc and 6\,kpc, and was assumed to have a constant flux density of $30 \,\rm \upmu Jy \, beam^{-1}$, with a spectral index $\alpha = 0.8$. A jet age of $10^7\,\rm yr$ was adopted, chosen to be consistent with the age of the stellar population of MO~\citepalias{Croft2006}. Lower and upper Lorentz factors for the electrons were fixed to $10^2$ and $10^5$ respectively. These parameters yield a jet power of approximately $1.2 \times 10^{43}\,\rm erg \, s^{-1}$.

%%%%%%%%%%%%%%%%%%%%%%%%%%%%%%%%%%%%%%%%%%%%%%%%%%
\section{Observations and data reduction}\label{sec: paper 3: Observations and data reduction}
We observed MO with the Wide-Field Spectrograph \citep[WiFeS; ][]{Dopita2007,Dopita2010} on the Australian National University 2.3\,m telescope at Siding Spring Observatory, NSW, Australia.
WiFeS is an image slicer integral field spectrograph comprising 25 1''-width slices with 0.5'' sampling along the image, binned to 1'' to better match the seeing, providing 1''\,$\times$\,1''  spaxels over a 25''\,$\times$ 38'' field-of-view. 

MO was observed on the 29th of November and 1st of December 2016 (PI Banfield, proposal ID 4160092) using the low-resolution B3000 (3200\,\AA--5000\,\AA, $R \sim 3000$, $\Delta v \approx 100 \,\rm km\,s^{-1}$) and the high-resolution R7000 (5290\,\AA--7060\,\AA, $R \sim 7000$, $\Delta v \approx 40 \,\rm km\,s^{-1}$) gratings with the RT560 beam splitter, using 1200\,s exposures. Due to the large angular extent of MO, the observations were dithered in an E-W direction using an offset of 12'', as shown in the bottom panel of Fig.~\ref{fig: paper 3: HST overlay}. 
The total effective exposure time was 9600\,s. 
The stars HD26169, HD16031, HD44007 and HD9051 were used as flux and telluric standards \citep{Bessell1999}.

%%%%%%%%%%%%%%%%%%%%%%%%%%%%%%%%%%%%%%%%%%%%%%%%%%
\subsection{Data reduction}
The observations were reduced in the standard way using \textsc{Pywifes}, the  data reduction pipeline for WiFeS~\citep{Childress2014}.
Cu-Ar and Ne-Ar arc lamp exposures were used to derive the wavelength solution, and exposures of the coronagraphic wire mask were used to calibrate the spatial alignment of the slits.
Quartz lamp and twilight flat exposures were used to correct for wavelength and spatial variations in the instrument response respectively.
Exposures of standard stars were used to correct for telluric absorption and for flux calibration.
Two data cubes were generated for each exposure, corresponding to the blue (B3000) and red (R7000) arms of the spectrograph respectively.

To maximise time on-source, we did not take sky frames during our observing campaign. Regions within the field-of-view with no source signal were instead used to extract the sky spectrum which was then subtracted from the data cube. Sky subtraction residuals were minimised by scaling the flux of the sky spectrum to the measured intensity of a subset of sky lines in each spaxel. 

The variety in atmospheric transmission and seeing conditions during our observing campaign led to subtle multiplicative offsets in the fluxes amongst individual exposures. Optimal scaling factors for each cube were computed using a least-squares minimisation technique between overlapping regions of the individual data cubes.

Mosaics were created by spatially shifting the scaled data cubes and combining them by taking the sigma-clipped median of each pixel. $1\sigma$ errors for each pixel value in the final mosaic were derived from the standard deviation of the pixels in the individual cubes.

\begin{figure*}
	\centering
	\includegraphics[width=1\linewidth]{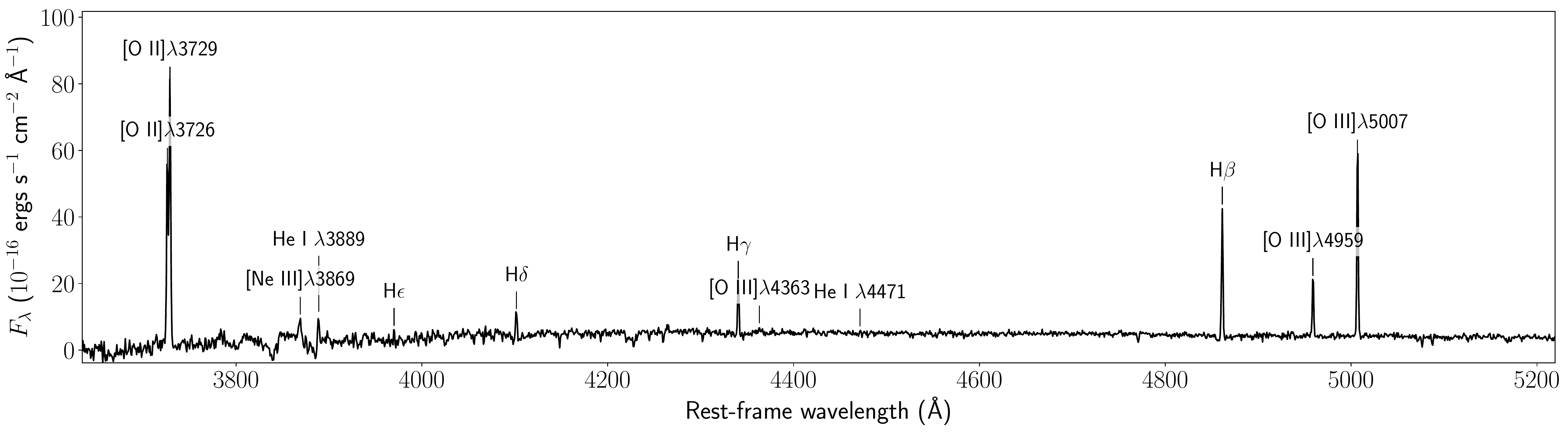}
	\includegraphics[width=1\linewidth]{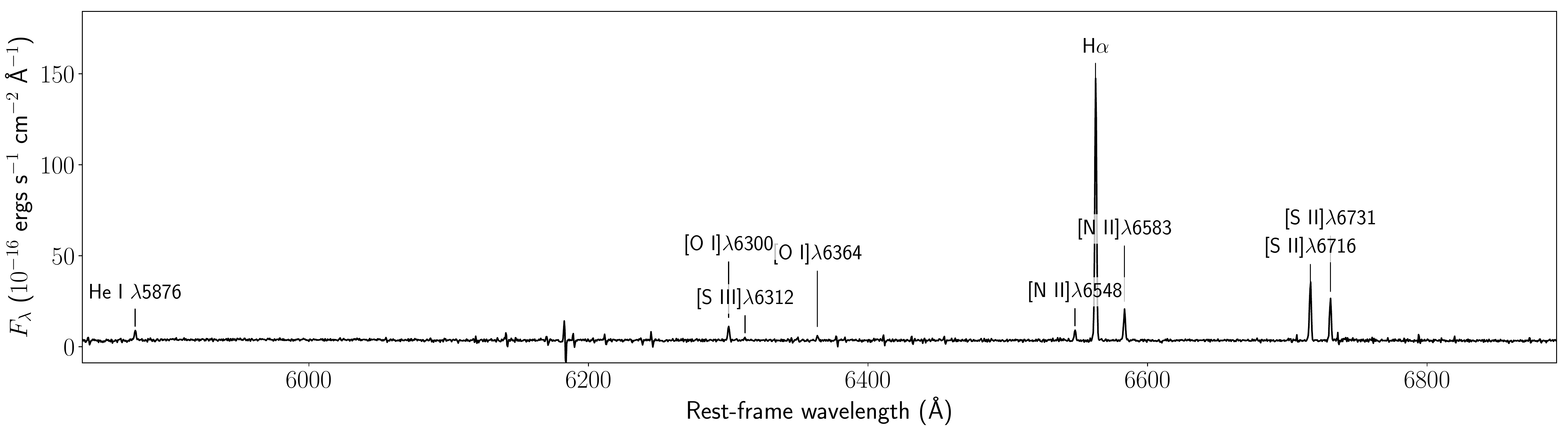}
	\caption{Integrated spectra (including the stellar continuum) extracted from spaxels in the WiFeS B3000 (top) and R7000 (bottom) data cubes in which the \ha{} line has S/N $>$ 3 and a flux greater than $10 \times 10^{-16} \,\rm erg \, s^{-1} \, cm^{-2}$.}
	\label{fig: paper 3: WiFeS spectra}
\end{figure*}

Finally, the data cubes were corrected for Galactic extinction using the extinction map of \citet{Schlafly&Finkbeiner2011} which gives $A_V = 0.1198$. To apply the correction to our WiFeS data we used reddening curve of \citet{Fitzpatrick&Massa2007} with $R_V = 3.1$. 
% Data quality
%The mean signal-to-noise ratio (S/N) in the continuum of the spectrum integrated over the full field-of-view was approximately $10$ in both the B3000 and R7000 data cubes.
Fig.~\ref{fig: paper 3: WiFeS spectra} shows 1D spectra extracted from the resulting data cubes.

%%%%%%%%%%%%%%%%%%%%%%%%%%%%%%%%%%%%%%%%%%%%%%%%%%
\section{Stellar population analysis}\label{sec: paper 3: Stellar population analysis}

We used the \textsc{python} implementation of Penalized Pixel Fitting (\ppxf{})\footnote{Available \url{https://www-astro.physics.ox.ac.uk/~mxc/software/\#ppxf}.}~\citep{Cappellari&Emsellem2004,Cappellari2017} to analyse the stellar population by fitting combinations of simple stellar population (SSP) templates to the stellar continuum. This is a refinement of the work of \citetalias{Croft2006}, who carried out a similar analysis using broad-band photometry.

Due to the predominantly young stellar population of MO \citepalias{Croft2006}, we used the SSP models of \citet{GonzalezDelgado2005} which have a fine age resolution at young stellar ages, with a resolution of 1\,Myr for ages up to 10\,Myr, 5\,Myr for ages from 10\,Myr to 100\,Myr and 100\,Myr from 100\,Myr to 1\,Gyr.
The SSP templates were generated using both the Geneva and the Padova isochrones with a Salpeter IMF, spanning an age range 1\,Myr to 10\,Gyr divided into 46 bins with five metallicities for the Geneva isochrones ($0.05\rm Z_\odot$, $0.2\rm Z_\odot$, $0.4\rm Z_\odot$, $1\rm Z_\odot$ and $2 \rm Z_\odot$), and three for the Padova isochrones ($0.2\rm Z_\odot$, $0.4\rm Z_\odot$ and $0.95\rm Z_\odot$). 
%We note that the assumption of a conventional IMF may not hold for MO, as discussed in Section~\ref{sec: paper 3: On the formation of Minkowski's Object}.

% Justification: Geneva isochrones
Because MO is thought to have a low gas-phase metallicity \citepalias{Croft2006,Lacy2017}, we used the Geneva isochrones for which the stellar templates extend to lower metallicities than the Padova isochrones.
Although the Geneva isochrones do not include stellar evolution along the red giant branch, we do not believe this to be of critical importance due to the predominantly young stellar population of MO. We confirmed that this did not bias our results by repeating our analysis using the Padova isochrones, which yielded very similar results (see Appendix~\ref{appendix: Additional PPXF fits}).

% Justification: binning
As can be seen in Fig.~\ref{fig: paper 3: WiFeS spectra}, there are very few stellar absorption features in the R7000 wavelength range; hence we only analysed our B3000 data with \ppxf{}.
Due to the low continuum surface brightness of MO, our data cubes lacked the S/N to recover ages and metallicities in each spaxel.
Because the line emission in MO has a young stellar population it is reasonable to assume that the majority of stars are located in the regions with bright emission line flux.
We therefore created an integrated spectrum by summing the spaxels in which the \ha{} flux exceeded $1.5 \times 10^{-15} \rm \, erg\,s^{-1}\,cm^{-2}$. 
This flux limit was chosen as the resulting region is similar to the 8'' aperture used by \citetalias{Croft2006} in their SSP analysis using SED fitting, enabling a direct comparison of our analysis to their results. 
The median S/N in the resulting spectrum was approximately 7.5 per spectral pixel.

To obtain the best-fit age and metallicity, the stellar continuum and the emission lines were fitted simultaneously, adopting independent velocity components for each, where emission lines were fit using Gaussian profiles.
A 4th-order multiplicative polynomial was included in the fit to compensate for extinction and calibration errors. 
\textit{Regularisation} was used to bias the best-fit template weights towards the smoothest solution consistent with the data. 
The regularisation parameter \texttt{regul} was determined using the standard method \citep[e.g.,][]{Boardman2017}. We performed a fit firstly with \texttt{regul} $ = 0$, and secondly where the noise on the input spectrum was multiplied by the reduced-$\chi^2$ of the first fit. \texttt{regul} was gradually increased until the difference in the reduced-$\chi^2$ for the regularised and non-regularised fits $\Delta \chi^2 \simeq \sqrt{2 N}$, where $\Delta \chi^2 = N \left(\chi^2/\textrm{DOF} - 1 \right)$, and $N$ is the number of data points in the input spectrum and $\textrm{DOF}$ is the number of degrees-of-freedom.

% Integrated spectrum
Fig.~\ref{fig: paper 3: ppxf integrated fit, age & metallicity} shows the best-fit spectrum and template weights.
There are two clear stellar populations in MO, with approximate ages of $4-5 \,\rm Myr$ and $1 \,\rm Gyr$.
The younger population is directly associated with MO, whereas the older population may represent stars associated with the stellar bridge, or with the object itself. These results are consistent with the broad-band SED analysis of \citetalias{Croft2006}, who found the best-fit stellar population is dominated by a component approximately $7.5 \, \rm Myr$ old, with up to $20$ per cent of the light attributed to a component at $1 \,\rm Gyr$. 

Simultaneously constraining both the age and metallicity of the stellar population is challenging due to the lack of strong metal absorption features in the stellar continuum. We therefore repeated our analysis using templates with fixed metallicities, yielding similar results, as detailed in Appendix~\ref{appendix: Additional PPXF fits}.

To further check whether MO is dominated by a young stellar population, we individually fitted each SSP template in age and metallicity (for both the Padova and Geneva isochrones), including Gaussian emission lines and a multiplicative polynomial to correct for calibration errors, and calculated the resulting reduced-$\chi^2$. For all template metallicities, the minimum reduced-$\chi^2$ occurs at ages $\lessapprox 15 \rm \, Myr$, with the $\chi^2$ gradually increasing with template age. Thus we conclude the stellar population in MO is indeed young.

\begin{figure*}
	\centering
	\includegraphics[width=1\linewidth]{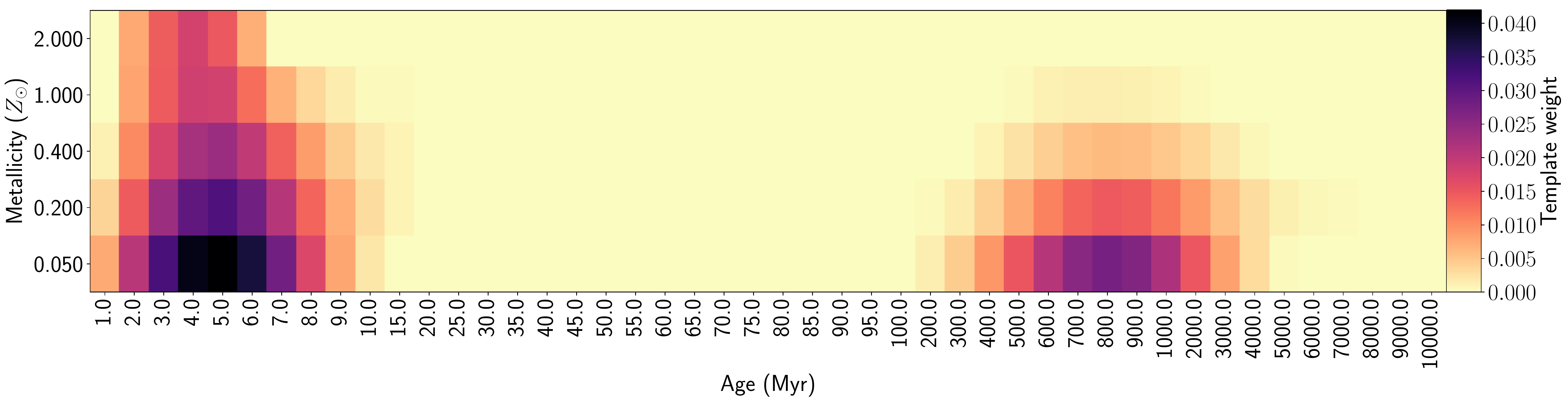}
	\includegraphics[width=1\linewidth]{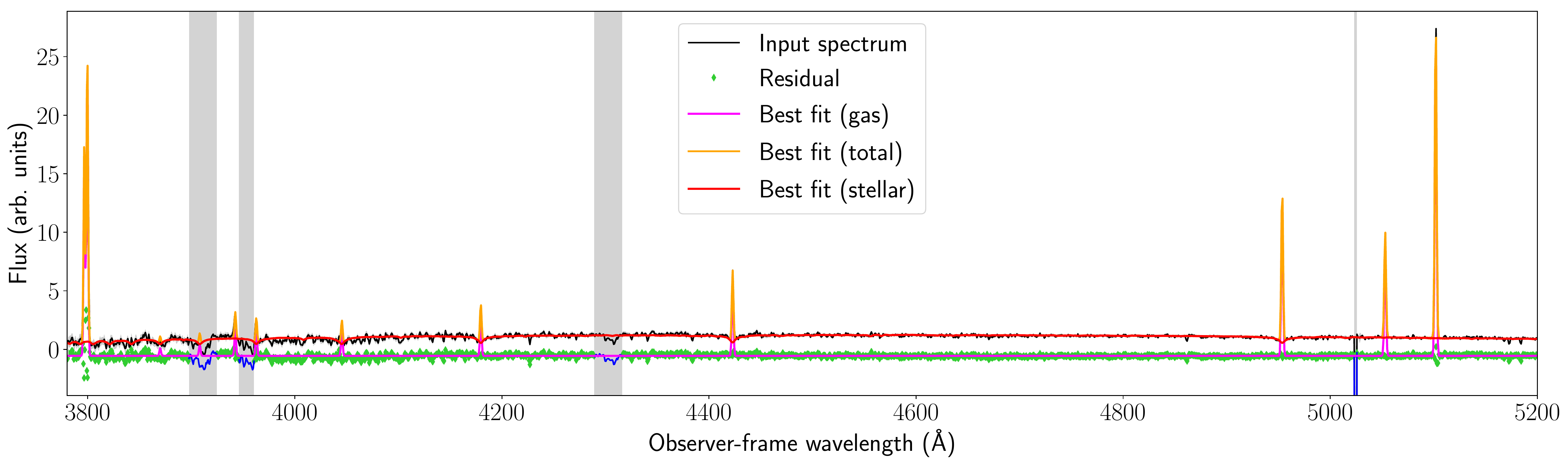}
	\caption{
		Results of the \ppxf{} fit used to estimate the age and metallicity of the stellar population using the Geneva isochrones. The top panel shows a 2D histogram of the best-fit age and metallicity of the stellar population. The bottom panel shows the integrated spectrum (corrected for Galactic extinction), overlaid with the best-fit spectrum with the stellar and emission line components are shown separately. $1\sigma$ errors are indicated in grey, and the green dots indicate the residual to the fit. The grey shaded regions indicate spectral windows dominated by residuals from the sky subtraction, and were not included in the fit.
	}
	\label{fig: paper 3: ppxf integrated fit, age & metallicity}
\end{figure*}

%%%%%%%%%%%%%%%%%%%%%%%%%%%%%%%%%%%%%%%%%%%%%%%%%%
\section{Emission line gas}\label{sec: paper 3: Emission line gas}

%------------------------------------------------%
\subsection{Emission line fitting}\label{subsec: paper 3: Emission line fitting}
We used \textsc{mpfit} \citep{Markwardt2009}, a \textsc{python} implementation of the Levenberg-Marquardt algorithm \citep{More1978} developed by M. Rivers\footnote{Available \url{http://cars9.uchicago.edu/software/python/mpfit.html}.} to estimate line fluxes in each spaxel by simultaneously fitting a single Gaussian profile to each emission line and a linear component to fit the stellar continuum. Each emission line was fitted over a small window centred on the systemic velocity of MO. All emission lines were fitted simultaneously and were constrained to have the same systemic velocity and velocity dispersion.

Due to the lack of prominent stellar absorption features and the high equivalent widths of the Balmer emission lines compared to the corresponding stellar absorption features (see Fig.~\ref{fig: paper 3: WiFeS spectra}), and because we lacked the S/N to fit the stellar continuum in each spaxel, we did not subtract the best-fit stellar continuum from spaxels before fitting emission lines.
To check whether this would effect our emission line fluxes, we used the equivalent width (EW) of the \hb{} line (approximately $7.6$\,\AA), estimated from our \ppxf{} fit, to estimate the stellar absorption flux in each spaxel by measuring the continuum flux in two windows adjacent to the \hb{} line. In most spaxels, the stellar absorption is less than $10\,\rm per \, cent$ of the measured \hb{} emission line flux.
To investigate the effect on the \ha{} absorption, we used a solar-metallicity \textsc{Starburst99} model~\citep{Leitherer1999} with an instantaneous burst star formation history and a Salpeter IMF to estimate the \ha{} EW because it was not measured in our \ppxf{} analysis. At an age of $10^7 \,\rm yr$ the \ha{} EW $\approx 16\,$\AA. Using the same method as for \hb{}, we again found the stellar absorption accounts for less than $10\,\rm per \, cent$ of the measured \ha{} emission line flux in most spaxels, confirming that this has a minimal effect on our reported emission line fluxes. As such, this systematic error has not been included in the reported line fluxes.

Unless specified otherwise, for the remainder of our analysis, we only include spaxels in which the fit has a reduced-$\chi^2 < 2$ and $\rm S/N > 3$ in the \ha{}, \hb{}, \forb{O}{ii}, \forb{O}{iii}, \forb{N}{ii} and \forb{S}{ii} lines. Upper limits for the weakest line used in our analysis, \forb{O}{i}$\uplambda 6300$, were calculated in each spaxel assuming a Gaussian amplitude equal to $3\sigma$, where $\sigma$ is the standard deviation in a window centred on the emission line, and the same width as the other lines in that spaxel.
In all quoted linewidths, the instrumental resolution was accounted for by subtracting the width of the line spread function (LSF) in quadrature from the width of the fitted Gaussian, where the width of the LSF was estimated by fitting Gaussian profiles to sky lines.

Integrated line fluxes for emission lines in MO are shown in Table~\ref{tab: paper 3: emission line fluxes}, which were estimated by summing the fluxes in each spaxel, except for very faint lines, such as He\,\textsc{i} $\uplambda3889$, for which fluxes were estimated by fitting a single Gaussian profile to the integrated spectrum.

\begin{table}
	\centering
	\caption{Total emission line fluxes. The $1\sigma$ errors shown represent the formal uncertainty on the emission line fit.}
	\begin{tabular}{c c}
		\hline
		\textbf{Emission line}              & \textbf{Flux} ($\rm 10^{-15} \, erg \, s^{-1} \, cm^{-2}$) \\ \hline
		[O\,\textsc{ii}]$\uplambda3726$          & $ 15.1 \pm 0.3$ \\
		{}[O\,\textsc{ii}]$\uplambda3729$         & $ 24.0 \pm 0.4$ \\
		\hb{}                       & $ 17.3 \pm 0.2$ \\
		{}[O\,\textsc{iii}]$\uplambda\uplambda4959,5007$ & $ 64.8 \pm 0.6$ \\
		{}[Ne\,\textsc{ii}]$\uplambda3869^\dagger$    & $ 1.7 \pm 0.3$ \\
		He\,\textsc{i} $\uplambda3889^\dagger$      & $ 2.1 \pm 0.3$ \\
		\he{}$^\dagger$                  & $ 1.3 \pm 0.3$ \\
		\hd{}$^\dagger$                  & $ 2.5 \pm 0.2$ \\
		\hg{}$^\dagger$                  & $ 5.3 \pm 0.2$ \\
		{}[O\,\textsc{i}]$\uplambda6300^\ddagger$         & $\leq 1.1$ \\
		\ha{}                       & $ 47.9 \pm 0.1$ \\
		{}[N\,\textsc{ii}]$\uplambda6548,6583$     & $ 5.60 \pm 0.06$ \\
		{}[S\,\textsc{ii}]$\uplambda6716$         & $ 8.87 \pm 0.07$ \\
		{}[S\,\textsc{ii}]$\uplambda6731$         & $ 5.79 \pm 0.06$ \\
		He\,\textsc{i} $\uplambda5876^\dagger$      & $ 1.13 \pm 0.05$ \\
		{}[O\,\textsc{i}]$\uplambda6364^\dagger$     & $ 0.39 \pm 0.05$ \\ \hline
		\multicolumn{2}{p{5.5cm}}{
			\textsuperscript{$\dagger$}\footnotesize{Flux estimated by fitting a single Gaussian profile to the integrated spectrum.}
			\textsuperscript{$\ddagger$}\footnotesize{$3\sigma$ upper limit.}
		}
	\end{tabular}
	\label{tab: paper 3: emission line fluxes}
\end{table}

%%%%%%%%%%%%%%%%%%%%%%%%%%%%%%%%%%%%%%%%%%%%%%%%%%
\subsection{Kinematics and morphology}

\begin{figure}
	\centering
	\includegraphics[height=0.30\textheight]{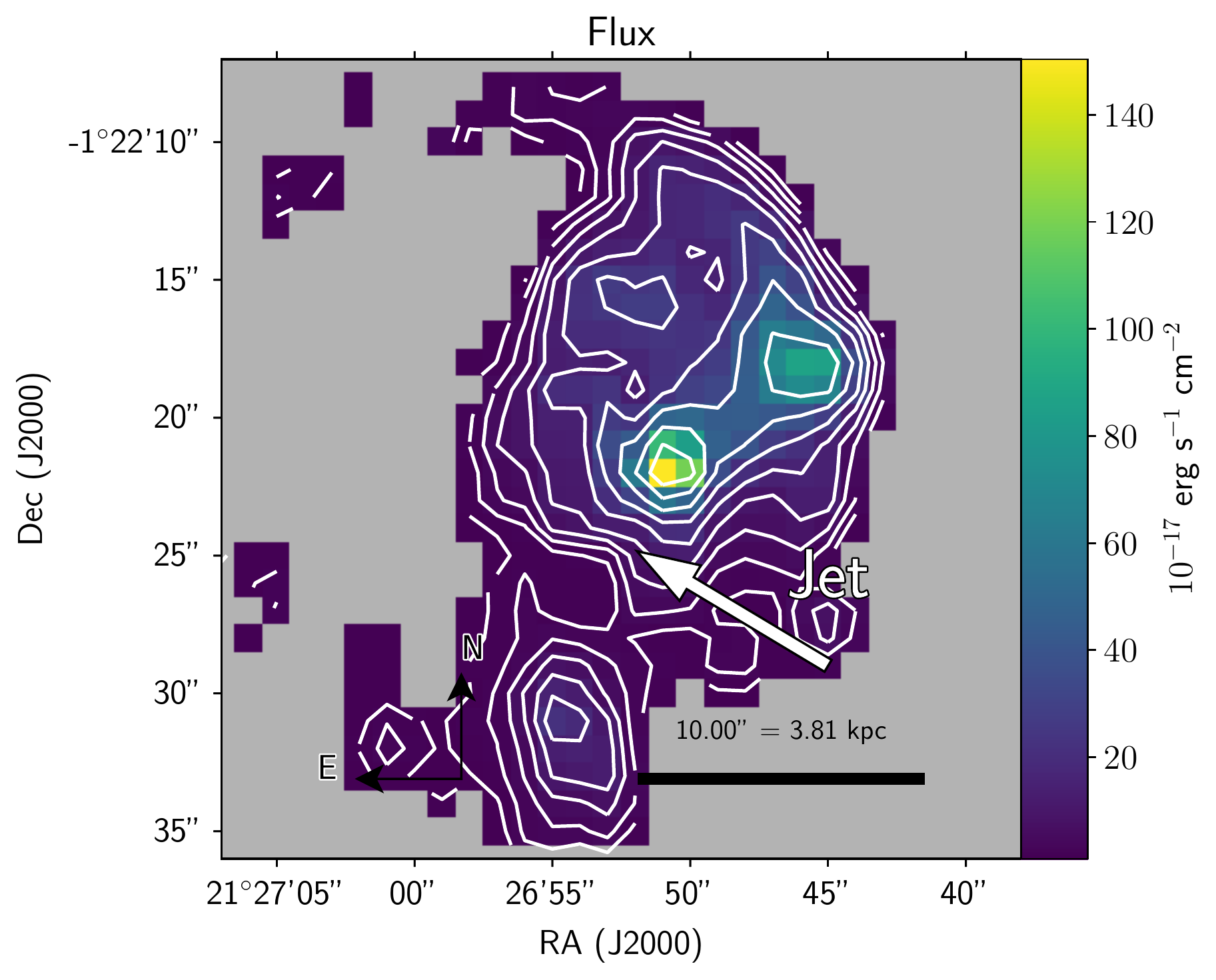}
	\includegraphics[height=0.30\textheight]{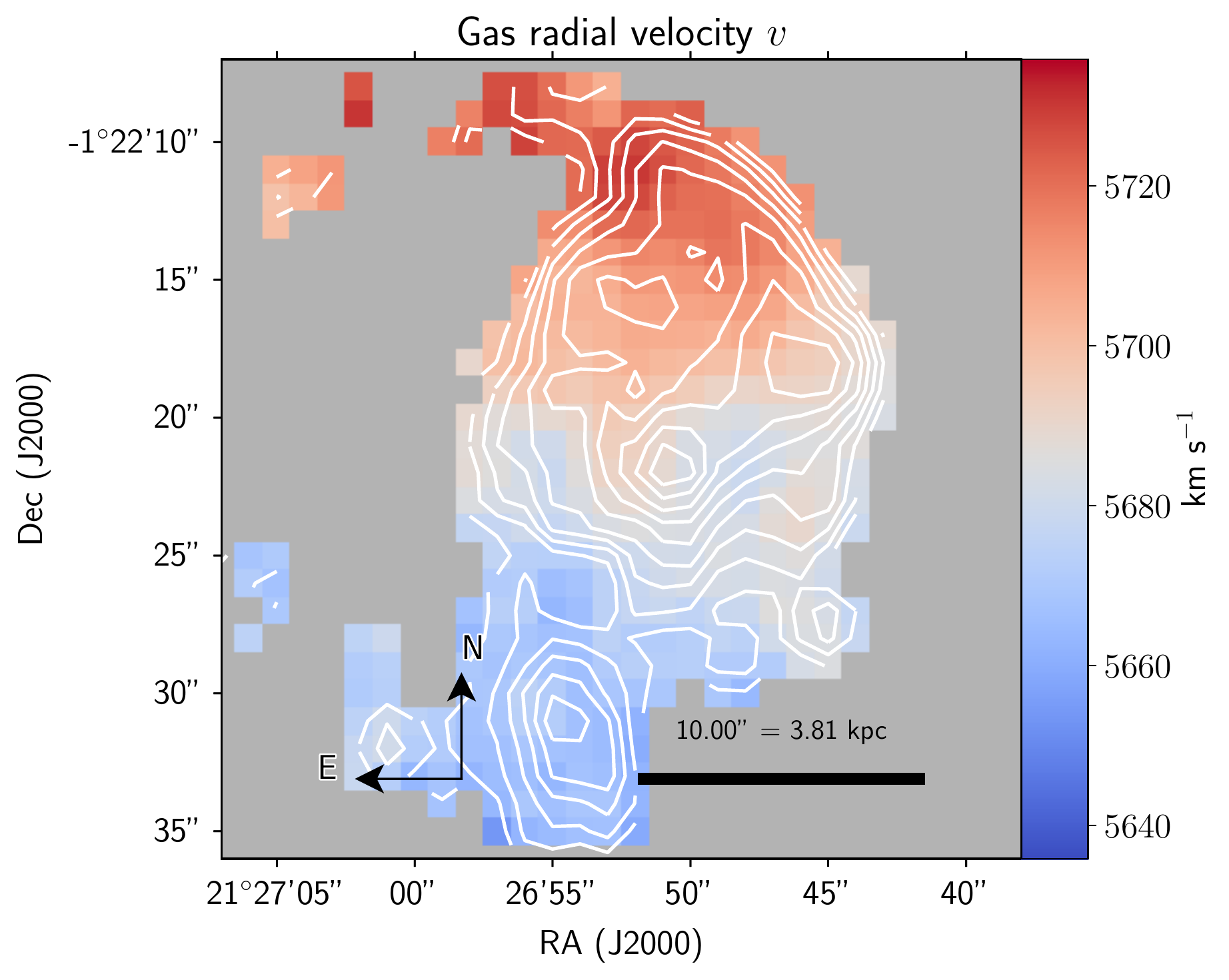}
	\includegraphics[height=0.30\textheight]{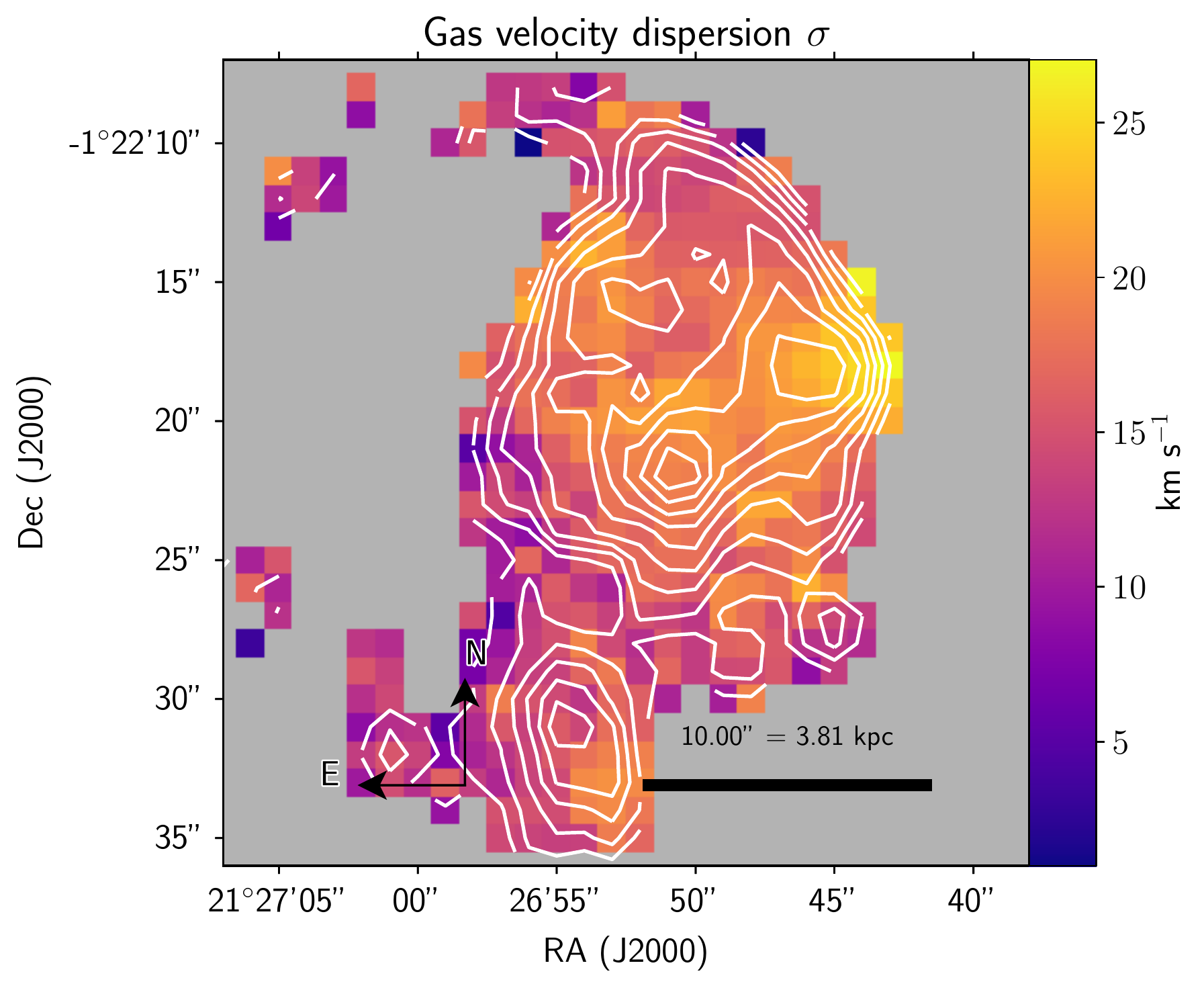}
	\caption{
		Top: \ha{} flux, where the approximate direction of the radio jet is shown; middle: radial velocity (with respect to the local standard of rest); bottom: velocity dispersion (corrected for instrumental broadening) from our Gaussian line fit. The white contours indicate the \ha{} flux on a log scale. All spaxels with \ha{} S/N > 3 are shown.
	}
	\label{fig: paper 3: Halpha kinematics and morphology}
\end{figure}

Fig.~\ref{fig: paper 3: Halpha kinematics and morphology} shows maps of the \ha{} flux, radial velocity and velocity dispersion.
The radial velocities increase from South to North, which is consistent with those of the CO and \hi{} \citepalias{Lacy2017}. 
The velocity dispersion is low, with values around $15\,\rm km\,s^{-1}$ in most parts of the object, peaking in the West with $\sigma \approx 30 \,\rm km\,s^{-1}$. 

The \ha{} emission appears to follow the stellar distribution, which can be seen in the bottom panel of Fig.~\ref{fig: paper 3: HST overlay}; the emission is concentrated along the central ``bar'' of the object running South-East to North-West, with fainter \ha{} emission visible along the extended filamentary structures. 
The two brightest CO clouds~\citepalias[see fig.\,1 of ][]{Lacy2017} are located along the filaments to the North-East, whereas the fainter two CO clouds overlap with the two bright \ha{} knots in the bar, perhaps indicating that star formation has been ongoing in the bar for a longer period than in the filaments. This scenario would be consistent with the jet triggering star formation as it propagates through the object from South-West to North-East. The bulk of the \hi{} being located further downstream~\citepalias[see fig. 2 of ][]{Lacy2017} also supports this hypothesis.

%%%%%%%%%%%%%%%%%%%%%%%%%%%%%%%%%%%%%%%%%%%%%%%%%%
\subsection{Reddening}

\begin{figure}
	\centering
	\includegraphics[height=0.30\textheight]{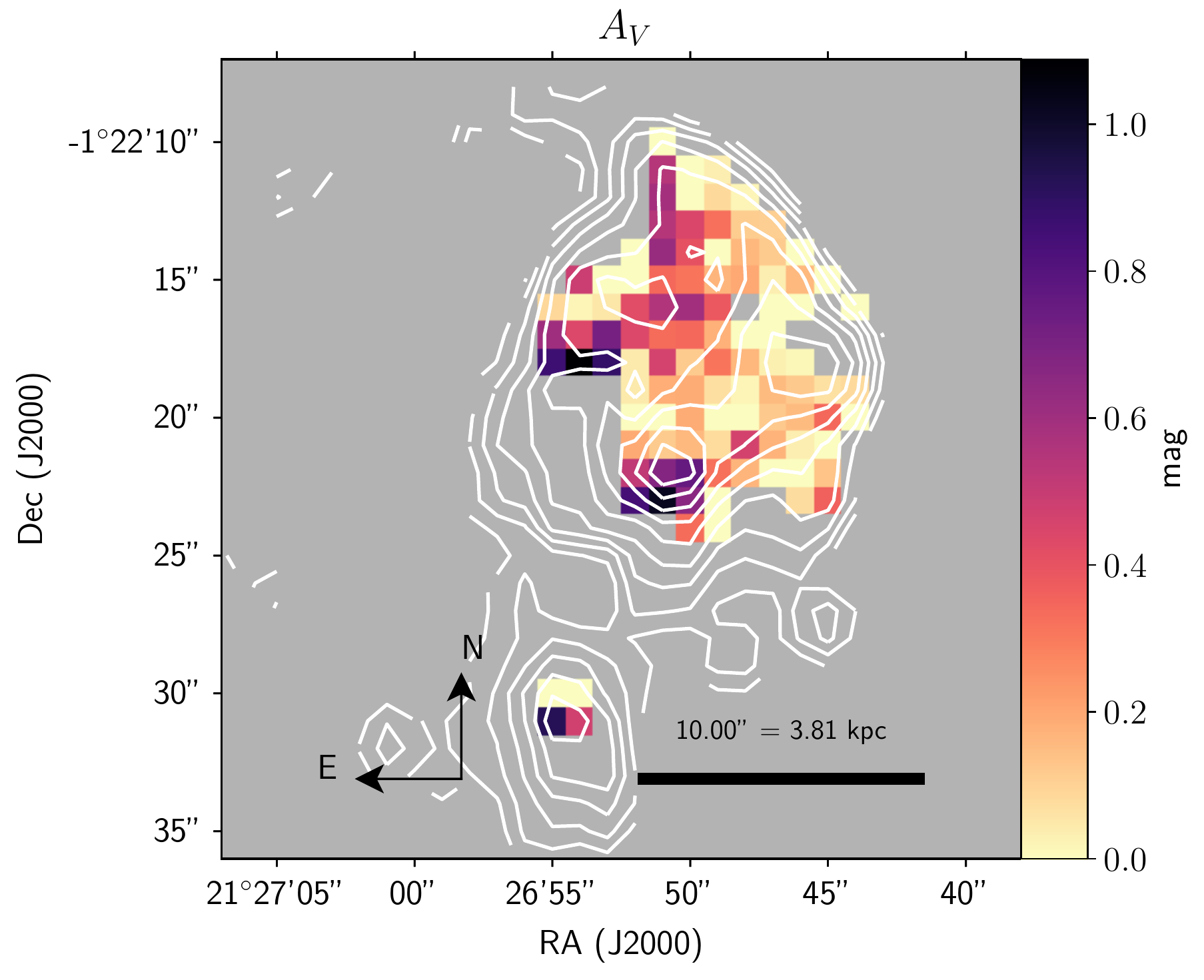}
	\includegraphics[height=0.30\textheight]{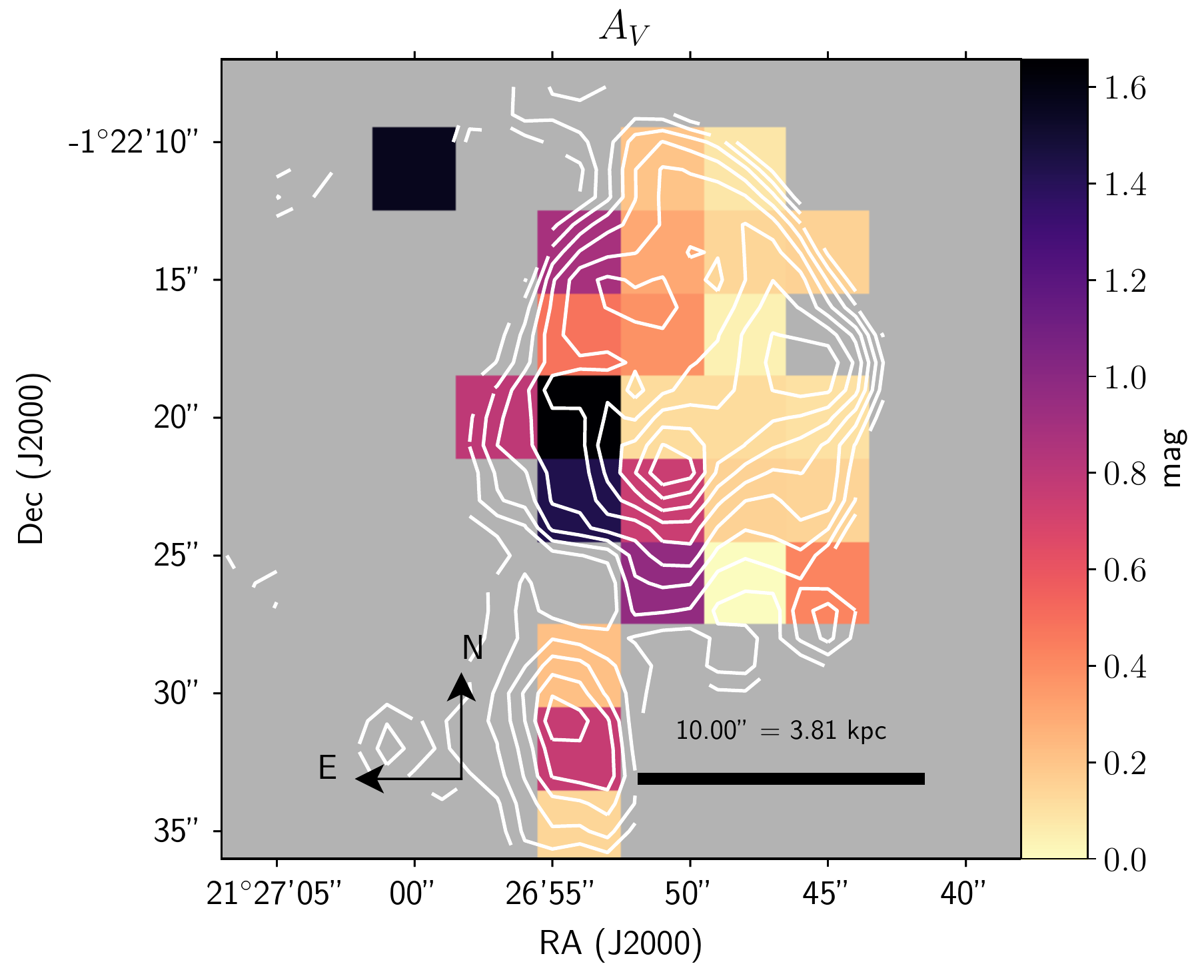}
	\caption{The $A_V$ estimated in each spaxel using the extinction curve of \citet{Fitzpatrick&Massa2007} with $R_V = 3.1$ in the spaxels of the unbinned data cube (top) and spatially binned data cube (bottom), where we only show spaxels in which the S/N in the \ha{}/\hb{} ratio exceeds 3. Spaxels with $A_V < 0$ but with values within $1\sigma$ of zero are indicated as having $A_V = 0$. Spaxels with $A_V < 0$ by an amount greater than $1\sigma$ from zero are not shown. The white contours indicate the \ha{} flux on a log scale.}
	\label{fig: paper 3: reddening}
\end{figure}

% Global
After correcting for foreground extinction, we used the \ha{} and \hb{} fluxes to estimate the extinction. The $A_V$ in each spaxel is shown in the upper panel of Fig.~\ref{fig: paper 3: reddening}, where we have used the reddening curve of \citet{Fitzpatrick&Massa2007} with $R_V = 3.1$. Only spaxels in which the S/N in the \ha{}/\hb{} ratio exceeds 3 are shown.
The mean $A_V = 0.17 \pm 0.04$, which is slightly higher than the value of \citetalias{Croft2006}, who reported negligible reddening within MO. However, our spatially resolved extinction map reveals significantly higher values of $A_V$ in several clumpy regions to the East, although as discussed in Section~\ref{subsec: paper 3: Excitation mechanism}, this may be a result of an intrinsically enhanced Balmer decrement due to X-ray photoionisation.

Because the S/N was insufficient to measure $A_V$ in all spaxels in which line emission was detected, we repeated the above process after spatially binning the data cube by a factor of three; the resulting extinction map is shown in the lower panel of Fig.~\ref{fig: paper 3: reddening}.
Emission line fluxes were corrected for intrinsic reddening by applying the $A_V$ estimated in the corresponding spaxel in the binned data cube; in those spaxels in which $A_V$ could not be measured or $A_V < 0$, no extinction correction was applied. 

%%%%%%%%%%%%%%%%%%%%%%%%%%%%%%%%%%%%%%%%%%%%%%%%%%
\subsection{Electron density}\label{subsec: paper 3: Electron density}

To estimate the electron density in each spaxel, we used the ratio $R$ of the emission lines in the [S\,\textsc{ii}]$\uplambda\uplambda 6716,6731$ doublet and eqn. 3 of \citet{Proxauf2014}. 
We detected no significant variation in $R$ across the object, with the mean $R  = 1.35 \pm 0.03$, placing MO on the cusp of the low-density limit ($n_e \lesssim 10 \,\rm cm^{-3}$), which is consistent with the findings of \citetalias{Croft2006}. 

%%%%%%%%%%%%%%%%%%%%%%%%%%%%%%%%%%%%%%%%%%%%%%%%%%
\subsection{Excitation mechanism}\label{subsec: paper 3: Excitation mechanism}

\begin{figure*}
	\centering
	\includegraphics[width=1\textwidth]{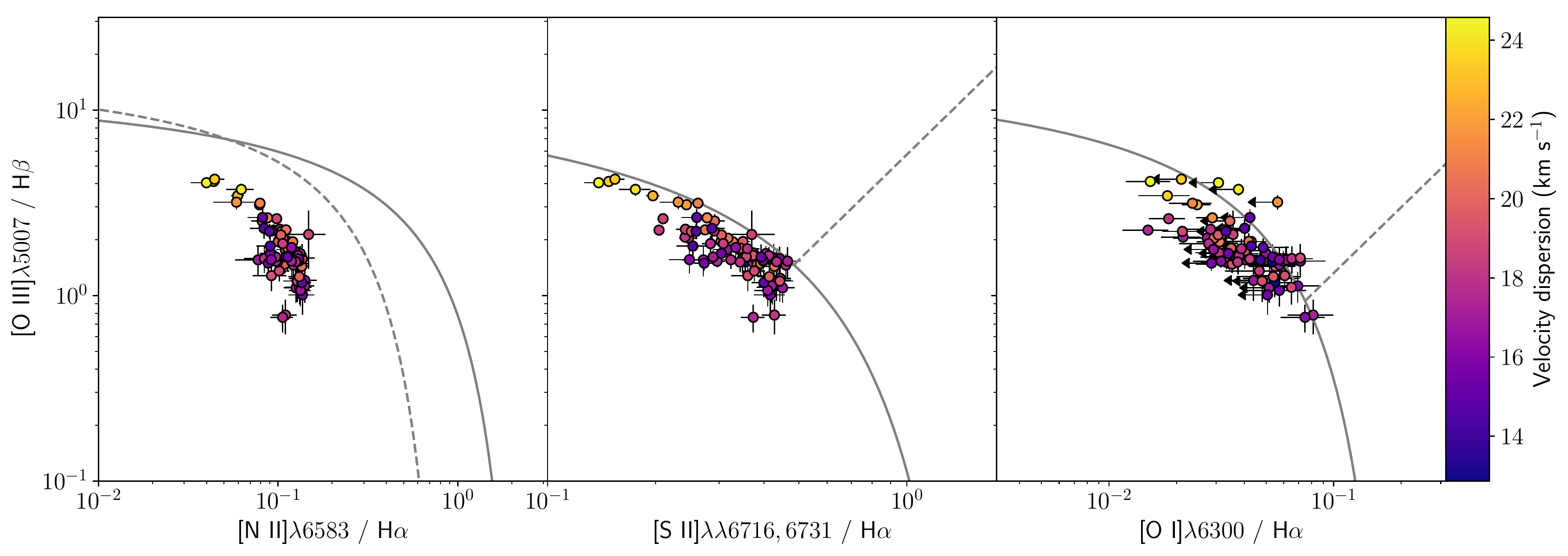}
	\includegraphics[width=1\textwidth]{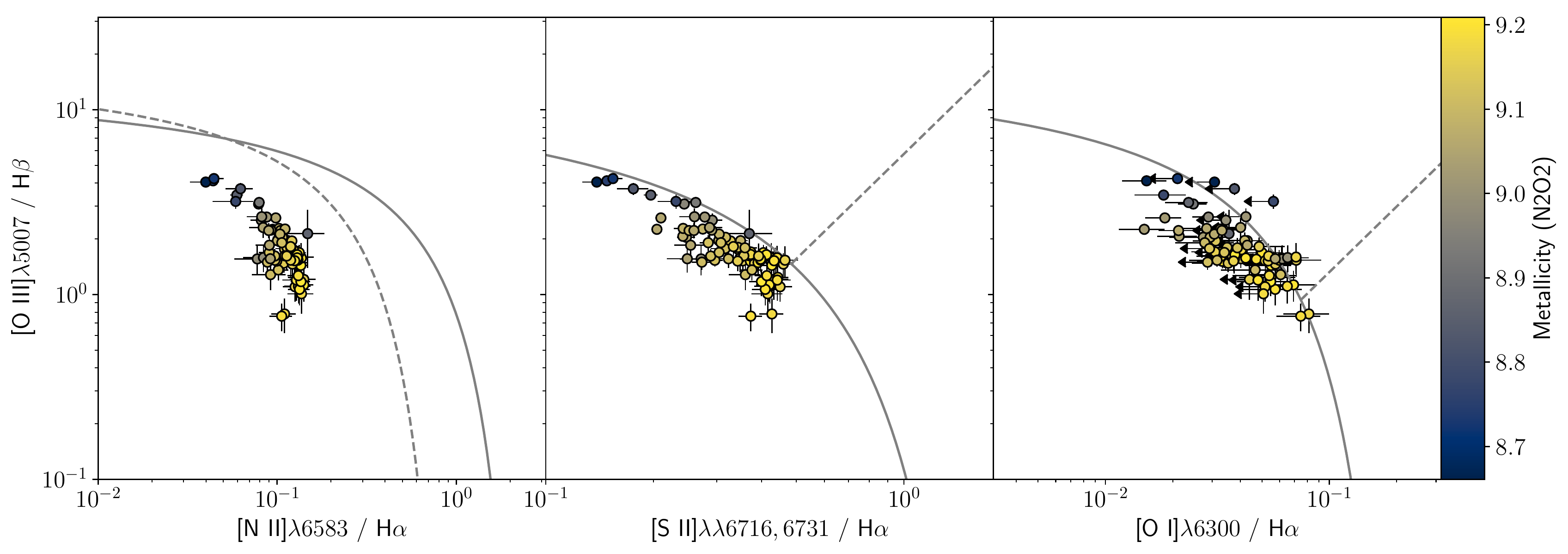}
	\includegraphics[width=1\textwidth]{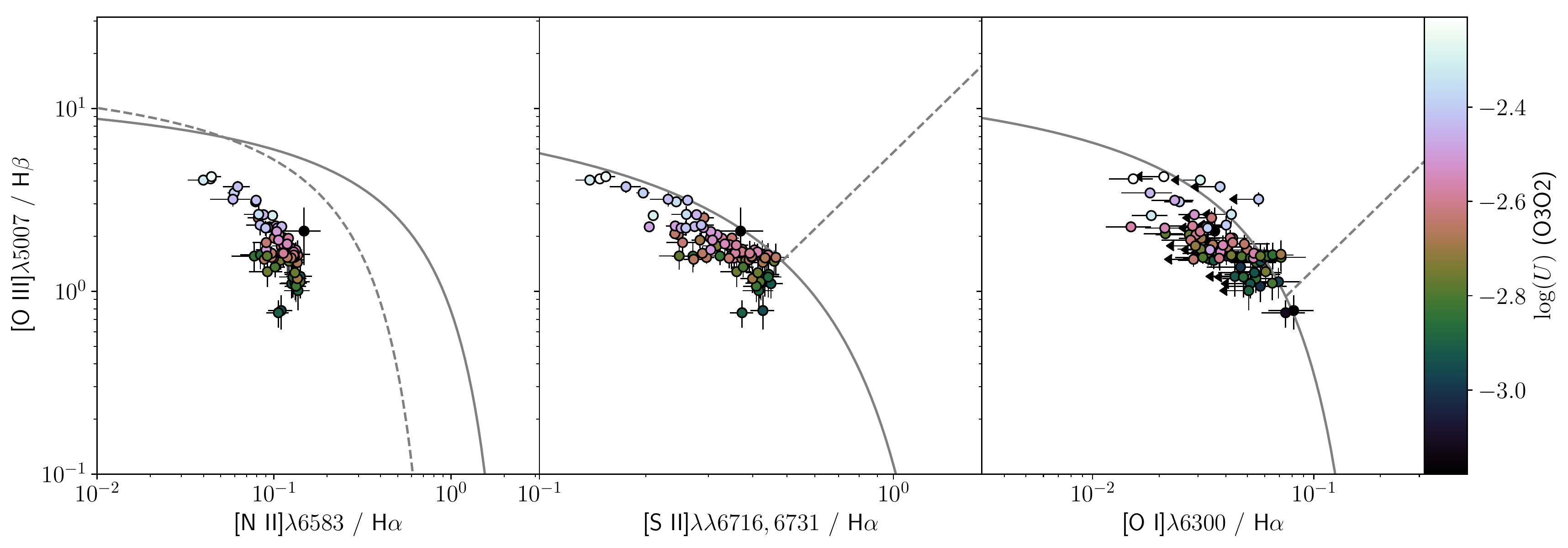}
	\caption{ODDs of MO, where each point represents an individual spaxel with $1\sigma$ error bars, except for points with arrows, which represent $3\sigma$ upper limits. The spaxels are colour coded by velocity dispersion (top), metallicity (middle) and ionisation parameter (bottom). The metallicity and ionisation parameter have been estimated using the N2O2 and O3O2 diagnostics respectively (see Fig.~\ref{fig: paper 3: metallicity & ionisation parameter}).
	The solid grey lines represent the maximum [O\,\textsc{iii}]/\hb{} ratio that can arise from star formation alone, derived from photoionisation models~\protect\citep{Kewley2001}.
	In the left panels, the dashed line is the equivalent empirical relation of \protect\citet{Kauffmann2003} which separates star-forming galaxies and AGN hosts. 
	In the middle and right panels, the dashed lines of \protect\citet{Kewley2006} separate Seyfert-like (above) and LINER-like ratios (below the line).}
	\label{fig: paper 3: ODD diagrams}
\end{figure*}

To determine the excitation mechanism of the emission line gas, we use optical diagnostic diagrams\,\citep[ODDs;][]{Baldwin1981,Veilleux&Osterbrock1987,Kewley2001,Kauffmann2003,Kewley2006}, in which the \forb{O}{iii}$\uplambda 5007$/\hb{} ratio is plotted as a function of the \forb{N}{ii}$\uplambda 6583$/\ha{}, \forb{S}{ii}$\uplambda\uplambda6716,6731$/\ha{} and \forb{O}{i}$\uplambda 6300$/\ha{} ratios. ODDs for each spaxel in MO are shown in Fig.~\ref{fig: paper 3: ODD diagrams}.

There is a broad spread in the \forb{S}{ii}/\ha{} and \forb{O}{i}/\ha{} ratios within the object, which is likely due to an intrinsic metallicity variation within the object. Indeed, inspection of the middle panel of Fig.~\ref{fig: paper 3: ODD diagrams} shows that the metallicity calculated using the N2O2 diagnostic of \citet{KewleyNichollsSutherland2019} increases smoothly as a function of the \forb{N}{ii}/\ha{}, \forb{S}{ii}/\ha{} and \forb{O}{i}/\ha{} ratios, following a trajectory very similar to the metallicity sequence observed in SDSS galaxies~\citep[e.g.,][]{Kewley2006}.

\begin{figure*}
	\includegraphics[width=0.75\linewidth]{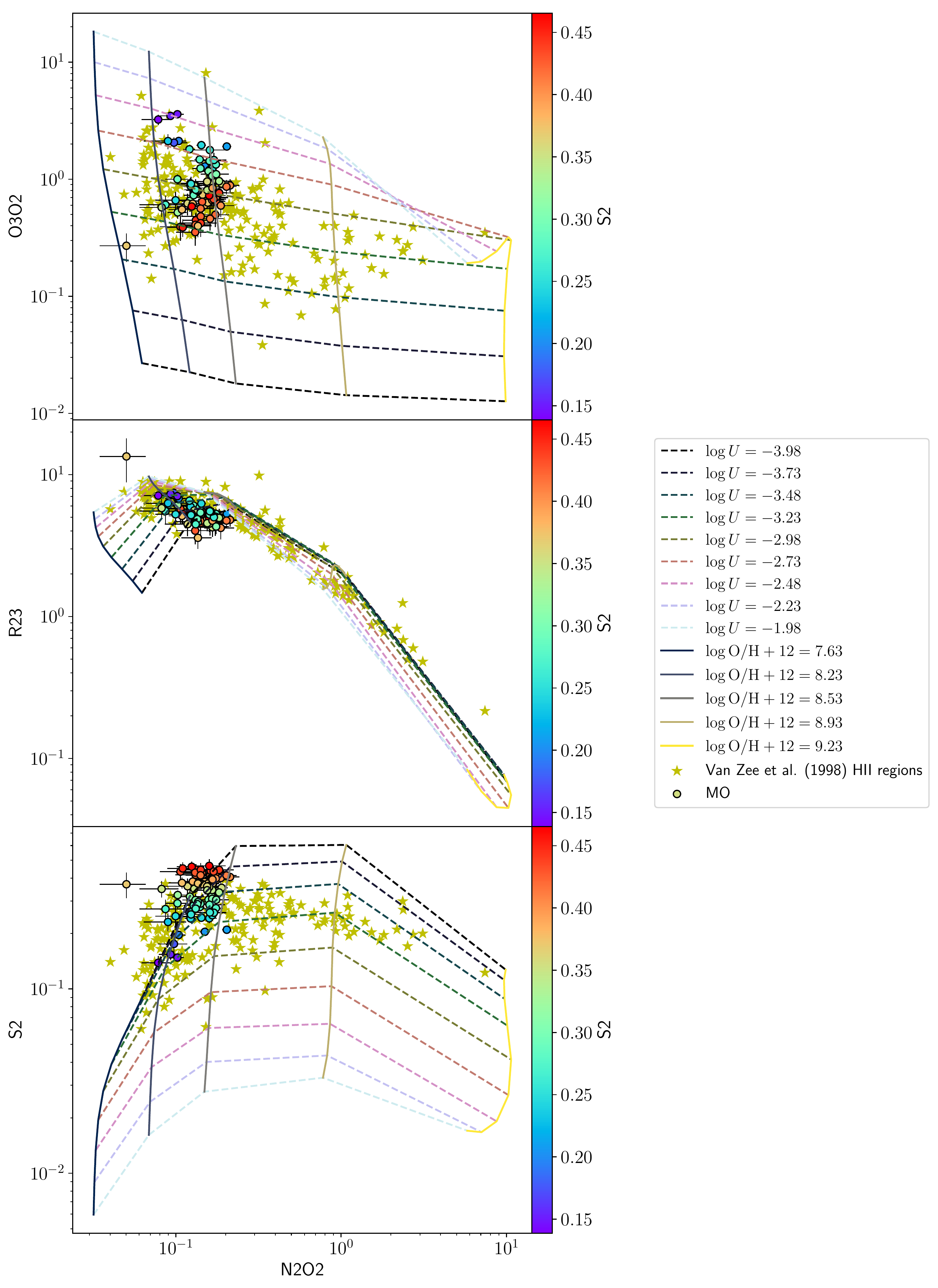}
	\caption{O3O2 (top), R23 (middle) and S2 (bottom) ratios as a function of the N2O2 ratio for spaxels in MO (circles with $1\sigma$ errors, coloured by S2 ratio) and for the \citet{vanZee1998} \hii{} regions (gold stars). The grids indicate line ratios from \textsc{Mappings} \hii{} region models, varying in metallicity (solid lines) and ionisation parameter (dashed lines).
	}
	\label{fig: paper 3: Line ratio diagrams w. Van Zee models}
\end{figure*}

Although most spaxels lie beneath the line of maximal star formation, there are some spaxels that lie beyond this line, indicating a contribution from ionising sources other than star formation.
In Fig.~\ref{fig: paper 3: Line ratio diagrams w. Van Zee models} we present additional line ratio diagrams, where we include the sample of `typical' \hii{} regions in nearby spiral galaxies from \citet{vanZee1998}. Grids from \citet{Kewley2019} representing \textsc{Mappings V}~\citep{Sutherland2018} \hii{} region plane-parallel models with $\log(P/k) = 5.0$, varying in ionisation parameter and metallicity, are shown for comparison.
Although the spaxels in MO are similar to the \hii{} regions in the R23 ((\forb{O}{ii}$\uplambda \uplambda 3726,3729$ + \forb{O}{iii}$\uplambda \uplambda 4959,5007$)/\hb{}) vs. N2O2 (\forb{N}{ii}$\uplambda\uplambda 6548,6583$/\forb{O}{ii}$\uplambda\uplambda 3726,3729$) and O3O2 (\forb{O}{iii}$\uplambda\uplambda 4959,5007$/\forb{O}{ii}$\uplambda\uplambda 3726,3729$) vs. N2O2 plots, a significant number of spaxels have S2 (\forb{S}{ii}$\uplambda\uplambda 6716,6731$/\ha{}) ratios much higher than both the \hii{} regions and \textsc{Mappings} models, suggesting the presence of secondary excitation mechanisms.
%As shown in Fig.~\ref{fig: paper 3: S2 ratio spatial distribution}, the spaxels with the highest S2 ratios are located away from the main star-forming regions. 

\begin{figure}
	\centering
	\includegraphics[height=0.29\textheight]{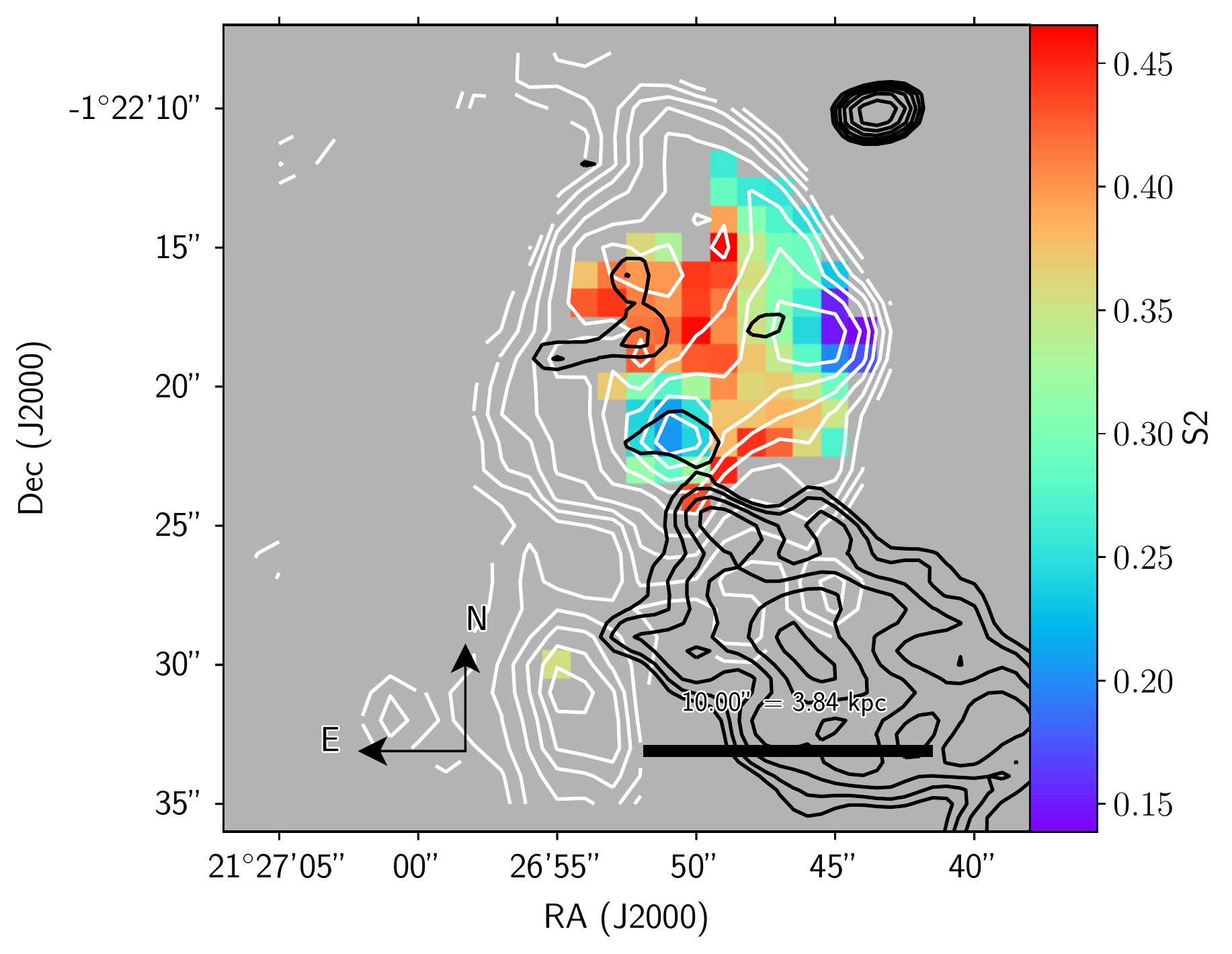}
	\caption{The S2 ratio in each spaxel. The white contours indicate the \ha{} flux on a log scale, and the black contours show the ALMA 106\,GHz continuum from the jet \citepalias{Lacy2017}. The contours represent 10 logarithmically spaced intervals from $3\sigma \,\rm Jy \, beam^{-1}$ to $1.6 \times 10^{-4} \,\rm Jy \, beam^{-1}$, where $\sigma = 9.2 \times 10^{-6} \, \rm Jy \, beam^{-1}$ is the measured rms noise in the image.}
	\label{fig: paper 3: S2 ratio spatial distribution}
\end{figure}

% NB modelling
As an additional check for whether the line ratios in MO are consistent with star formation alone, we used \textsc{NebulaBayes} \citep{Thomas2018}, a \textsc{Python} package that uses a Bayesian method to estimate ISM conditions from observed emission line fluxes using input model grids.
\textsc{NebulaBayes} was used with the same \hii{} region model grid generated using \textsc{Mappings}, however with $\log(P/k)$ varying from $\log(P/k) = 4.0$ to $\log(P/k) = 9.0$. 
In most spaxels, the best fit model had a very poor reduced-$\chi^2$ and $\log(P/k) \approx 6.5$. Such a high pressure is inconsistent with both the estimated conditions within the jet plasma in MO \citep[$P/k \approx 1.2 \times 10^5 \, \rm K \, cm^{-3}$,][]{vanBreugel1985}, and with the low electron density predicted by the \forb{S}{ii} ratio ($n \lesssim 10 \,\rm cm^{-3}$, Section~\ref{subsec: paper 3: Electron density}). 
Using \textsc{NebulaBayes} with a log-normal prior on the ISM pressure centred on $\log(P/k) \leq 5.0$ yielded similarly poor reduced-$\chi^2$ values in most spaxels, providing further evidence for excitation mechanisms other than star formation.

In the following sections, we investigate shock excitation, diffuse ionised gas (DIG) and X-ray photoionisation from the nucleus of NGC\,541 as possible ionising sources in MO.

%%%%%%%%%%%%%%%%%%%%%%%%%%%%%%%%%%%%%%%%%%%%%%%%%%%%%%%
\subsubsection{Shock excitation}\label{subsubsec: paper 3: Shock excitation}

% What about shocks?
Given that the jet appears to be passing through the centre of MO (Fig.~\ref{fig: paper 3: HST overlay}), shocks are likely to be present.
In Appendix~\ref{appendix: Estimating the shock velocities in MO}, we estimate the shock velocities induced by a typical transonic FR\,I-type jet interacting with an \hi{} cloud to be $\leq 20 \, \rm km \, s^{-1}$. This is consistent with the observed \hi{} velocity dispersion~\citepalias{Lacy2017}.
Such slow shocks would not produce any detectable optical line emission; however, faster shocks may be present due to its inhomogeneous morphology.

\begin{figure*}
	\centering
	\subcaptionbox{\label{fig: paper 3: ODD w. DS17 models}}{\includegraphics[width=1\linewidth]{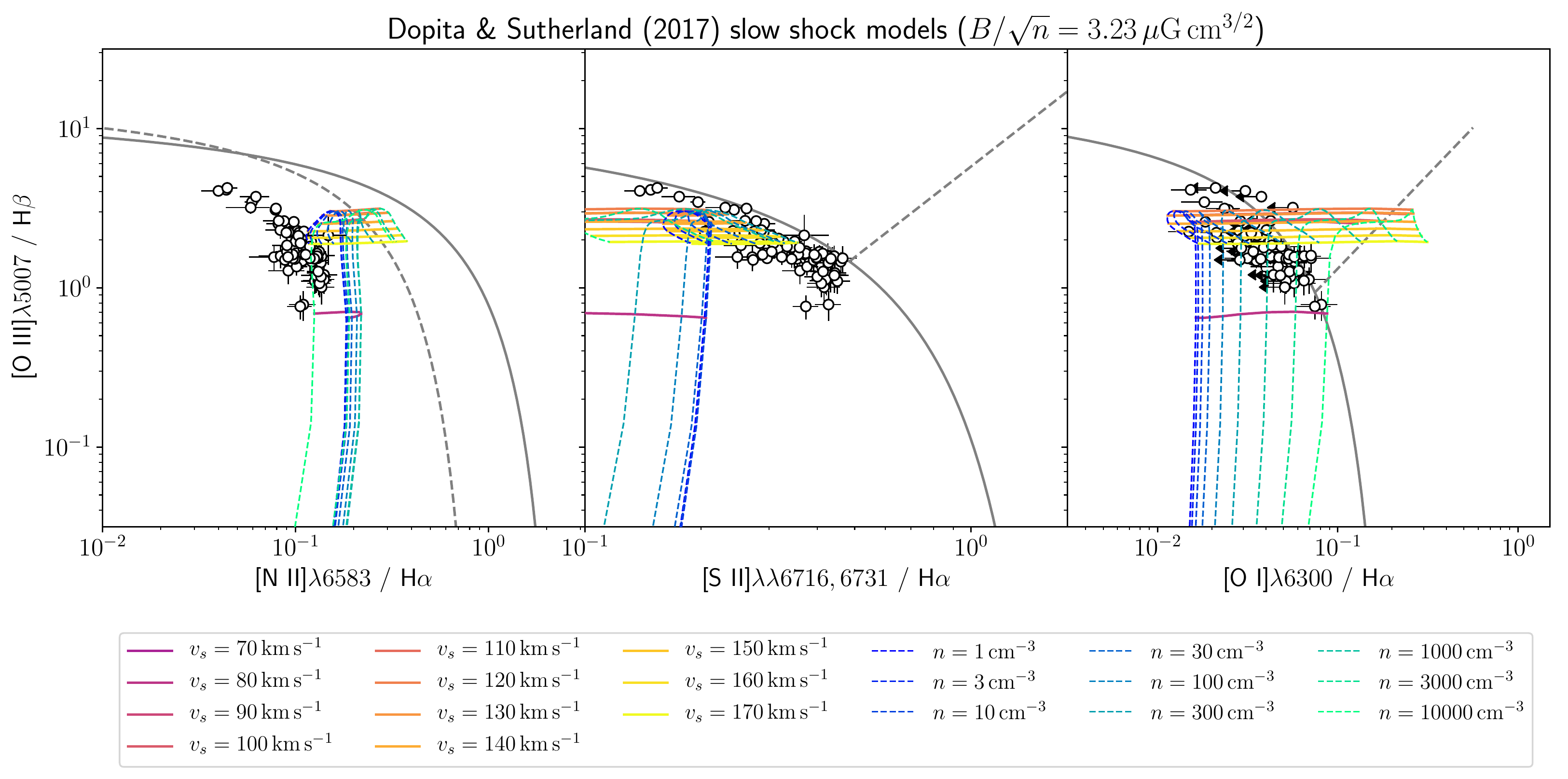}}
	\subcaptionbox{\label{fig: paper 3: ODD w. A08 models}}{\includegraphics[width=1\linewidth]{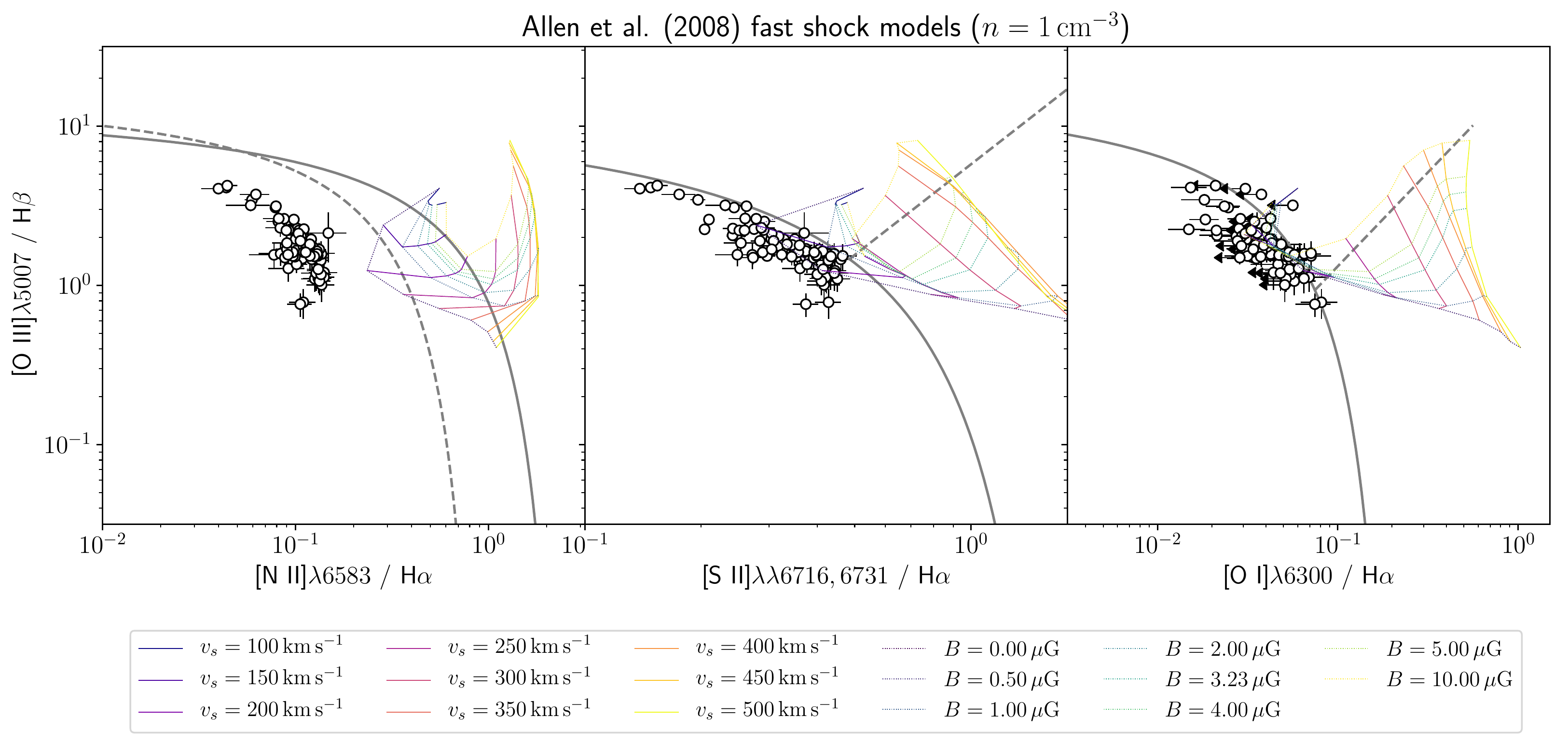}}
	\caption{
		ODDs overlaid with (a) the slow shock models of \citet{Dopita&Sutherland2017} and (b) the fast shock models of \citet{Allen2008}. White circles correspond to individual spaxels in MO with $1\sigma$ error bars, except for points with arrows, which represent $3\sigma$ upper limits. The grey solid and dashed lines are as in Fig.~\ref{fig: paper 3: ODD diagrams}.
	} 
	\label{fig: paper 3: ODD diagram w/ shock models}
\end{figure*}

% Fast shocks
To determine whether shocks could cause the enhanced \forb{S}{ii} and \forb{O}{i} emission, in Figs.~\ref{fig: paper 3: ODD w. DS17 models} and \ref{fig: paper 3: ODD w. A08 models} we show ODDs overlaid with the slow and fast shock models of \citet{Dopita&Sutherland2017} and \citet{Allen2008} respectively. 
The slow shock models are shown for shock speeds $v_s = 70 - 170 \,\rm km\,s^{-1}$ and pre-shock densities $n = 1 - 10000 \,\rm cm^{-3}$. In these models, the shock propagates into a solar metallicity nebula with a fixed magnetic parameter of $B / \sqrt{n} = 3.23 \rm \, \upmu G \, cm^{3/2}$, corresponding to equipartition.
The fast shock models are shown for shock speeds $v_s = 100 - 500 \,\rm km\,s^{-1}$ and magnetic field strengths $B = 0 - 10 \,\rm \upmu G$. These models correspond to a solar metallicity nebula with a pre-shock density $n = 1\,\rm cm^{-3}$, and include line emission from both the post-shock gas and the photoionised precursor.

\begin{figure*}
	\centering
	\includegraphics[width=0.75\linewidth]{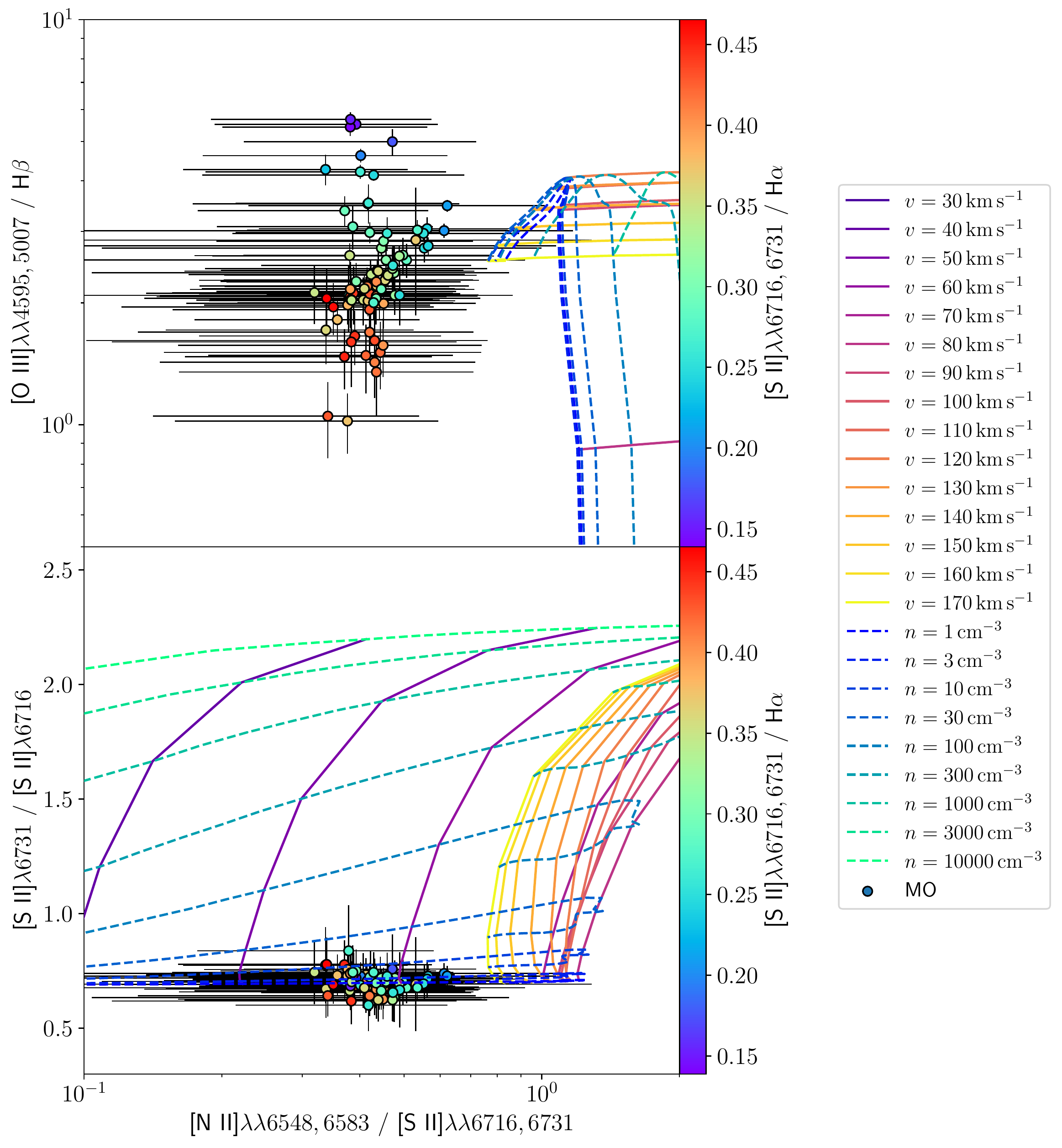}
	\caption{
			\forb{O}{iii}$\uplambda \uplambda 4959,5007$/\hb{} (top) and \forb{S}{ii}$\uplambda 6731$/\forb{S}{ii}$\uplambda 6716$ (bottom) ratios as a function of \forb{N}{ii}$\uplambda \uplambda 6548,6583$/\forb{S}{ii}$\uplambda \uplambda 6716,6731$ for spaxels in MO (circles with $1\sigma$ error bars, coloured by S2 ratio). The grids indicate line ratios from the slow shock models of \citet{Dopita&Sutherland2017}, varying in shock speed (solid lines) and pre-shock density (dashed lines).
	} 
	\label{fig: paper 3: DS17 shock model line ratio diagrams}
\end{figure*}

Although the \citet{Dopita&Sutherland2017} models appear to overlap with the data in the ODDs, slow shocks with pre-shock densities of $n = 1-10 \, \rm cm^{-3}$ models (consistent with the observed electron density estimated in Section\,\ref{subsec: paper 3: Electron density}) would tend to reduce the \forb{S}{ii} and \forb{O}{i} whilst enhancing the \forb{N}{ii}, contrary to our observations.
Furthermore, as shown in Fig.~\ref{fig: paper 3: DS17 shock model line ratio diagrams}, whilst the \forb{N}{ii}/\forb{S}{ii} ratio is consistent with shock speeds of $v_s \approx 55 \,\rm km\,s^{-1}$ and a pre-shock density of $n = 1\,\rm cm^{-3}$, shocks of this speed do not generate any measurable \forb{O}{iii} emission. Rather, the measured \forb{O}{iii}/\hb{} ratios are consistent with $v_s \gtrapprox 70 \,\rm km\,s^{-1}$; shocks of this speed over-predict the observed \forb{N}{ii}/\forb{S}{ii} ratios by a factor $\approx 2-3$.
Similarly, none of the fast shock models of \citet{Allen2008} would produce the enhanced \forb{S}{ii}/\ha{} and \forb{O}{i}/\ha{} line ratios whilst keeping the \forb{N}{ii}/\ha{} within the range of observed values.

% ODDs 

% Other line ratio diagrams

Although simultaneously low \forb{N}{ii}/\ha{} and high \forb{O}{iii}/\hb{} could theoretically arise from a mixture of shock speeds resulting from the shock passing through an inhomogeneous medium, any significant contribution from shocks sufficiently fast to cause detectable \forb{O}{iii} would produce much higher line widths than observed ($\sigma \lessapprox 30 \,\rm km\,s^{-1}$; see Fig~\ref{fig: paper 3: Halpha kinematics and morphology}). 
We therefore conclude that shocks are highly unlikely to be the primary cause of the enhanced \forb{S}{ii}/\ha{} and \forb{O}{i}/\ha{} ratios.
However, we cannot rule out the presence of shocks that are too slow to contribute to the optical line emission. Further observations in the near-IR may reveal ro-vibrational \hh{} emission, which traces shocks too slow to dissociate molecular gas \citep[e.g.,][]{Flower&PineauDesForets2010,Zovaro2019a,Zovaro2019b}.

%%%%%%%%%%%%%%%%%%%%%%%%%%%%%%%%%%%%%%%%%%%%%%%%%%%%%%%

\subsubsection{Diffuse Ionised Gas}\label{subsubsec: paper 3: Diffuse Ionised Gas}

We now investigate whether the line ratios in MO are consistent with contamination from diffuse ionised gas (DIG), the low-density ionised medium outside \hii{} regions that is photoionised by escaped radiation from nearby star-forming regions. DIG is characterised by strong emission in low-ionisation species such as \forb{O}{i}, \forb{N}{ii} and \forb{S}{ii}, and weak emission in high-ionisation species such as \forb{O}{iii} \citep[e.g.,][]{Rand1998,Haffner2009}. 

The diffuse structure of MO implies that DIG is highly likely to contribute to the line emission, particularly in the regions in between the filaments visible in  \textit{HST} imaging (see the bottom panel of Fig.~\ref{fig: paper 3: HST overlay}). Indeed, as shown in Fig.~\ref{fig: paper 3: S2 ratio spatial distribution}, the spaxels with the highest \forb{S}{ii}/\ha{} ratios are located away from the brightest star forming regions, and in the areas away from the bar and in between the filaments, as expected of DIG.

\begin{figure*}
	\includegraphics[width=1\linewidth]{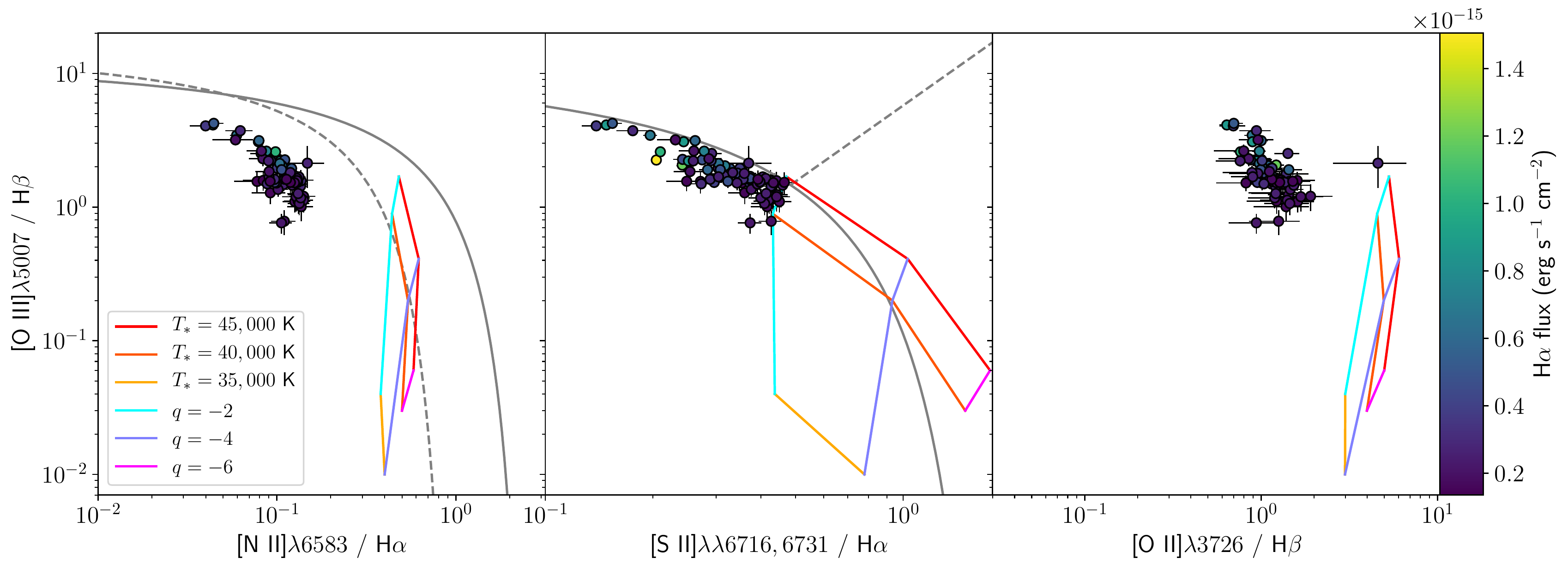}
	\caption{
		\forb{O}{iii}$\uplambda 5007$/\hb{} ratio as a function of the \forb{N}{ii}$\uplambda 6583$/\ha{}, \forb{S}{ii}$\uplambda \uplambda 6716,6731$/\ha{} and \forb{O}{ii}$\uplambda 3726$/\hb{} ratios in each spaxel in MO (circles with $1\sigma$ error bars, coloured by the \ha{} flux). The grids indicate the DIG photoionisation models of \citet{Mathis1986} for different combinations of the parameter $q$ (a proxy for the ionisation parameter) and the temperature of the ionising star $T_*$. The grey solid and dashed lines are as in Fig.~\ref{fig: paper 3: ODD diagrams}.
	}
	\label{fig: paper 3: ODD w. DIG models}
\end{figure*}

In Fig.~\ref{fig: paper 3: ODD w. DIG models} we present a ODD diagram overlaid with the DIG photoionisation models of \citet{Mathis1986}, where the \forb{O}{i}$\uplambda 6300$/\ha{} ratio has been replaced by the \forb{O}{ii}$\uplambda 3726$/\hb{} ratio because \forb{O}{i}$\uplambda 6300$ is not included in the models.
The DIG models are parametrised by the temperature $T_*$ of the central O-type star in the model \hii{} region, and the parameter $q$, which is a proxy for the ionisation parameter; $q \approx 1$ in regular \hii{} regions, whereas for the models used here, $-6 \leq q \leq -2$.
The spaxels with line ratios most similar to the DIG models tend to have low \ha{} surface brightnesses and lie outside the main star-forming regions round the outskirts of the object, or in between the filaments (see Fig.~\ref{fig: paper 3: Halpha kinematics and morphology}), as expected for DIG. However, the models are unable to explain the enhancement in the \forb{S}{ii} relative to the \forb{N}{ii} and \forb{O}{ii}. We therefore rule out DIG as the main contributor to the unusual line ratios in MO.

%%%%%%%%%%%%%%%%%%%%%%%%%%%%%%%%%%%%%%%%%%%%%%%%%%%%%%%

\subsubsection{X-ray photoionisation}

% Signatures of X-ray photoionistaion
We now consider whether X-ray photoionisation from the nucleus of NGC\,541 can explain the observed line ratios.
X-ray photons produced via inverse Compton scattering in the corona of the AGN have a long mean-free-path length in the ISM, resulting in a very hard radiation field in the outskirts of the nebula that creates an extended, weakly ionised transition region \citep{Dopita&Sutherland2003}.
Although this region has a low ionisation parameter, inner-shell photoionisations from X-ray photons create high-energy electrons that trigger cascades of collisional excitations, leading to enhanced forbidden line emission in species with ionisation potentials lower that that of \hi{}. In particular, \forb{S}{ii} and \forb{O}{i} are enhanced, whereas \forb{N}{ii} and \forb{O}{iii} are relatively weak, as is the case in MO. 

Due to enhanced collisional excitation, X-ray photoionisation results in a Balmer decrement larger than the case B recombination value of 2.85 \citep{Dopita&Sutherland2003}. Therefore, if X-ray photoionisation from the AGN of NGC\,541 is the dominant secondary excitation mechanism, the $A_V$, which was estimated assuming the case B recombination Balmer decrement, will be correlated with the S2 ratio, which is indeed observed (see Figs.~\ref{fig: paper 3: reddening} and \ref{fig: paper 3: S2 ratio spatial distribution}). 

Due to its significant distance from the AGN, existing AGN photoionisation models, which generally simulate nebulae close to the AGN in the host galaxy, cannot be applied to MO. 
%We are therefore unable to conduct a quantitative analysis without first creating a set of AGN photoionisation models better-suited to the conditions in MO.
A program such as \textsc{Mappings} \citep{Sutherland2018} can be used to simulate a photoionised nebula placed an arbitrary distance from the AGN, where the AGN SED can be simulated using existing models \citep[e.g., those of][]{Jin2012}. 
However, such models require numerous parameters to constrain the properties of the nebula and AGN SED, many of which cannot be reliably estimated from existing observations of MO and NGC\,541. 
%For example, in MO, the \forb{S}{ii} ratio only provides an upper limit to the ISM density ($n \lesssim 10 \,\rm cm^{-3}$). 
%%, and we did not detect the \forb{O}{iii}$\uplambda 4363$ line which is required to make a direct temperature estimate.
%Meanwhile, the AGN SED is controlled by the Eddington ratio, the presence of a soft X-ray excess, the radius of the hot corona, and other parameters that cannot be estimated from existing observations of NGC\,541. 
%In particular, the jet axis of NGC\,541 suggests that the accretion disc's axis of rotation is transverse to the line-of-sight, obscuring our view of the accretion disc, thereby meaning we cannot directly measure the X-ray properties of the AGN. 
%This is apparent from the measured X-ray luminosity of NGC\,541 \citep[$L_{\rm X} = 10^{40.83} \,\rm erg\,s^{-1}$,][]{Hudaverdi2006}, which is consistent with a combination of low-mass X-ray binaries and hot ISM alone \citep{Matsushita2001}. 
Hence, determining whether AGN photoionisation can reproduce the observed line ratios in MO requires a detailed exploration of the model parameter space, which will be the subject of future work.

%%%%%%%%%%%%%%%%%%%%%%%%%%%%%%%%%%%%%%%%%%%%%%%%%%
% Concluding sentence
\vspace{\baselineskip}
In conclusion, star formation most likely dominates the line emission in MO, but the enhanced \forb{S}{ii}/\ha{} and \forb{O}{i}/\ha{} ratios indicate the presence of other additional ionising sources. 
Neither shock nor DIG models can simultaneously produce the enhanced \forb{S}{ii}/\ha{} and \forb{O}{i}/\ha{} ratios whilst keeping the \forb{N}{ii}/\ha{} ratios within the range of observed values. 
X-ray photoionisation from the AGN of NGC\,541 may contribute, but further modelling is required to confirm this. 
Given the unusual location and structure of MO, it is possible that the observed line ratios may be a result of gas photoionised by a combination of star formation, AGN and DIG, which is then shocked by the passage of the jet. 
Modelling such a complex system is beyond the scope of this paper; additional observations will be required to constrain the free parameters in such a model.

The presence of both an intrinsic metallicity variation, and possible ionisation from a range of sources other than star formation, makes MO too complex for us to quantify the contribution of shocks, DIG or AGN photoionisation to the line emission using a simple `mixing' analysis~\citep[e.g.,][]{Davies2017}. 
However, \citet{Kewley&Ellison2008} found that only an AGN photoionisation contribution greater than $15 \,\rm per \, cent$ can produce line ratios lying above the maximum starburst line of \citet{Kewley2001} in all three ODDs shown in Fig.~\ref{fig: paper 3: ODD diagrams}. Since the harder radiation fields of AGN have a similar effect on the line ratios as shocks and DIG, and because there are very few spaxels with line ratios above the maximum starburst line by greater than $1\sigma$, we therefore place an upper limit of $15 \,\rm per \, cent$ on the contribution of non-stellar ionisation sources in MO. 

%%%%%%%%%%%%%%%%%%%%%%%%%%%%%%%%%%%%%%%%%%%%%%%%%%
\subsection{Star formation rate}\label{subsec: paper 3: Star formation rate}

We calculated the global SFR using the same calibration used by \citetalias{Croft2006} (${\rm SFR(H\upalpha)} = (7.0 - 7.9) \times 10^{-42}L( {\rm H\upalpha})$), which has a systematic uncertainty of approximately $30 \rm \, per \, cent$ due to uncertainty in the shape of the IMF at the high-mass end.
Using the total \ha{} flux gives $\rm SFR = 0.28 \pm 0.09 \,\rm M_\odot\,yr^{-1}$. As discussed in Section~\ref{subsec: paper 3: Excitation mechanism}, the \ha{} flux is mildly contaminated by shocks, DIG and/or X-ray photoionisation; hence this should be treated as an upper limit. We therefore estimate a conservative lower limit by assuming 15\,per\,cent of the \ha{} flux is due to ionisation sources other than star formation, yielding a SFR in the range $(0.24-0.28) \pm 0.09 \,\rm M_\odot\,yr^{-1}$.

% Depletion time
This corresponds to a \hi{} depletion time $t_{\rm dep, \, H\,I} \sim 1.7 - 2.0 \,\rm Gyr$~\citepalias[using the estimated \hi{} mass of ][]{Croft2006}. Using the reported \hh{} masses of \citetalias{Lacy2017} using $\alpha_{\rm CO}$ values appropriate for half-solar- and MW-like metallicity galaxies yields \hh{} depletion times of $t_{\rm dep, \, H_2} \sim 0.64 - 0.75 \, \rm Gyr$ and $t_{\rm dep, \, H_2} \sim 0.11 - 0.12 \, \rm Gyr$ respectively.

Our SFR estimate and total \ha{} flux ($\left(4.79 \pm 0.1\right) \times 10^{-14}\rm\,erg\,s^{-1}\,cm^{-2}$) are approximately half that of \citetalias{Croft2006} ($0.5 \,\rm M_\odot \, yr^{-1}$ and $8.6 \times 10^{-14}\rm\,erg\,s^{-1}\,cm^{-2}$ respectively). The reason for this discrepancy is unclear. Our flux estimate is closer to that of \citet{vanBreugel1985}, who reported a combined \ha{}+\forb{N}{ii} flux of $3.2 \times 10^{-14} \rm\, erg \, s^{-1} \, cm^{-2}$.

%Moreover, the SFR is derived assuming that MO has a conventional initial mass function (IMF).
\citetalias{Lacy2017} claimed that the recent type-II supernova (SN) explosion 2010ib~\citep{2010CBET.2464....1C,2010CBET.2490....2G} is evidence that MO may have a top-heavy IMF due to the improbability of observing a SN explosion in an object with such a low SFR. 
However, the best-fit star formation history (SFH) from our \ppxf{} analysis (Section~\ref{sec: paper 3: Stellar population analysis}) indicates that the majority of the stellar mass formed during a period of $\sim 1-2 \,\rm Myr$ approximately $10^7 \,\rm yr$ ago, during which the SFR would have been much greater than the present-day value. 
Modelling MO as a solar metallicity $10^7\,\rm M_\odot$ SSP with an instantaneous burst SFH and a Salpeter IMF using \textsc{Starburst99}~\citep{Leitherer1999}, the expected SN rate at $10^7\,\rm yr$ is approximately $0.03 \,\rm yr^{-1}$, corresponding to one SN explosion every $\sim 30\,\rm yr$. Therefore, SN explosion 2010ib is consistent with MO having a conventional IMF.

%A top-heavy IMF would also result in shocks driven by stellar winds from massive stars, and may partially explain the unusual line ratios.

% Note about top-heavy IMF...

%I had a look at the SN rate plot from SB99, and for an ~10^7 Msun instantaneous burst, the SN rate peaks at ~10^7 yr with a frequency of ~0.03/yr, so based on this, we should expect to see 1 SN explosion every 30 years - therefore the occurrence of an SN explosion in MO isn’t all that unexpected. 

%%%%%%%%%%%%%%%%%%%%%%%%%%%%%%%%%%%%%%%%%%%%%%%%%%
\subsection{Metallicity and ionisation parameter}\label{subsec: paper 3: Metallicity and ionisation parameter}

We followed the iterative method of \citet{Kewley&Dopita2002} to simultaneously estimate the gas-phase oxygen abundance \logoh{} (henceforth referred to as metallicity) and dimensionless ionisation parameter \logu{} in each spaxel using the theoretical diagnostics provided by \citet{KewleyNichollsSutherland2019} assuming a constant pressure $\log(P/k) = 5.0$, which is appropriate for typical \hii{} regions.

Strong-line abundance and ionisation parameter diagnostics are derived from models of standard \hii{} regions, and may yield spurious results if the line emission is contaminated by processes other than star formation. 
As discussed in Section~\ref{subsec: paper 3: Excitation mechanism}, we estimate the contribution to the line emission from sources other than star formation to be less than $15 \,\rm per \, cent$. At this level, \citet{Kewley&Ellison2008} found that the N2O2 (\forb{N}{ii}$\uplambda\uplambda 6548,6583$/\forb{O}{ii}$\uplambda\uplambda 3726,3729$) metallicity diagnostic is robust to contamination from AGN photoionisation. We expect the harder radiation fields from shocks and DIG to have a similar effect on the \forb{N}{ii}/\forb{O}{ii} ratio.
If this is the case, then the N2O2 diagnostic can be used reliably in MO; however, we note that our metallicity estimates are uncertain because we are unable to quantify the amount of contamination from sources other than star formation. 

\begin{figure}
	\centering
	\includegraphics[height=0.30\textheight]{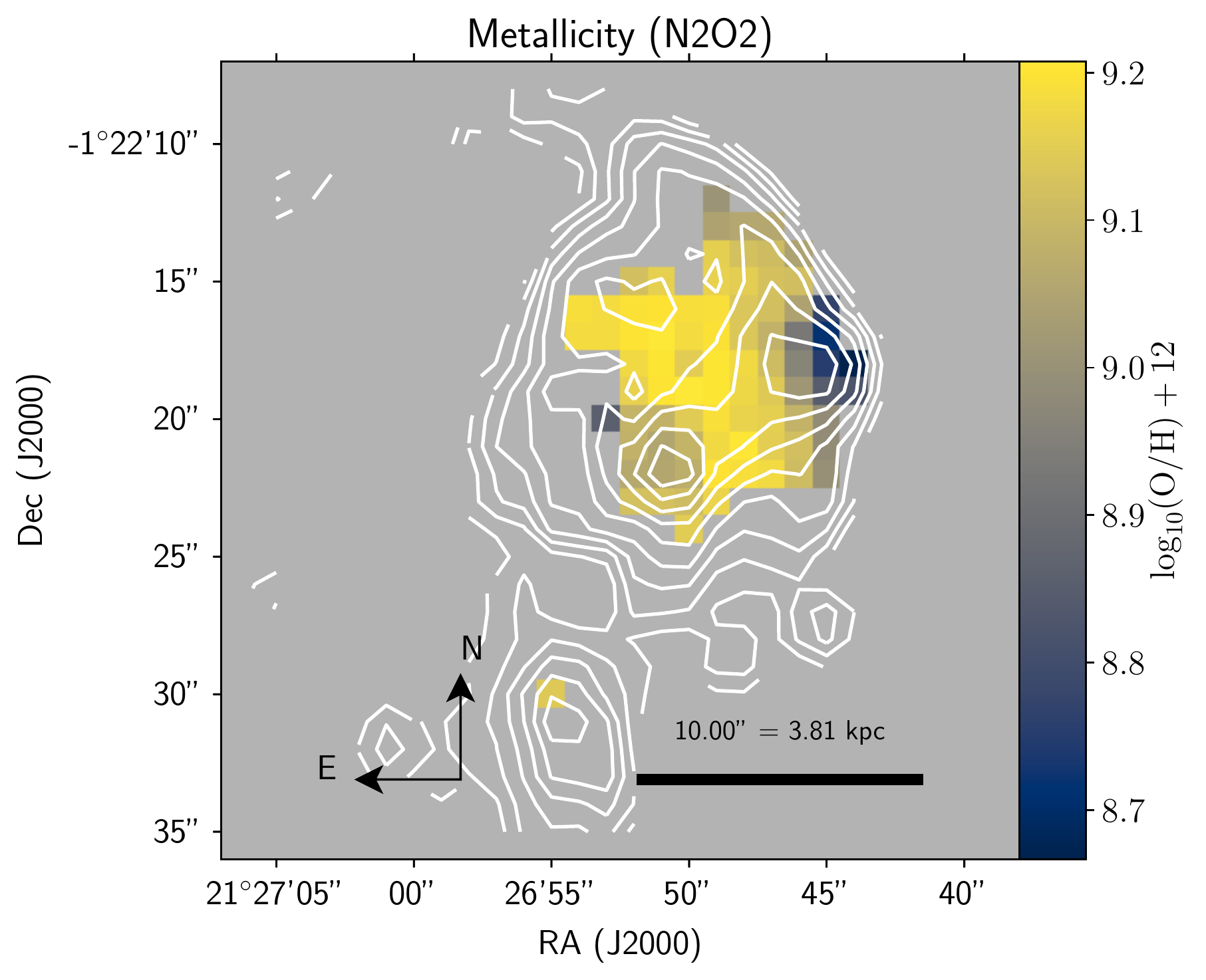}
	\includegraphics[height=0.30\textheight]{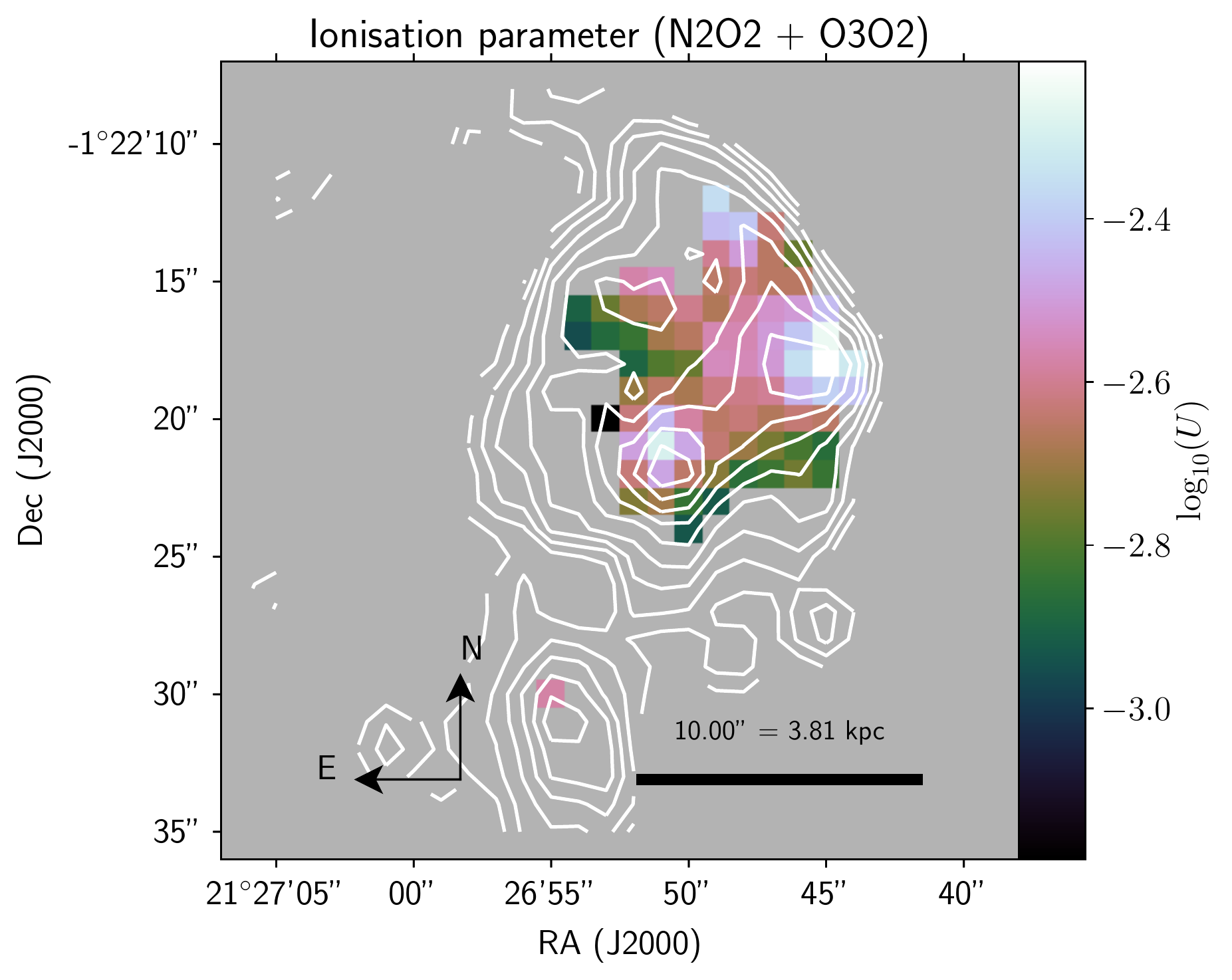}
	\caption{Gas-phase metallicity of MO (top) calculated using the N2O2 diagnostic of \citet{KewleyNichollsSutherland2019} and the dimensionless ionisation parameter (bottom) calculated using the O3O2 diagnostic of \citet{KewleyNichollsSutherland2019}. The white contours show the \ha{} flux on a log scale.}
	\label{fig: paper 3: metallicity & ionisation parameter}
\end{figure}

The bottom panel of Fig.~\ref{fig: paper 3: metallicity & ionisation parameter} shows $\logu{}$ estimated using the O3O2 (\forb{O}{iii}$\uplambda\uplambda 4959,5007$/\forb{O}{ii}$\uplambda\uplambda 3726,3729$) diagnostic of \citet{KewleyNichollsSutherland2019}.
The typical uncertainty in each spaxel is $0.02 \,\rm dex$.
In most parts of the object $\logu{} \approx -3.0$, which is consistent with the finding of \citetalias{Croft2006}; however, $\logu{} \approx -2.0$ in some regions, exceeding the values found in typical H~\textsc{ii} regions.

The top panel of Fig.~\ref{fig: paper 3: metallicity & ionisation parameter} shows the metallicity estimated using the N2O2 metallicity diagnostic.
The typical uncertainty in each spaxel is $0.01 \,\rm dex$.
There is significant variation in the metallicity across the object, with the lowest-metallicity regions lying near the Eastern and Western edges of the object, and $\logoh{}$ increasing by up to $\sim 0.5\,\rm dex$ in the centre. 
Metallicity estimates using other diagnostics including the N2 (\forb{N}{ii}$\uplambda 6583$/\ha{}), O2S2 (\forb{O}{ii}$\uplambda \uplambda 3726,3729$/\forb{S}{ii}$\uplambda \uplambda 6716,6731$), O3N2 ((\forb{O}{iii}$\uplambda 5007$/\hb{})/(\forb{N}{ii}$\uplambda 6583$/\ha{})) and R23 ((\forb{O}{ii}$\uplambda \uplambda 3726,3729$ + \forb{O}{iii}$\uplambda \uplambda 4959,5007$)/\hb{}) are presented in Appendix \ref{appendix: Additional metallicity diagnostics}, revealing similar metallicity variations within the object. The magnitudes of the measured metallicity variations for the N2O2, R23, O2S2, N2 and O3N2 diagnostics are 0.55\,dex, 0.43\,dex, 0.62\,dex, 0.33\,dex and 0.36\,dex respectively; we note that diagnostics other than N2O2 are more susceptible to contamination from sources other than star formation, and may therefore under- or over-estimate the intrinsic variation in \logoh{}.

We caution that strong-line diagnostics do not provide accurate estimates of the absolute metallicity due to systematic uncertainties associated with the models used to derive these diagnostics~\cite{Kewley&Ellison2008,Kewley2019}. This effect is apparent from the significant variations in the absolute metallicities produced by the different diagnostics. None the less, these diagnostics can be used to provide accurate estimates of the variation in \logoh{} within the object.

As noted previously, contamination from sources other than star formation may bias our metallicity estimates.
As shown in Fig.~\ref{fig: paper 3: ODD diagrams}, the N2O2 metallicity is strongly correlated with the S2 ratio. Comparison of Fig.~\ref{fig: paper 3: S2 ratio spatial distribution} and Fig.~\ref{fig: paper 3: metallicity & ionisation parameter} also shows that the higher-metallicity and higher-S2 regions are located away from the main star-forming regions, and would therefore have line ratios more strongly contaminated by processes other than star formation. It is therefore possible that the metallicity is being over-estimated in these regions. 

%%%%%%%%%%%%%%%%%%%%%%%%%%%%%%%%%%%%%%%%%%%%%%%%%%
\section{Possible explanations for the metallicity variation}\label{sec: paper 3: The cause of the metallicity variation}

In this section, we explore several scenarios that could produce the large metallicity variation observed within MO.

\subsection{Enrichment from star formation}\label{subsubsec: paper 3: Star formation?}
Star formation proceeding from SE-NW could potentially cause the metallicity variation, as regions with more recent star formation would have had less time to enrich the ISM with metals via supernova explosions and stellar winds.
Although we were unable to spatially resolve MO in our \ppxf{} analysis (Section~\ref{sec: paper 3: Stellar population analysis}), the global best-fit stellar population is dominated by a young population $\lesssim 10\,\rm Myr$ old, with a spread of only $1-2$ Myr (Fig.~\ref{fig: paper 3: ppxf integrated fit, age & metallicity}).

To determine whether an age spread of a few Myr in the stellar population could explain the observed metallicity variation, we used an SSP model to estimate the oxygen yield per stellar mass for stars of a given age and metallicity. 	
The SSP model considers enrichment contributions from various channels. Yields from type II supernovae \citep[including a prescription for Ibc;][]{Kobayashi2004} are calculated as a function of age and metallicity, where the age determines the main-sequence turn off mass using the metallicity-dependent lifetimes of \citet{Kodama&Arimoto1997}. 
A metallicity-dependent hypernova (HN) fraction is also included, with a greater contribution from HN at low metallicity \citep{Kobayashi&Nakasato2011}.
Winds from asymptotic giant branch stars further contribute \citep{Kobayashi2011}.
Type Ia supernovae are included in the model \citep{Kobayashi&Nomoto2009}, but do not contribute on the short timescales considered here.
We assume a \citet{Kroupa2001} IMF (slope $\alpha = 1.3$) in the range $0.01-120\,\rm M_\odot$, with an upper mass limit for core-collapse supernovae of $50\,\rm M_\odot$. 

\begin{figure}
	\centering
	\includegraphics[width=1\linewidth]{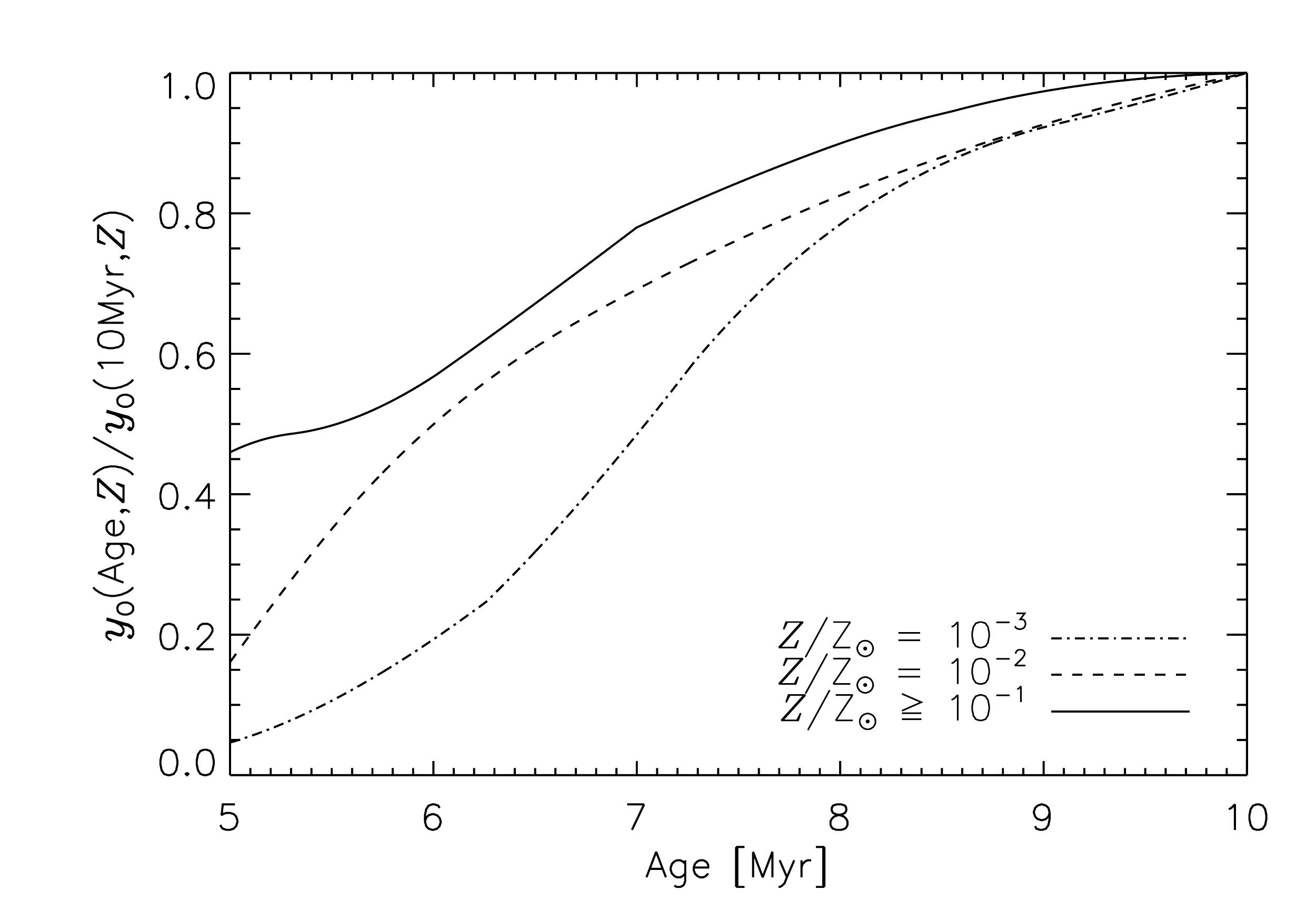}
	\caption{Oxygen yield per stellar mass (normalised to the yield at 10 Myr) as a function of stellar age, estimated using the chemical enrichment model described in Section~\ref{subsubsec: paper 3: Star formation?}. The solid, dashed and dash-dotted lines show yields for stellar metallicities of $10^{-3}$, $10^{-2}$ and $ \geq 10^{-1} \,\rm Z_\odot$ respectively.}
	\label{fig: paper 3: MO yield}
\end{figure}

As shown in Fig.~\ref{fig: paper 3: MO yield}, for stellar metallicities $\geq 0.1 \,\rm Z_\odot$ (consistent with the gas-phase metallicity), the oxygen yield increases by only a factor of 2 over a period of 5 Myr, insufficient to explain the metallicity enhancement of $\sim 0.5 \,\rm dex$ in the centre of the object.
Stellar enrichment alone is therefore unable to explain the metallicity variation.

% Explanation 3: Entrainment of enriched gas along the jet
\subsection{An enriched jet-driven outflow from NGC\,541}
A metallicity variation in MO could have arisen if some of the gas from which it formed originated in a jet-driven outflow from NGC\,541.
X-ray studies of clusters have revealed metallicity enhancements of up to 0.2\,dex occurring along the path of jets from BCGs \citep[][]{Kirkpatrick2009,Kirkpatrick2011}, indicating that jets can transport enriched material up to several $100\,\rm kpc$ from the host. 
Hence it is plausible for enriched gas to be found along the jet of NGC\,541.
However, it is highly unlikely that any metal-enriched gas from NGC\,541 would become bound to MO given its very low stellar mass ($\sim 10^7 \,\rm M_\odot$). We therefore rule out this scenario.

\subsection{Pre-existing enriched gas in the stellar bridge}
MO may have formed from dense clumps of gas residing in the stellar bridge connecting NGC\,541 and NGC\,545/547, as we discuss further in Section~\ref{sec: paper 3: On the formation of Minkowski's Object}. 
This gas will have been stripped from the interacting galaxies, and would therefore be enriched relative to the ICM. Moreover, the recent type Ia supernova explosion 2018ctv \citep{2018ATel11762....1C,2018ATel12313....1M,2018ATel12315....1P} in the stellar bridge indicates ongoing stellar enrichment in the ICM of Abell\,194.

To check whether this is plausible, we compared the gas-phase metallicities in MO and in NGC\,541.
\citet{Hamer2016} observed NGC\,541 with VLT/VIMOS and obtained spatially resolved emission line fluxes, which we used to estimate the gas-phase metallicity.
%%%%%%%%%%%%%%%%%%%%%%%%%%%%%%%%%%%%%%%%%%%%%%%%%%%%%%%%%%%%%%%%
% THESIS ONLY
%%%%%%%%%%%%%%%%%%%%%%%%%%%%%%%%%%%%%%%%%%%%%%%%%%%%%%%%%%%%%%%%
%In Table~\ref{tab: paper 3: NGC 541 line fluxes}, we show the $\logoh{}$ computed from the line fluxes reported in their appendix F integrated over the entire extent of the galaxy (`total'), in the nucleus (`central') and the fluxes calculated as the difference between these two estimates (`extended'). 
%%%%%%%%%%%%%%%%%%%%%%%%%%%%%%%%%%%%%%%%%%%%%%%%%%%%%%%%%%%%%%%%
Due to the limited number of emission lines in the wavelength range of their observations, we used the N2 diagnostic, adopting a fiducial $\logu{} = -3.0$, appropriate for normal \hii{} regions. The metallicity in the outer regions of NGC\,541 ($\logoh \approx 8.8$) is approximately 0.2\,dex higher than that observed in MO using the same diagnostic (see Appendix \ref{appendix: Additional metallicity diagnostics}), meaning that the ISM in NGC\,541 is sufficiently enriched to explain the high observed metallicities in MO. However, this scenario cannot explain the metallicity variation of 0.5 within the object. Moreover, the N2 diagnostic is subject to contamination by the AGN; further observations of NGC\,541 are therefore required to observe emission lines enabling us to apply a metallicity diagnostic more robust to AGN contamination, such as N2O2. 

%%%%%%%%%%%%%%%%%%%%%%%%%%%%%%%%%%%%%%%%%%%%%%%%%%%%%%%%%%%%%%%%
% THESIS ONLY
%%%%%%%%%%%%%%%%%%%%%%%%%%%%%%%%%%%%%%%%%%%%%%%%%%%%%%%%%%%%%%%%
%\begin{table}
%	\centering
%	\caption{Gas-phase metallicity in different regions of NGC\,541 estimated using VLT/VIMOS emission line fluxes from \citet{Hamer2016}, where we have used the N2 diagnostic of \citet{KewleyNichollsSutherland2019} assuming $\logu{} = -3.0$.}
%	\begin{tabular}{c c}
%		\hline
%		\textbf{Region} & \textbf{$\logoh{}$} \\
%		\hline
%		Total & 8.9682 \\
%		Central & 9.2353 \\
%		Extended & 8.7814 \\ 
%		\hline
%	\end{tabular}
%	\label{tab: paper 3: NGC 541 line fluxes}
%\end{table}
%%%%%%%%%%%%%%%%%%%%%%%%%%%%%%%%%%%%%%%%%%%%%%%%%%%%%%%%%%%%%%%%

% Smackdown
\vspace{\baselineskip}
Having ruled out star formation, an enriched jet-driven outflow and enrichment of gas in the stellar bridge as the cause of the metallicity variation, its origin is uncertain, although we note that the magnitude of the variation may be over-estimated due to contamination from sources other than star formation.

%%%%%%%%%%%%%%%%%%%%%%%%%%%%%%%%%%%%%%%%%%%%%%%%%%
\section{On the formation of Minkowski's Object}\label{sec: paper 3: On the formation of Minkowski's Object}

%%%%%%%%%%%%%%%%%%%%%%%%%%%%%%%%%%%%%%%%
% NEW 

% Intro sentence: here we discuss possible formation theories etc.
%Whilst FR\,I-type radio galaxies are relatively common in local clusters, objects like MO are extremely rare, suggesting there is something unusual about the conditions in Abell\,194 that have enabled positive feedback to take place. 
Here, we discuss possible scenarios for the formation of MO.

% Condensation directly from the hot ICM
\citetalias{Croft2006} suggest that MO formed from the hot, tenuous ICM that underwent runaway cooling due to shocks induced by the passage of the jet. 
However, simulations by \citet{Fragile2004} suggest that MO would have required a pre-collision cloud density of approximately $10 \,\rm cm^{-3}$ to form via a jet interaction.
This density is substantially higher than estimated ICM densities in local clusters \citep[$\sim 10^{-3} - 10^{-2} \,\rm cm^{-3}$; e.g.,][]{Mohr1999}; it therefore appears unlikely that MO formed directly from the ICM.
MO may instead have formed from pre-existing dense clouds in the path of the jet. 
As shown in Fig.~\ref{fig: paper 3: Abell 194}, Abell\,194 harbours a stellar bridge connecting NGC\,541 and NGC\,545/547, which is aligned with the jet from NGC\,541 in projection. 
The stellar bridge may contain \hi{} stripped from the interacting galaxies, akin to the intra-group \hi{} observed in groups such as Stephan's Quintet~\citep{Sulentic2001}. MO is indeed associated with a large \hi{} cloud with mass $4.9 \times 10^8 \,\rm M_\odot$; such pre-existing \hi{} may have been sufficiently dense for jet-induced star formation to occur.

One of the only other known examples of jet-induced star formation, the string of OB associations in the Northern jet of Centaurus A, is believed to have formed due to a jet interaction with an \hi{} cloud that drifted into the jet's path.
Crucially, this cloud has been detected \textit{outside} the jet's path, and appears to be part of a larger ring of \hi{} clouds that may have formed during a past merger event~\citep{Schiminovich1994,Oosterloo&Morganti2005}, indicating that this gas existed prior to the jet interaction.
In contrast, all of the \hi{} associated with MO is spatially coincident with the jet; \citetalias{Croft2006} claim this is evidence that the jet triggered the formation of the cloud. However, \citetalias{Croft2006} remark that additional \hi{} could exist at velocities outside the low bandwidth ($500 \,\rm km\,s^{-1}$) of their observations, and that the \hi{} cloud may be part of a larger pre-existing \hi{} complex.

We note that although MO harbours 2--10 times the molecular gas mass of the Cen A filaments ($1.4 \times 10^7 \,\rm M_\odot$ in the Cen A filaments, \citet{Charmandaris2000}; $(0.3 - 1.8) \times 10^8 \,\rm M_\odot$ in MO, \citetalias{Lacy2017}), the Cen A filaments have a total $\rm SFR = 4 \times 10^{-3}\,\rm M_\odot \, yr^{-1}$~\citep{Salome2016a}, almost 100 times lower than that of MO. It is unclear why the star-formation efficiency should differ so drastically between the two objects. Unfortunately, as discussed in Section~\ref{subsec: paper 3: Metallicity and ionisation parameter}, we are unable to further constrain the molecular gas of MO mass with an updated CO-to-\hh{} conversion factor because we cannot reliably estimate the absolute gas-phase metallicity from our observations. The true star formation efficiency of MO therefore remains unclear, although we note that our new SFR estimate, which is approximately half that of \citetalias{Croft2006}, places MO significantly closer to the Kennicutt-Schmidt relation~\citep{Kennicutt1998} regardless of the assumed $\alpha_{\rm CO}$~\citepalias[see fig. 7 of][]{Lacy2017}.

% Jet interaction with cooling filaments 
A second possibility is that MO formed from pre-existing filaments in the ICM.
\ha{}-emitting filaments are regularly observed in cool-core clusters, and may trace cooling gas flowing into the BCG~\citep[e.g.,][]{Heckman1989}. Many filaments exhibit line emission with LINER-like line ratios, attributed to in-situ star formation and shocks~\citep[e.g.,][]{McDonald2012}. 
Jet-induced shocks could lead to enhanced star formation in these filaments. However, it seems unlikely that Abell\,194 would harbour such filaments, because it lacks the localised, enhanced X-ray emission associated with cooling gas~\citep{Nikogossyan1999,Sakelliou2008}.
Moreover, these filaments generally exhibit $\log($\forb{N}{ii}$\uplambda 6583$/\ha{}$) \sim -0.5 - 0.5$, much higher than observed in MO ($\log($\forb{N}{ii}$\uplambda 6583$/\ha{}$) \sim -1$), in addition to much broader line widths~\citep[$\sim 100 \,\rm km\,s^{-1}$; e.g.,][]{Heckman1989,Sabra2000,Farage2010,McDonald2012,Hamer2015}. We therefore conclude that it is unlikely for MO to have formed from such filaments.

% Pre-exiting DG
% What was it that Leisman+2017 said about the old stellar pops of hi-rich DGs?
The final scenario we consider is that MO is a gas-rich dwarf galaxy that is currently falling into Abell\,194, and has a temporarily elevated SFR due to the jet interaction.
In recent years, a population of isolated \hi{}-rich ultra-diffuse dwarf galaxies (HUDs) have been discovered, characterised by blue colours, extended ($\sim 10-30 \,\rm kpc$) clumpy morphologies and high gas fractions, with $M_{\rm H\,I} \sim 10^8 - 10^9 \,\rm M_\odot$ and $M_* \sim 10^7 - 10^8 \,\rm M_\odot$~\citep{Leisman2017}, similar to MO. However, with a surface brightness $\mu_g = 21.8 \,\rm mag \, arcsec^{-2}$, MO is much brighter than most HUDs, which by definition have $\mu_g \geq 24 \,\rm mag \, arcsec^{-2}$. 
If MO truly originated from a HUD, its current surface brightness may be temporarily elevated due to jet-induced star formation.

Under this formation scenario, we may speculate on the fate of MO.
As they fall into clusters, gas-rich dwarfs are predicted to rapidly lose their gas due to ram-pressure stripping or tidal interactions, eventually transforming into ultra-diffuse galaxies~\citep{Leisman2017,Roman&Trujillo2017}: extended, low surface-brightness, quiescent galaxies which are numerous in local clusters~\citep[e.g.,][]{vanDokkum2015,Koda2015}. 
%Therefore, if MO is indeed a HUD, it may eventually transform into one of these ultra-diffuse galaxies.
%To test this hypothesis, we predict the future surface brightness of MO assuming passive evolution.
To model the quenching of MO as it falls into Abell\,194, we used a \textsc{Starburst99}~\citep{Leitherer1999} solar metallicity SSP model with a Salpeter IMF and an instantaneous SFH. 
Given the present-day $V$ band surface brightness of $\mu_V = 22.7 \,\rm mag \, arcsec^{-2}$, we predict MO to fade to $\mu_V = 24.2 \,\rm mag \, arcsec^{-2}$ and $\mu_V = 25.7 \,\rm mag \, arcsec^{-2}$ in approximately 100 Myr and 1 Gyr respectively, firmly placing MO into the surface brightness regime of ultra-diffuse galaxies. The correspondingly low stellar mass surface densities suggests that MO will eventually become disrupted as it falls into the cluster.
% Visibility?

If MO is truly a gas-rich dwarf that has wandered into the path of the jet, we may be observing it during a brief phase where it is bright enough to see in shallow observations. It is therefore possible that there are many objects similar to MO that have now faded, and that MO is perhaps not as unique as is currently believed.

% DM fraction
Establishing whether MO resides in its own dark matter (DM) halo would enable us to distinguish between these different formation scenarios. If MO formed from the ICM, then it would not be associated with a DM halo, whereas ultra-diffuse galaxies are known to span a range of DM halo masses~\citep{Zaritsky2017}. Unfortunately, the irregular morphology of MO makes it difficult to measure its inclination using our WiFeS observations, meaning we cannot constrain the gravitational mass of MO from its rotation curve. Further observations and modelling of MO and Abell\,194 will be required to distinguish between these different formation scenarios.

%%%%%%%%%%%%%%%%%%%%%%%%%%%%%%%%%%%%%%%%%%%%%%%%%%
\section{Conclusion}\label{sec: paper 3: Conclusion}

% Keep it short; 1-2 paragraphs
We conducted a study of Minkowski's Object (MO), an peculiar star-forming dwarf galaxy located along the path of a jet from the nearby radio galaxy NGC\,541, using optical integral field spectroscopy. 

% Stellar pop
Analysis of the stellar continuum with \ppxf{} confirms that MO is dominated by a predominantly young stellar population $\sim 10\,\rm Myr$ old, indicating that the bulk of its stellar mass formed during a recent jet interaction. A secondary population at $\sim 1 \,\rm Gyr$ is also present, and may represent stars in the stellar bridge between NGC\,541 and NGC\,545/547 or a pre-existing stellar population associated with MO.

In most parts of the object, the emission line ratios are consistent with star formation, although some regions exhibit enhanced \forb{S}{ii}/\ha{} and \forb{O}{i}/\ha{} ratios suggesting the presence of non-stellar ionising sources. 
A complex combination of shocks, DIG and X-ray photoionisation from the AGN of NGC\,541 may explain the line ratios, although further modelling is required to determine whether this is plausible.	

Strong-line metallicity diagnostics indicate a gas-phase metallicity that varies by $\sim 0.5\,\rm dex$ across the object, which cannot be explained by enrichment from in-situ star formation, enriched gas entrained in the jet from NGC\,541, nor the presence of pre-existing enriched gas in the stellar bridge, although we note the magnitude of the variation may be over-estimated due to contamination from sources other than star formation.

We explored several formation scenarios for MO, and concluded that it formed from either (a) pre-existing \hi{} clouds in the stellar bridge between NGC\,541 and NGC\,545/547 that were stripped from these galaxies during an interaction or (b) a gas-rich dwarf galaxy that wandered into the path of the jet. Further observations and modelling may enable us to distinguish between these two scenarios.

\section*{Acknowledgements}

The authors would like to thank Julie Banfield for leading the initial observing programme, Adam D. Thomas for providing assistance with \textsc{NebulaBayes}, and Chiaki Kobayashi for assistance with the SSP chemical enrichment modelling. We also thank the anonymous reviewer for their insightful comments and suggestions which have improved this work.

This research has made use of the NASA/IPAC Extragalactic Database, which is funded by the National Aeronautics and Space Administration and operated by the California Institute of Technology.

This research made use of QFitsView\footnote{\url{https://www.mpe.mpg.de/~ott/QFitsView/}}, a software package for reducing astronomical data written by Thomas Ott, 
and the \textsc{python} packages \textsc{NumPy}\footnote{\url{https://numpy.org/}}\,\citep{Harris2020}, \textsc{SciPy}\footnote{\url{http://www.scipy.org/}}\,\citep{Scipy2001}, and \textsc{Astropy},\footnote{\url{http://www.astropy.org}} a community-developed core package for Astronomy\,\citep{Astropy2013,Astropy2018}. 

Model generated with \textsc{Sed@.0} code\footnote{\textsc{Sed@} is a synthesis code included in the {\it Legacy Tool project} of the {\it Violent Star Formation Network}; see {\it \textsc{Sed@} Reference Manual} at \url{http://www.iaa.es/~mcs/sed@} for more information.} with the following inputs: IMF from \cite{Salpeter1955} in the mass range $0.1-120 \rm\, M_\odot$; High Resolution library from \citet{Martins2004,GonzalezDelgado2005} based on atmosphere models from \textsc{Phoenix} \citep{Hauschildt&Baron1999,Allard2001}, \textsc{Atlas9} \citep{Kurucz1991} computed with \textsc{Spectrum} \citep{Gray&Corbally1994}, \textsc{Atlas9} library computed with \textsc{Synspec} \citep{Hubeny&Lanz2011}, and \textsc{Tlusty} \citep{Lanz&Hubeny2003}. 
Geneva isochrones computed with the isochrone program presented in \cite{Meynet1995} and following the prescriptions quoted in \cite{Cervino2001} from the evolutionary tracks from \cite{Schaller1992} at $ Z=0.001$/$ Z=0.020$; \cite{Charbonnel1993} at $ Z=0.004$; \cite{Schaerer1993a} at $ Z=0.008$ and \cite{Schaerer1993b} at $ Z=0.040$.
Padova isochrones presented in \cite{Girardi2002}\footnote{Available at \url{http://pleiadi.pd.astro.it/}} based on the (solar scaled mixture) tracks from \cite{Girardi2000,Bertelli1994} that includes overshooting and a simple synthetic evolution of TP-AGB \cite{Girardi&Bertelli1998}.

Parts of this research were supported by the Australian Research Council Centre of Excellence for All Sky Astrophysics in 3 Dimensions (ASTRO 3D), through project number CE170100013. L. J. K. gratefully acknowledges the support of an ARC Laureate Fellowship (FL150100113). 

\section*{Data availability}

The data underlying this article will be shared on reasonable request to the corresponding author.


% The best way to enter references is to use BibTeX
\bibliographystyle{mnras}
\bibliography{bibliography.bib}

%%%%%%%%%%%%%%%%%%%% APPENDICES %%%%%%%%%%%%%%%%%%

\appendix
\section{Additional \ppxf{} fits}\label{appendix: Additional PPXF fits}

In Fig.~\ref{fig: A4: ppxf integrated fit, age & metallicity (Padova)} we show the histogram and spectrum of the best-fit combination of stellar templates generated using \ppxf{} with Padova isochrones. The results are very similar to those obtained using the Geneva isochrones.

We also repeated our analysis using stellar templates with constant metallicity using both the Geneva and Padova isochrones. The best-fit SFHs, shown in Fig.~\ref{fig: A4: ppxf 1D fits}, are similar regardless of the metallicity and isochrones used.

\begin{figure*}
	\centering
	\includegraphics[width=1\linewidth]{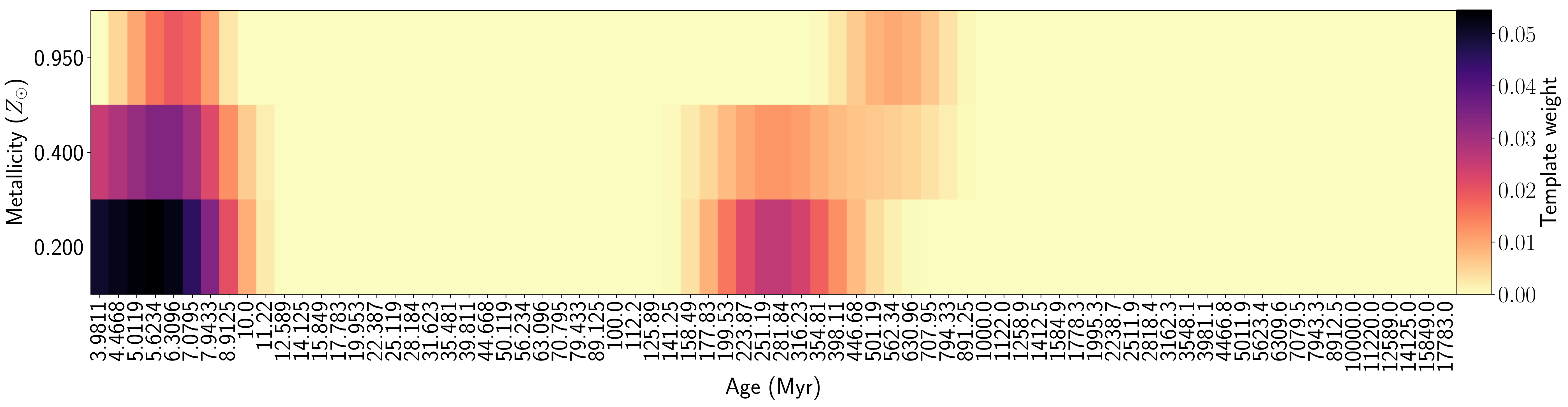}
	\includegraphics[width=1\linewidth]{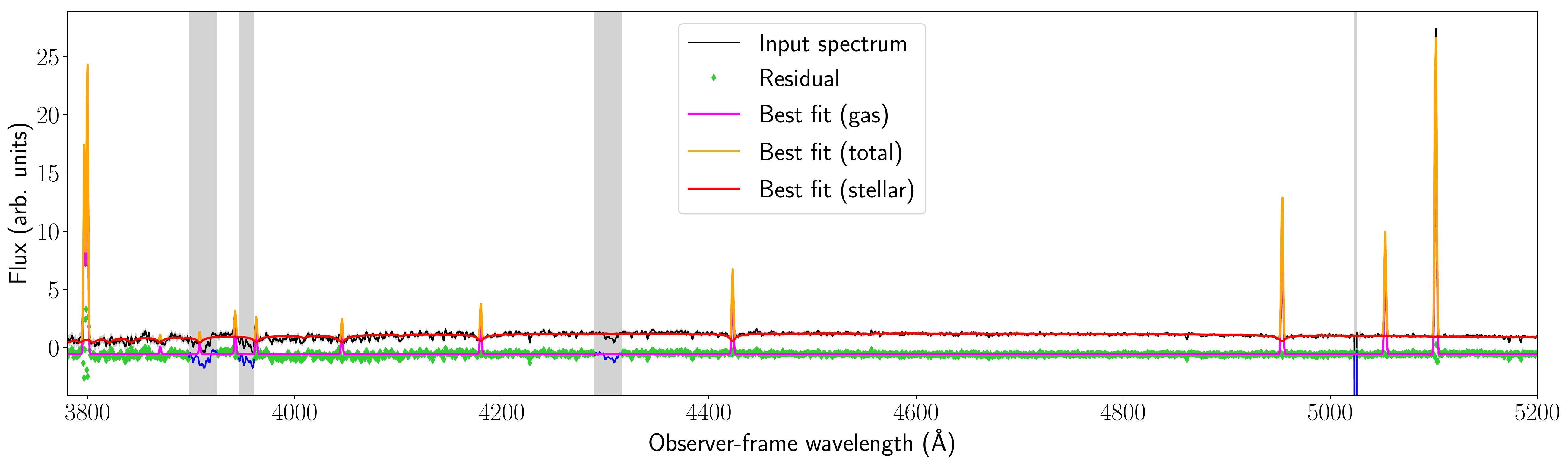}
	\caption{
		Results of the \ppxf{} fit used to estimate the age and metallicity of the stellar population using the Padova isochrones. Figure details are given in Fig.~\ref{fig: paper 3: ppxf integrated fit, age & metallicity}.
	}
	\label{fig: A4: ppxf integrated fit, age & metallicity (Padova)}
\end{figure*}
\begin{figure*}
	\centering
	\includegraphics[width=1\linewidth]{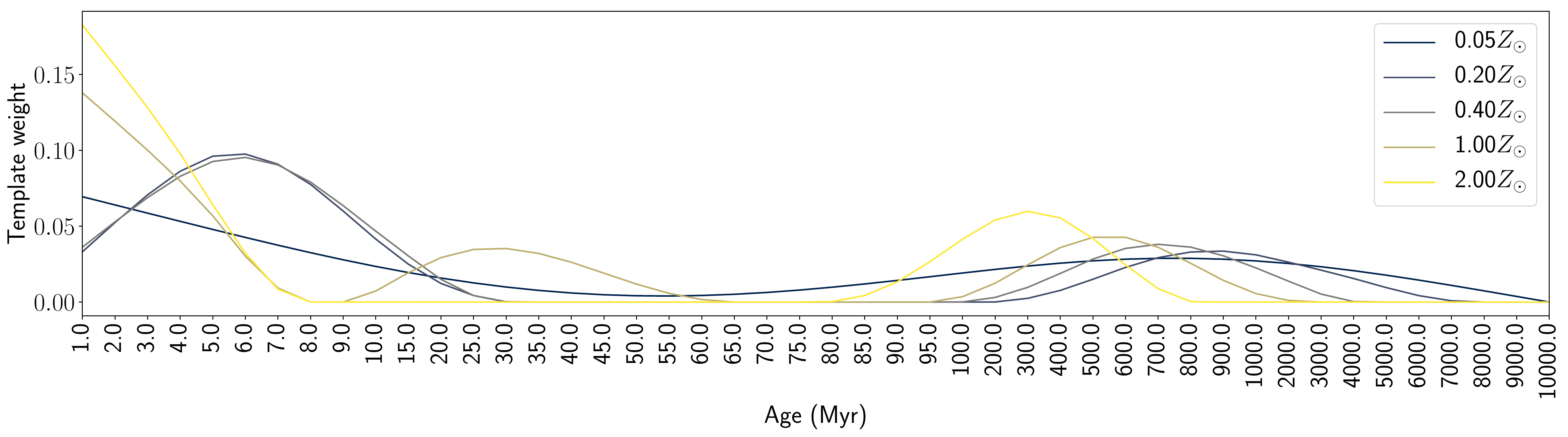}
	\includegraphics[width=1\linewidth]{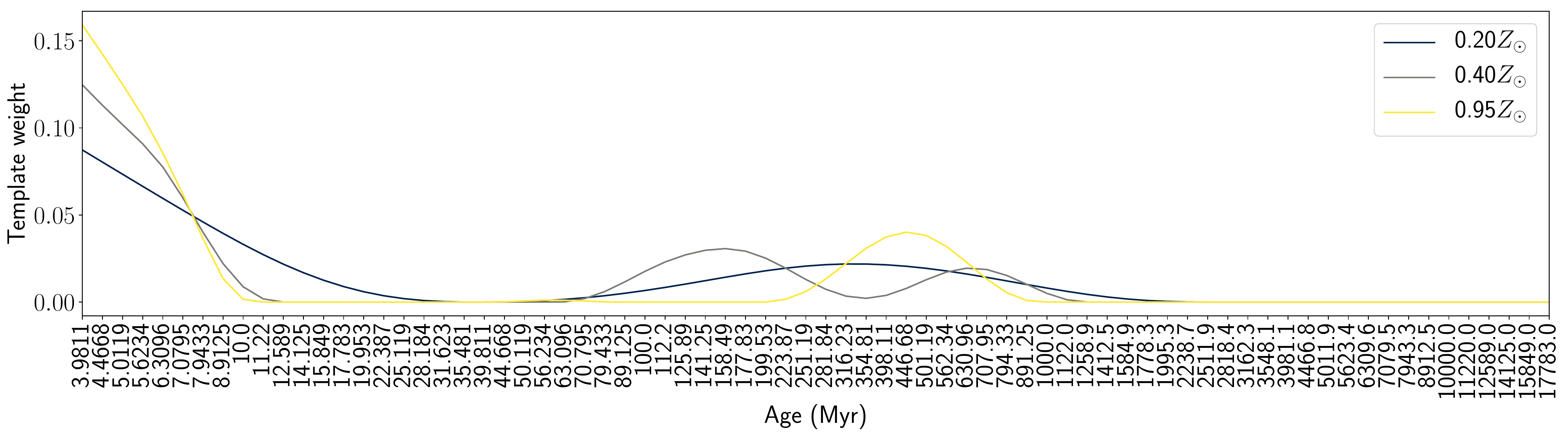}
	\caption{Best-fit SFHs generated using Geneva (top) and Padova (bottom) isochrones as a function of metallicity.}
	\label{fig: A4: ppxf 1D fits}
\end{figure*}

\section{Additional metallicity diagnostics}\label{appendix: Additional metallicity diagnostics}

\begin{figure*}
	\centering
	\includegraphics[height=0.23\textheight]{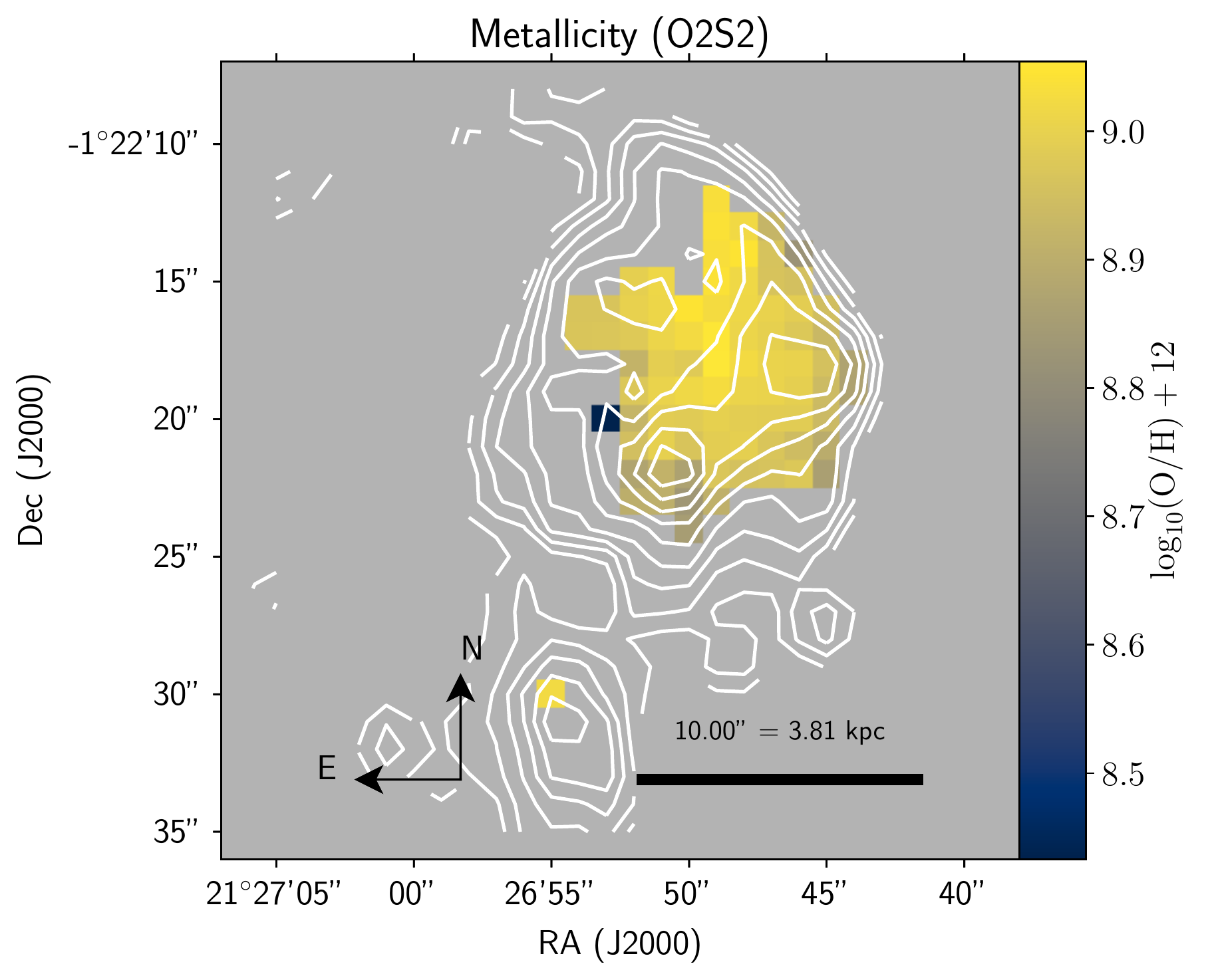}
	\includegraphics[height=0.23\textheight]{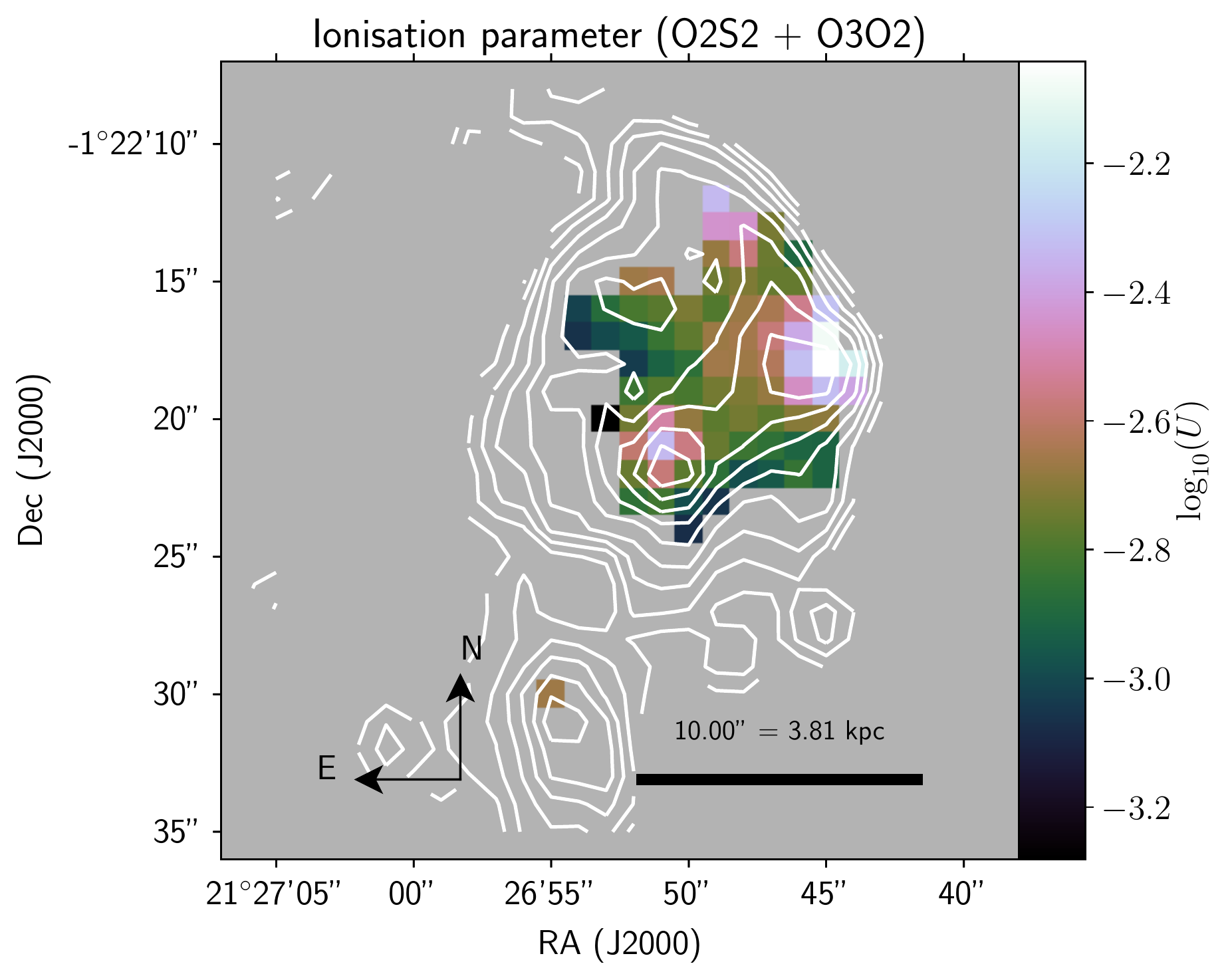}
	\includegraphics[height=0.23\textheight]{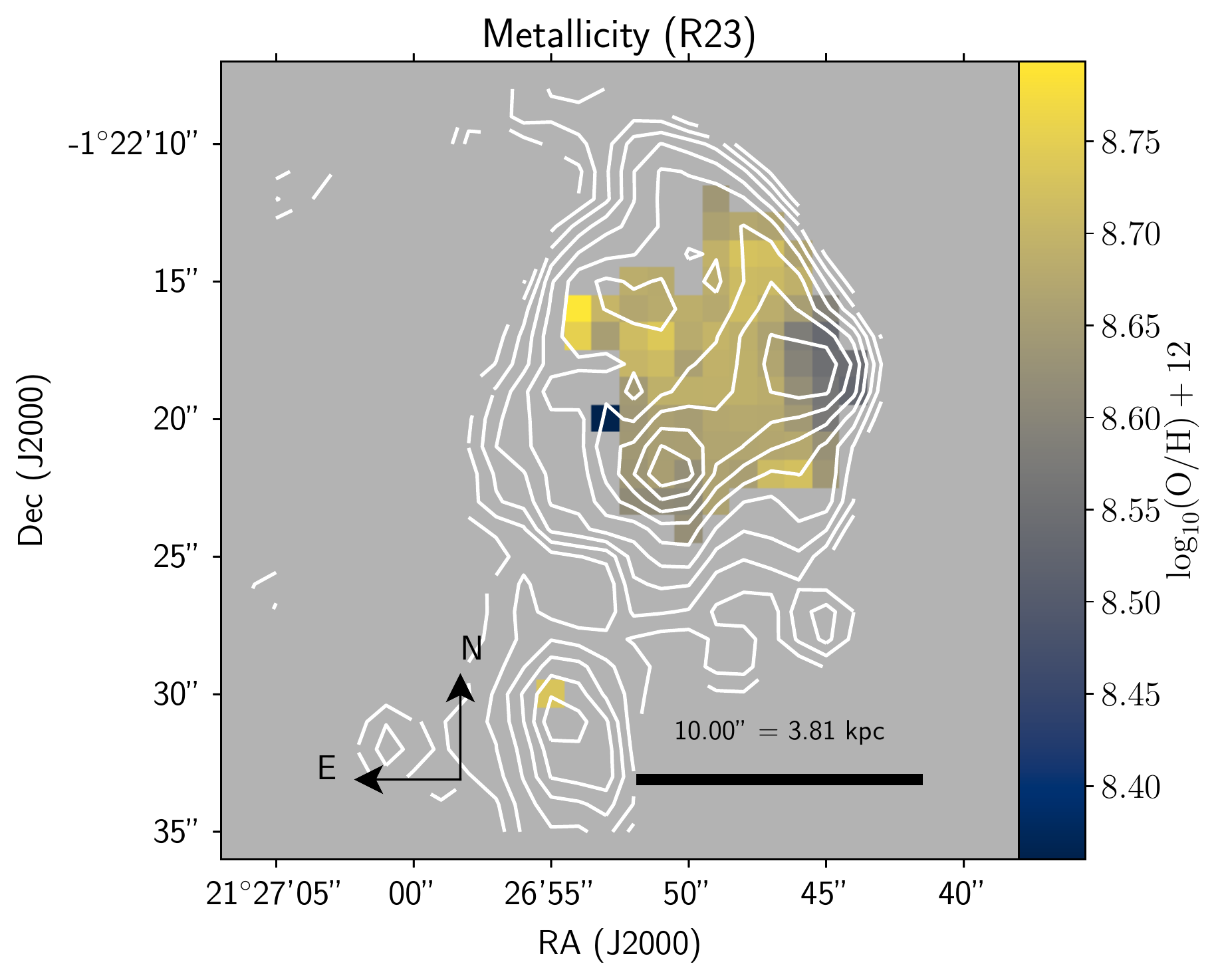}
	\includegraphics[height=0.23\textheight]{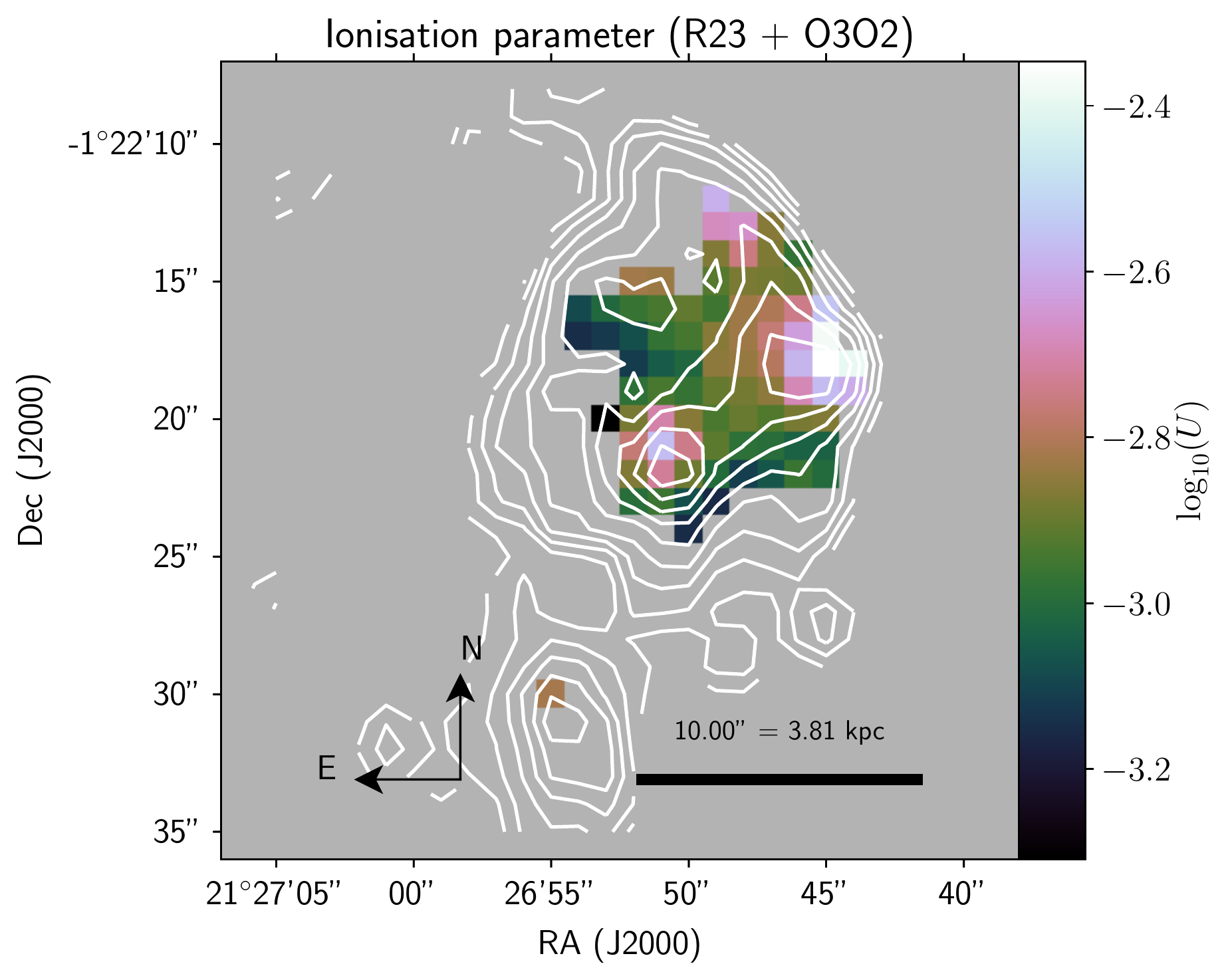}
	\includegraphics[height=0.23\textheight]{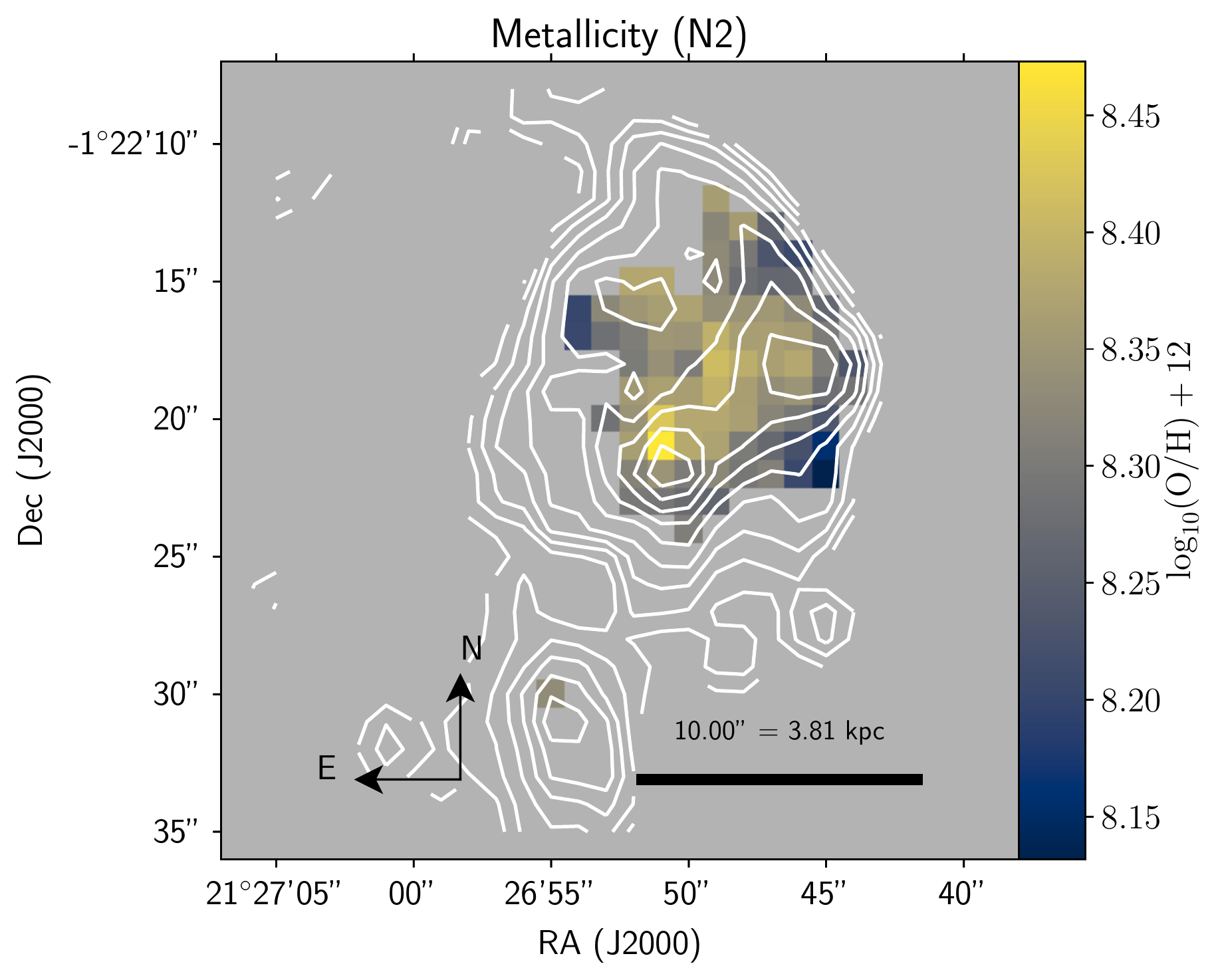}
	\includegraphics[height=0.23\textheight]{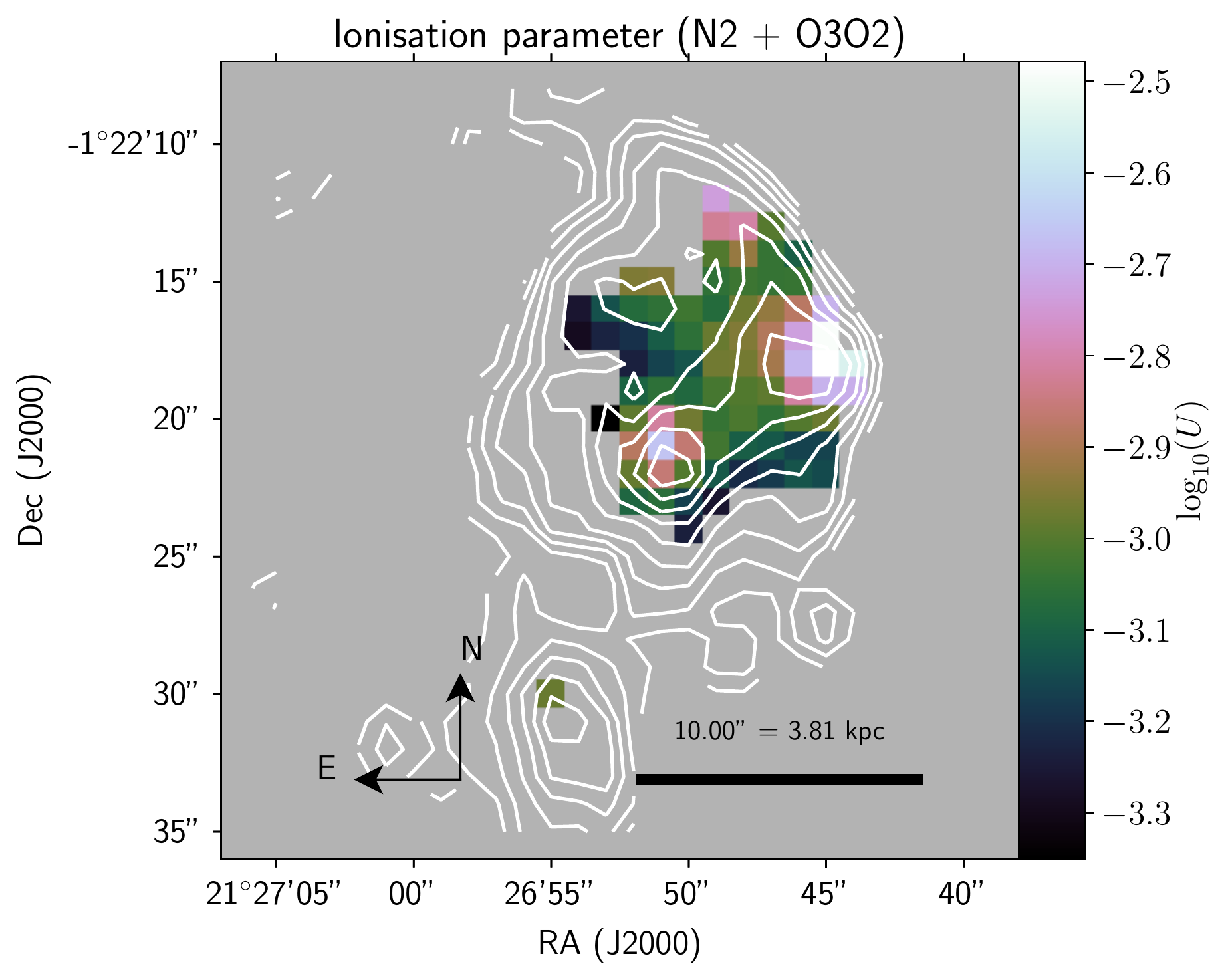}
	\includegraphics[height=0.23\textheight]{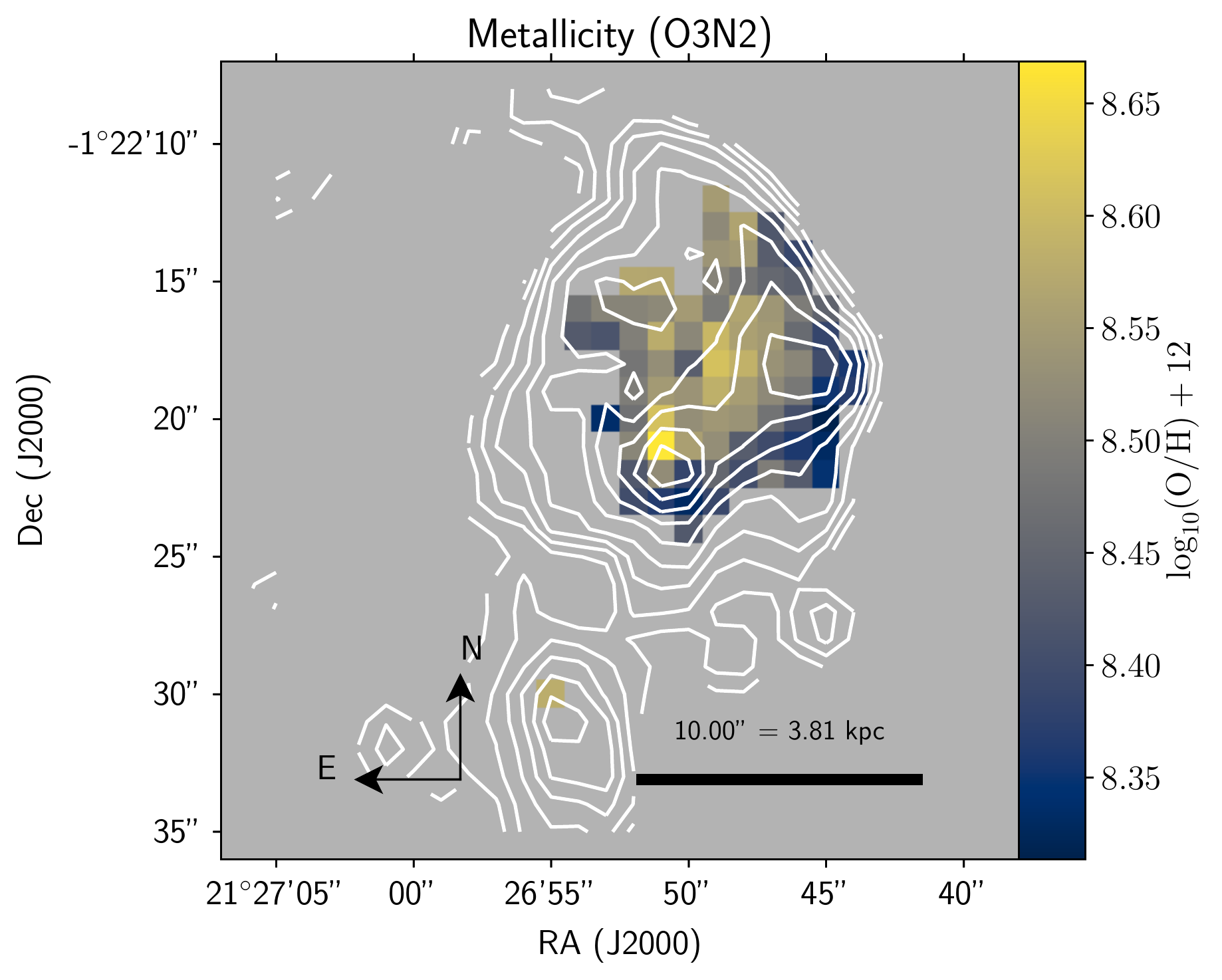}
	\includegraphics[height=0.23\textheight]{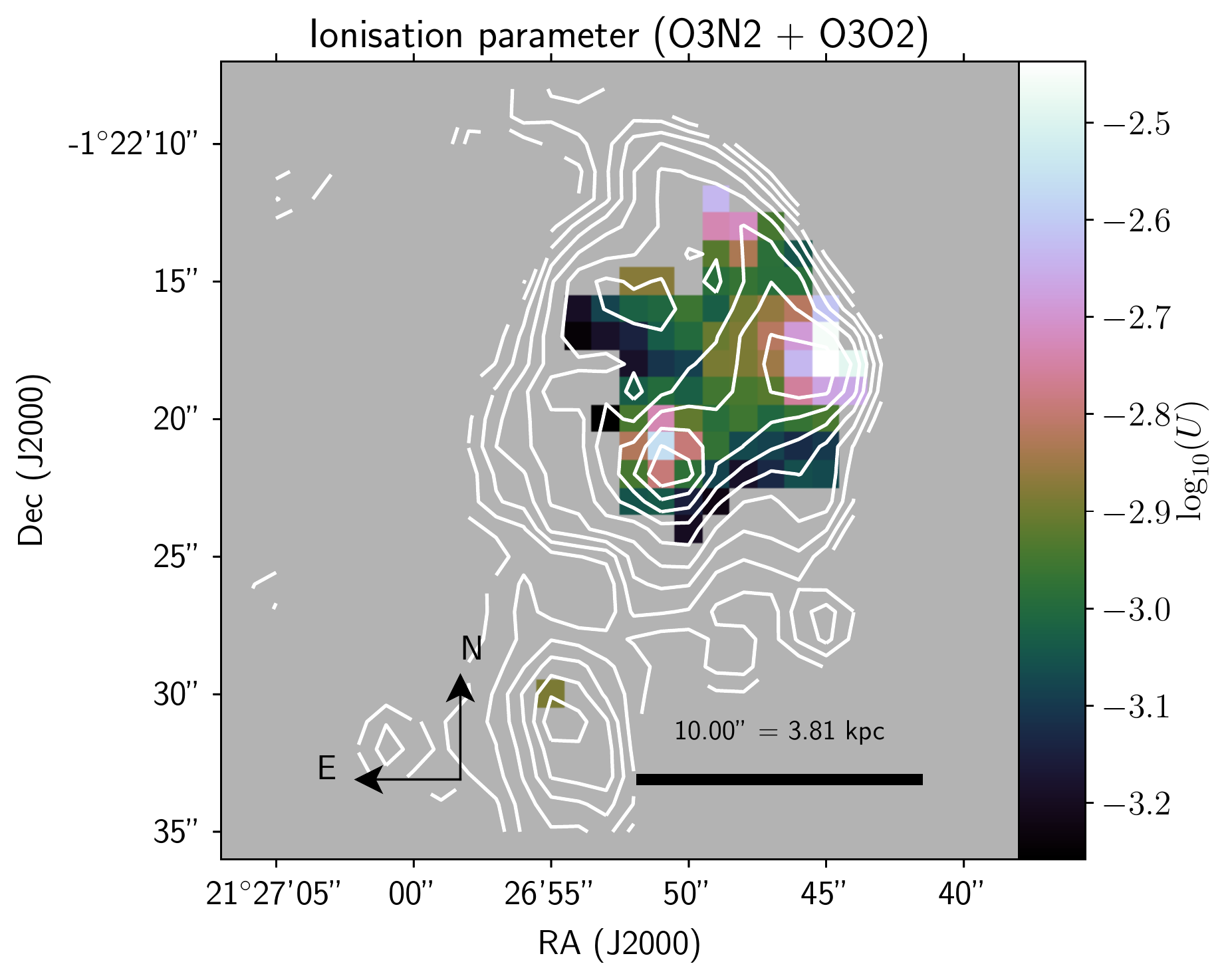}
	\caption{The left column shows the gas-phase metallicity $\logoh{}$, calculated using the O2S2 (\forb{O}{ii}$\uplambda \uplambda 3726,3729$/\forb{S}{ii}$\uplambda \uplambda 6716,6731$), R23 ([\forb{O}{ii}$\uplambda \uplambda 3726,3729$ + \forb{O}{iii}$\uplambda \uplambda 4959,5007$]/\hb{}), N2 (\forb{N}{ii}$\uplambda 6583$/\ha{}) and O3N2 ([\forb{O}{iii}$\uplambda 5007$/\hb{}]/[\forb{N}{ii}$\uplambda 6583$/\ha{}]) diagnostics, and dimensionless ionisation parameter $\logu{}$, calculated using the O3O2 diagnostic. The white contours show the \ha{} flux on a log scale.}
	\label{fig: A3: metallicity & U}
\end{figure*}

Fig.~\ref{fig: A3: metallicity & U} shows the gas-phase metallicity (left column) and ionisation parameter (right column) estimated using the O2S2, R23, N2 and O3N2 metallicity diagnostics and the O3O2 ionisation parameter diagnostic of \citet{KewleyNichollsSutherland2019}.

\section{Estimation of shock velocities}
\label{appendix: Estimating the shock velocities in MO}

Because the jet impacting MO is an FR\,I jet, the usual approximations derived from a hypersonic or supersonic jet impacting a relatively dense medium \citep[e.g.][]{Blandford&Rees1974} are not appropriate. In this appendix, we develop an estimate for the cloud shock velocity resulting from the impact of a transonic jet on a relatively dense cloud.

\begin{figure}
	\begin{center}
		\includegraphics[width=0.4\textwidth]{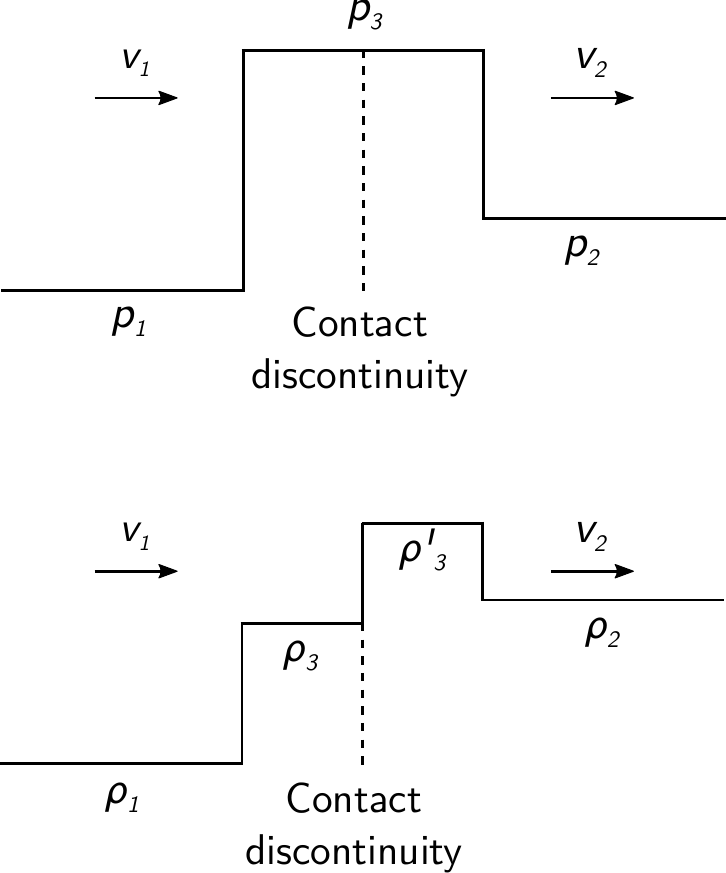}
		\caption{Configuration of pressure (top) and density (bottom) for a one-dimensional shock tube, following \citet{Landau1987}. Region 1 corresponds to the jet and region 2 to the ISM impacted by the jet. In the calculations presented here, we assume that $p_1 = p_2$.}
		\label{fig: A4: shock tube}
	\end{center}
\end{figure}

The dynamical setting, which is a classic shock tube configuration \citep[see][\S 100]{Landau1987} is indicated in Figure~\ref{fig: A4: shock tube}. The jet pressure and density are $p_1$ and $\rho_1$, and the cloud pressure and density are $p_2$ and $\rho_2$. The shock pressure for both the jet terminus and the shocked cloud is $p_3$, the shocked jet density is $\rho_3$ and the shocked cloud density is $\rho_3^\prime$. In the following we assume that, initially, both the cloud and jet are in pressure equilibrium with the interstellar medium so that $p_1 \approx p_2$. The pressure difference $p_3-p_1$ is denoted by $\Delta p$; we cannot assume that $\Delta p \gg p_1$. The polytropic indices for jet and cloud are $\gamma_1$ and $\gamma_2$ respectively, and the velocities of the jet and cloud in the frame of the contact discontinuity are $v_1$ and $v_2$, respectively.

The frame-independent relative velocity between jet and cloud is 
\begin{equation}
\Delta v = v_{\rm jet} + v_{\rm infall}
\end{equation}
where $v_{\rm jet}$ and $v_{\rm infall}$ are respectively the jet velocity and infall velocity of MO in the frame of the radio galaxy NGC\,541.
For the shock tube analysis:
\begin{equation}
\Delta v = v_1 - v_2
\end{equation}
where $v_1$ and $v_2$ are respectively the jet velocity and cloud velocity in the frame of the contact discontinuity separating the jet and cloud gas. We have:
\begin{eqnarray}
\Delta v = v_1 - v_2 = && \left( \frac {2}{\rho_1} \right)^{1/2} \>
\frac {p_3-p_1}{\left[ (\gamma_1-1) p_1 + (\gamma_1+1)p_3 \right]^{1/2}} 
\nonumber \\
+ && \left( \frac {2}{\rho_2} \right)^{1/2} \>
\frac {p_3-p_2}{\left[ (\gamma_2-1) p_2 + (\gamma_2+1)p_3 \right]^{1/2}}
\end{eqnarray}
Because the cloud density $\rho_2$ is much greater than the jet density $\rho_1$, we neglect the second term in this expression and write:
\begin{equation}
\Delta v = \left( \frac {2}{\rho_1} \right)^{1/2} \>
\frac {\Delta p}{\left[ 2 \gamma_1 p_1 + (\gamma_1+1) \Delta p \right]^{1/2}} 
\label{e:delta_v}
\end{equation}
Defining the Mach number, corresponding to $\Delta v$, by $M_1 = (\gamma_1 p_1/\rho_1)^{-1/2} \Delta v$ the solution of Eqn.~\ref{e:delta_v} for $\Delta p$ is
\begin{equation}
\frac {\Delta p}{p_1} = \frac {\gamma_1 (\gamma_1+1)}{2} M_1^2 f(M_1) 
\label{delta_p}
\end{equation}
where 
\begin{eqnarray}
f(M_1) &=& \frac {1}{2} \left\lbrace 1 + \left[ 1 + \frac {16}{(\gamma_1+1)^2} M_1^{-2}  \right]^{1/2}\right\rbrace \label{f(M1)}\\
&\rightarrow& 1 \quad \hbox{for} M_1 \gg 1
\end{eqnarray}

It follows from the Rankine Hugoniot relations that the velocity of the shock propagating from region~$3^\prime$ to region~2 (i.e. the cloud shock) is given by:
\begin{equation}
v_{\rm sh,cl}^2 = \frac {1}{2 \rho_2} \left[  (\gamma_2-1)p_2 + (\gamma_2+1) p_3 \right] 
\end{equation}
and for $p_1=p_2$ 
\begin{equation}
v_{\rm sh}^2 = c_{\rm s,2}^2\left[ 1 + \frac{(\gamma_2+1)}{2 \gamma_2} \frac{\Delta p}{p_1} \right]
\end{equation}
where $c_{\rm s,2}$ is the adiabatic sound speed in the cloud. 

Using Eqn.~\ref{delta_p} for $\Delta p$, we have for the velocity of the cloud shock
\begin{equation}
v_{\rm sh,cl} = 
c_{\rm s,2} M_{\rm sh,cl} 
\label{v_shock}
\end{equation}
where the shock Mach number 
\begin{equation}
M_{\rm sh,cl} = \left[ 1 + \frac {\gamma_1 (\gamma_1 +1)(\gamma_2+1)}{4 \gamma_2}  M_1^2 f(M_1) \right]^{1/2}
\end{equation}
and $f(M_1)$ is given by Eqn.~\ref{f(M1)} above. $M_{\rm sh,cl}$ is always greater than unity, but 
$\sim 1$, for low $M_1$.

In terms of the temperature, $T_2$, of the cloud, the shock velocity is 
\begin{eqnarray}
v_{\rm sh, cl}&=& \left( \frac {k T_2}{\mu m} \right)^{1/2}
M_{\rm sh,cl} \\
&\approx& 15 \> \left(\frac {T_2}{10^4}\right)^{1/2}
M_{\rm sh,cl} \quad \rm km \> s^{-1}
\end{eqnarray}
where $\mu \approx 0.62$ is the molecular weight and 
$m \approx 1.66 \times 10^{-24} \rm \> gm$ represents an atomic mass unit.

For jet Mach numbers $=(1,2,3)$ the values of $M_{\rm sh, cl} \approx (1.7, 2.6, 3.6)$. Hence, attributing the velocity dispersion in Minkowski's Object to a population of shocks, driven by a transonic ($M_1 \sim 1-2$) jet, it is apparent that a velocity dispersion $\leqslant 20 \> \rm km \> s^{-1}$ would arise for cloud temperatures $\sim 10^3 - 10^4 \> \rm K$ and a jet Mach number $\sim 1-2$.

% My text
The velocity dispersion resulting from such slow shocks are consistent with those measured in the \hi{} cloud associated with MO~\citepalias{Lacy2017}. Although they are also consistent with those observed in the ionised gas (see Fig.~\ref{fig: paper 3: Halpha kinematics and morphology}), we note that the resultant velocity dispersions are similar to those of \hii{} regions where the line widths are driven by internal processes~\citep[e.g.,][]{Rozas2006}. Hence we cannot attribute the observed line widths in our WiFeS observations to these shocks, although it does indicate that the jet does not significantly increase the turbulent velocities in the ionised gas.

% Don't change these lines
\bsp	% typesetting comment
\label{lastpage}
\end{document}